\documentclass[11pt,a4paper]{quantumarticle}
\pdfoutput=1

\usepackage{color}
\usepackage{graphicx}
\usepackage{wrapfig,enumerate,slashed}
\usepackage[utf8]{inputenc}
\usepackage{adjustbox}

\usepackage[english]{babel}
\usepackage[T1]{fontenc}
\usepackage{amsmath}
\usepackage{hyperref}

\usepackage{appendix}
\usepackage{placeins}

\usepackage{footmisc}
\usepackage{amsmath}
\usepackage{wasysym} 
\usepackage{graphicx}
\usepackage{color}
\usepackage{comment}
\usepackage{hyperref}
\usepackage{subcaption}
\usepackage{slashed}
\usepackage{booktabs}

\newcommand{\be}{\begin{eqnarray}}
\newcommand{\ee}{\end{eqnarray}}

\newcommand{\bee}{\begin{eqnarray}}
\newcommand{\eee}{\end{eqnarray}}
\newcommand{\beeq}{\begin{equation}}
\newcommand{\eeeq}{\end{equation}}

\usepackage{listings}
\usepackage{color}

\definecolor{dkgreen}{rgb}{0,0.6,0}
\definecolor{gray}{rgb}{0.5,0.5,0.5}
\definecolor{mauve}{rgb}{0.58,0,0.82}

\begin{document}

\title{Quantum Optimisation of Complex Systems with a Quantum Annealer}
%\preprint{}

\author{Steve Abel}
\email{steve.abel@durham.ac.uk}
\affiliation{Institute for Particle Physics Phenomenology, Durham University, Durham DH1 3LE, UK}
\affiliation{Department of Mathematical Sciences, Durham University, Durham DH1 3LE, UK}

\author{Andrew Blance}
\email{andrew.t.blance@durham.ac.uk}
\affiliation{Institute for Particle Physics Phenomenology, Durham University, Durham DH1 3LE, UK}
\affiliation{Department of Physics,  Durham University, Durham DH1 3LE, UK}
\affiliation{Institute for Data Science, Durham University, Durham, DH1 3LE, UK}

\author{Michael Spannowsky}
\affiliation{Institute for Particle Physics Phenomenology, Durham University, Durham DH1 3LE, UK}
\affiliation{Department of Physics, Durham University, Durham DH1 3LE, UK}
\email{michael.spannowsky@durham.ac.uk}

\maketitle

\begin{abstract}
{\small
We perform an in-depth comparison of  quantum annealing with several classical optimisation techniques, namely thermal annealing, Nelder-Mead, and gradient descent. 
We begin with a direct study of the 2D Ising model on a quantum annealer, and compare its properties directly with those of the thermal 2D Ising model. These properties  include 
an Ising-like phase transition that can be induced by either a change in ``quantum-ness'' of the theory (by way of the transverse field component on the annealer), or by a scaling 
the Ising couplings up or down. This behaviour is in accord with what is expected from the physical understanding of the quantum system. We then go on to demonstrate the 
efficacy of the quantum annealer at minimising several increasingly hard two dimensional potentials. For all the potentials we find the general behaviour that Nelder-Mead and gradient descent methods are 
very susceptible to becoming trapped in false minima, while the thermal anneal method is somewhat better at discovering the true minimum.  However, and despite current limitations on its 
size, the quantum annealer performs a minimisation very markedly better than any of these classical techniques. A quantum anneal can be designed so that the system almost never 
gets trapped in a false minimum, and rapidly and successfully minimises the potentials.  }
\end{abstract}

\flushbottom

%%%%%%%%%%%%%%%%%%%%%%%%%%%%%%%%%%%%%%%%%%%%%%%

\section{\label{Sec:Intro}Introduction}

Finding the ground state of a complex system is a task of utmost importance in many research fields. In biology for example, protein folding, namely the dynamical process whereby a protein chain folds into its characteristic three-dimensional structure is a well-known phenomenon. This unique structure corresponds to the energetic ground state of its configuration space \cite{Levinthal}. In chemistry and drug design, bindings between molecules result in an energy potential that leads the compounds to self-assemble in a specific configuration into their energy ground states \cite{Aspuru,Lanyon,elfving2020quantum}. In finance, quantitative optimisation of portfolios can yield increased profitability \cite{fintech:1,fintech:2,fintech:3}. And in physics, the scope of applications ranges from complex systems in classical mechanics \cite{vanBrunt}, through quantum mechanics \cite{fock,macdonald, white,drake94} and condensed matter systems \cite{bogdanov94,bogdanov06,segall, Buhrandt:2013uma,Schenk:2020lea} all the way to models in string theory \cite{Becker_2002, Kachru:2003aw, Balasubramanian_2004, Balasubramanian_2005, AbdusSalam:2020ywo} and high-energy physics \cite{Bardin:1992jc, Lafaye:2004cn, Flacher:2008zq,Criado:2020zwu,Balazs:2021uhg}. Further, broad classes of mathematical problems, for example finding the solution to a differential equation \cite{Raissi18,Piscopo:2019txs,Araz:2021hpx}, can by variational methods be rephrased as an optimisation task that ultimately corresponds to finding the extremum of a complicated system.

Due to the importance and prevalence of optimisation problems, many methods have been devised to find the extremum of a system. Such methods include various sampling algorithms for discretised or latticised systems \cite{Metropolis1953EquationOS}, optimisation algorithms for continuous systems \cite{10.1093/comjnl/7.4.308,Hestenes1952MethodsOC, 10.5555/865018} and, in recent years, machine learning algorithms \cite{rumelhart}, i.e. self-adaptive neural networks. See \cite{Balazs:2021uhg} for a review of several popular classical optimisation algorithms.

However, the performance of any such optimisation algorithm is characterised by the speed at which it can reliably locate the global extremum of the problem. Often, even {\it a posteriori}, it is impossible to assess whether an optimisation algorithm has actually managed to find the global optimum or whether it only settled in a local extremum. Surveying the entire configuration space of a problem can quickly become prohibitive and for NP-hard problems the difficulty typically increases exponentially with dimensionality. Some confidence can sometimes be regained by repeating the search from different starting conditions, but even then, classical optimisation algorithms are often overwhelmed by the local structure of the configuration space, such that it is hard to prevent them falling into the ``domain of attraction'' of a nearby local minimum. 

With the advent of powerful near-term quantum devices and quantum algorithms, it is natural to ask if these devices could provide a qualitatively different solution to the problem of finding the ground state of a complex system, specifically if they provide a novel avenue to find the global minimum of a system reliably and quickly. In fact, finding the ground state of a complex system was one of the first highly anticipated applications of quantum computers \cite{feynman1982}. 
While several different quantum computing paradigms have been proposed, 
one type of quantum computer, even though non-universal, has been designed with the very task in mind of minimising  a potential: namely a quantum annealer \cite{finilla94a,kadowaki98a,brooke99a,dickson13a,lanting14a,albash15a,albash16a,boixo16a,chancellor16b,Benedetti16a,Muthukrishnan2016,cervera}. 

The purpose of this paper is to demonstrate the qualitative difference between classical and quantum optimisation algorithms. Our focus is on quantum annealers, but we should remark that   our discussion remains valid for other quantum computing paradigms, such as gate quantum computers. However at the time of writing accessible quantum annealers, for example the devices provided by D-Wave Systems \cite{LantingAQC2017}, have the significant advantage over quantum gate computers that they offer systems with several thousand qubits \cite{dattani2019pegasus}.

The study we present is relatively straightforward, but due to its novel nature, and in order to provide arguments in support of the procedures we follow, our discussion will be relatively methodical. Thus we begin in Sec.~\ref{Sec:QA} by first comparing quantum annealers to their close thermal annealing cousins. We discus how both kinds of annealer sample the configuration space, and we outline the different ways that these systems find the energetic ground state. We then apply both methods to solve the 2-D Ising model in Sec.~\ref{Sec:Ising}. We choose the 2-D Ising model as the simplest non-trivial example of a latticised system that has a vast configuration space, scaling as $2^N$ with $N$ spins, and a highly degenerate energy landscape. The Ising model is of course of interest in and of itself, and we provide a novel analysis of the physics of its quantum incarnation. 

To go beyond the basic Ising model, we use the domain wall encoding (DWE). Using DWE one can encode a lattice approximation of a continuous function, as well as perform operations on such functions by employing the finite difference method. This extends the applicability of quantum annealers to continuous systems. In Sec.~\ref{Sec:Function} we first recap DWE, which we use for the quantum and thermal annealers.  We then study three example functions, in order to make a quantitative comparison between optimisation using the Gradient Descent Method (\ref{appendix:gdm}), Nelder-Mead Method (\ref{appendix:nmm}), Thermal Annealing (\ref{appendix:tam}) and Quantum Annealing algorithms. We find a clear qualitative and quantitative difference in the performance of quantum versus classical algorithms in solving such optimisation problems. We offer a summary and conclusions in Sec.~\ref{Sec:conc}.
\vspace{0.5cm}

\section{\label{Sec:QA}Quantum Annealing and the Ising Model Encoding}

\subsection{Quantum versus Classical Annealers}

The device that we will be studying in here is the quantum annealer \cite{farhi00a,neven2009nips}, a system of 
linked qubits with adjustable couplings.  Such devices made available by D-Wave \cite{LantingAQC2017} have been able successfully to simulate condensed matter systems, sometimes showing advantages over classical counterparts \cite{Kadowaki_1998,Heim,Harris:2018Science,King:2018Nature,King:2019arXiv,Albash_2021,oshiyama2021classical}.

Quantum annealers operate in a dissipative rather than fully coherent way. This means that they are not useful 
for studying for example interference properties that depend on coherence, but can in principle be very effective for studying long timescale quantum processes, for example those in which quantum tunnelling allows barrier penetration that would classically be forbidden \cite{Abel:2020qzm}. This makes them ideal for finding the energetic minimum in systems with 
complicated energetic landscapes and many barriers and minima. 

 The basic characteristics of the quantum annealer are embodied in its Hamiltonian, which takes the form of a 
 generalised Ising model:
\begin{align}
\mathcal{H}~=&~\,B(s)\left(\sum_{ij} J_{ij}\sigma_{i}^{z}\sigma_{j}^{z}~+~\sum_{i} h_{i}\sigma_{i}^{z}\right) \label{eq:ising}
\\ &\qquad\qquad\qquad\qquad\qquad ~+~A(s)\sum_{i}\sigma_{i}^{x}~,\nonumber 
\end{align}
where $i,j$ label the qubits, $\sigma_{i}^{z}$ are the $z-$spin
Pauli matrices, and $\sigma_{i}^{x}$ are the transverse field components,
while the couplings $h_{i}$ and $J_{ij}$ between the qubits are
set and kept constant. 

\begin{figure}
\centering{}\includegraphics[scale=0.3]{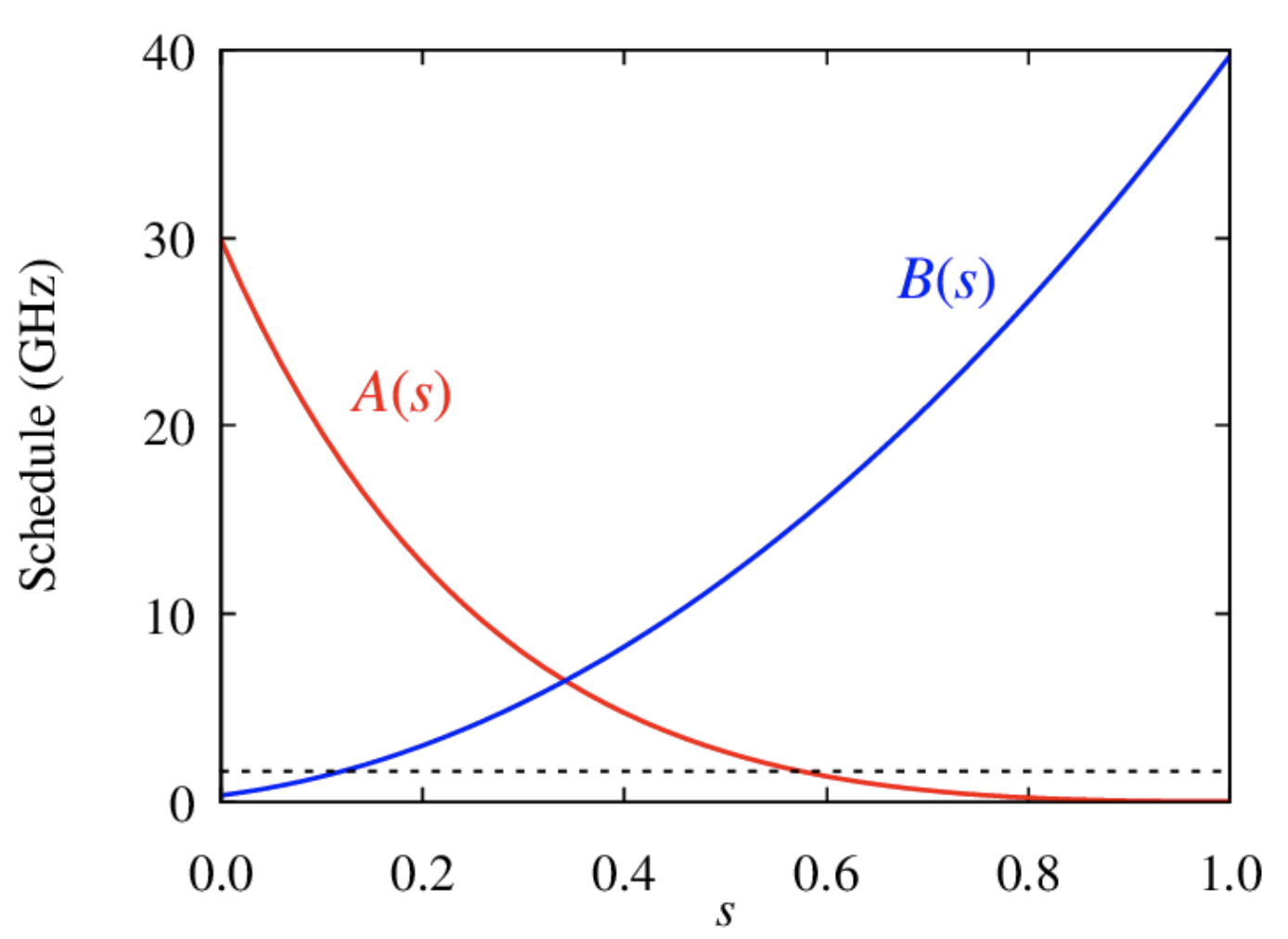}\caption{Anneal schedule parameters. The thermal contribution is shown as a
solid line, while $A$ and $B$ are the coefficients scaling the  transverse field  and classical
Ising contributions respectively.  \label{fig:Anneal-schedule-parameters}}
\end{figure}

The parameter $s(t)$ (with $t$ being time)
is a user-defined control-parameter that can be adjusted, while $A(s)$ and $B(s)$ describe the consequent change in the quantum
characteristics of the annealer. As shown in Fig.\ref{fig:Anneal-schedule-parameters}, smaller $s\in [0,1]$ means larger  
transverse field parameter $A$ compared to $B$, which induces more ``hopping'' of $\sigma^z$ spins, which 
overall means a system that is more characteristically ``quantum''. 

To perform the task
of finding a global optimisation, the first objective is to encode the problem to
be solved into the ``classical'' Ising model Hamiltonian represented by the $B$-terms, such that the 
energetic minimum would correspond to the desired solution.
One then adjusts $s$ to alter the relative sizes of the parameters $A,~B$ to perform
a so-called anneal, in the hope that the systems ends up in the global minimum. 

\begin{figure}[tbh]
	\centering
	\begin{subfigure}{0.95\linewidth} \centering
		\includegraphics[width=\textwidth]{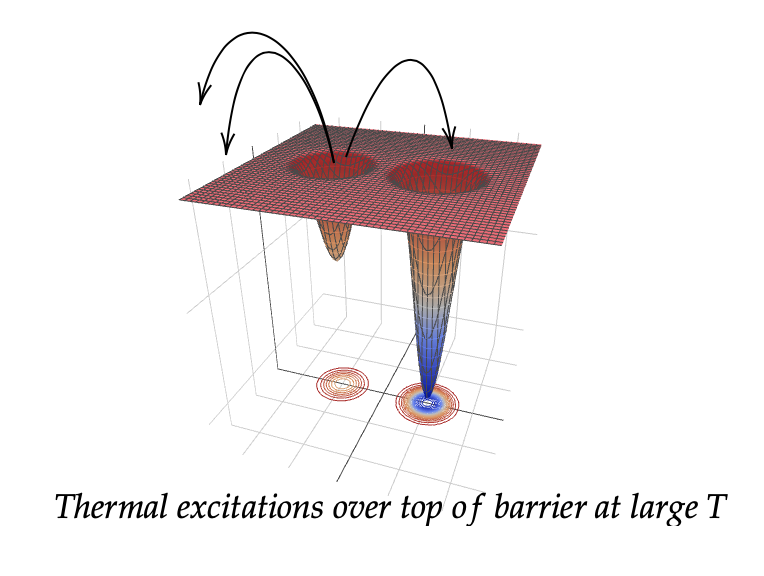}
	\end{subfigure}
	\begin{subfigure}{0.95\linewidth} \centering
		\includegraphics[width=\textwidth]{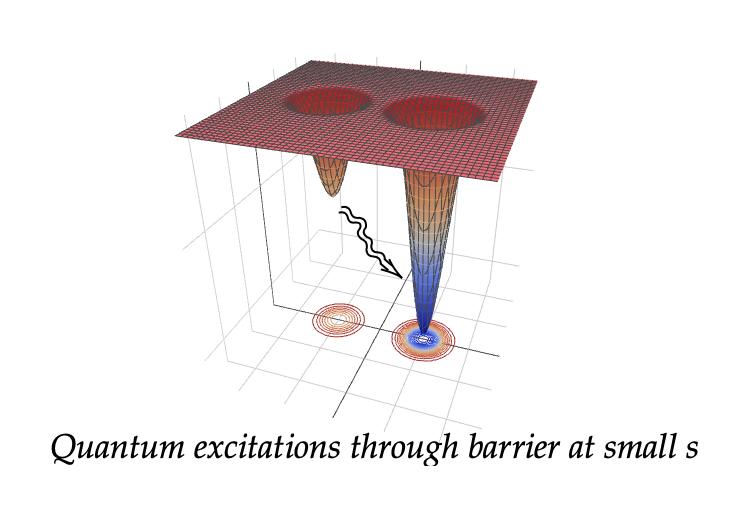}
	\end{subfigure}
	\caption{{Cartoon of thermal versus quantum annealing}\label{fig:generic}}
\end{figure}

Indeed although we will be comparing quantum optimisation with several classical optimisation methods, which as we said are summarised in Appendix \ref{appendix:classical_opt}, 
it is really thermal annealing which is the closest classical equivalent. For example both classical and quantum annealers  are typically based on the kind of Ising spin model displayed in Eq.~(\ref{eq:ising}). 
In the two panels of Fig.\ref{fig:generic} we show schematically  the different optimisation strategies in a system with two competing minima that has been encoded on such an Ising model. The strategy for thermal annealing 
with the Metropolis algorithm, see appendix \ref{appendix:tam}, is based on Boltzmann weighted excitation and is shown in the first panel of Fig.~\ref{fig:generic}. The generic approach is to perform many trials 
that begin with a raised temperature $T$ which is gradually reduced, with the probability of energetically unfavourable transitions being suppressed by a factor $e^{-\Delta E/kT}$ where $\Delta E $ is the would-be gain in energy. The finessing of the temperature as a function of time, i.e. the anneal schedule, is a crucial factor in the 
success or otherwise of the search strategy. Clearly the initial temperature should be high enough that the system can clear the barriers. It must then be cooled slowly enough so as to avoid the system  freezing into a suboptimal state. We can also appreciate from the figure what can go wrong with a thermal anneal. Aside from the possibility of getting stuck in a local minimum, a very tall thin barrier surrounding the global minimum is clearly hard for this technique to overcome. Moreover high dimensionality is also likely to be
a detrimental factor, as the excited state will have a large phase space to explore, and may never return to the domain of attraction of the global minimum. 

The second panel of Fig.~\ref{fig:generic} illustrates the contrasting  behaviour and search strategy for a quantum annealer. 
In quantum annealing it is the parameter $s(t)$ that is adjusted to perform the anneal. 
The anneal schedule typically begins by taking the system to a highly quantum state, i.e. low $s$. 
This allows the system to tunnel to the global minimum but its wavefunction will be very wide, so precision will at that point be low. (Note that we are here thinking of the wavefunction in the potential encoded in the Ising $\sigma^z$ spins, so the wavefunction is ``wide'' in the $Z$ direction: it is really the quantum hopping induced by the transverse $\sigma^x$ term that is dialling the quantumness in the Ising model up or down.) 

 Then by increasing $s$ in the anneal schedule, the system will  remain trapped in the deeper minimum with the wavefunction becoming narrower as the system becomes ``more classical'', and the precision increases. As in the thermal case the solution is found by performing many trials. In the case of a quantum annealer the tunnelling is very rapid when the barriers are thin and the wells are deep, but becomes less efficient when the barrier is wide compared to its depth. However it is also somewhat less affected by the dimensionality of the problem. Barrier penetration goes more or less in a straight line between the two minima shown (although the rate does have some dimension dependence). In addition it is less easy for the system to get lost in the phase space, for example if the flat directions surrounded the minima in Fig.~\ref{fig:generic} were to become large. 

We can appreciate this different behaviour from that fact that the rate at which quantum tunnelling takes place is roughly proportional to the instanton tunneling factor 
$e^{-\omega \sqrt{2m\Delta E}/\hbar}$, where here $\omega$ is a measure of the width of the barrier, $\Delta E$ is its height, and the parameters $m$ and $\hbar$ in this rate are implicit in the Schr\"odinger equation and are difficult to determine in the present context. 
Note the different expected response to tall, thin barriers: namely   the exponent increases as only by the square-root of the height, whereas the wide barriers are in principle more of an obstruction.  (Although as we shall see, not much in practice). 
A particular aspect of this rate that will be important is that the 
{\it height} of the barriers, $\Delta E$, clearly scales with the couplings in ${\cal H}$. Thus the parameter that corresponds most closely to the temperature  $T$ in the Metropolis algorithm is actually not $s$ itself but rather the square root of the typical inverse $J_{ij}$ coupling: it will be useful therefore to introduce an overall scaling of ${\cal H}$ by a factor $\lambda$, and then to compare the behaviour of the thermal system at different $T$ with the behaviour of the quantum system at different $\lambda$ but at constant $s$, with the expectation that in that case 
\be
T~\sim~ 1/\sqrt{\lambda}~.  
\ee
Note that by Eq.~(\ref{eq:ising}) this overall coupling factor is really just another way of changing the relative values of $A$ and $B$, but it has the distinct advantage that the relation to the barrier heights is better understood, and also that the barriers and the `quantumness'' of the system are not being simultaneously adjusted, so there is more control. In this sense the coupling and $s$ are not really separate parameters because $B(s)$ is itself changing as we vary $s$. For the quantum system there are then two distinct kinds of strategies: studying the behaviour of the system at constant $s$ for different values of  $\lambda $ provides the most direct comparison of the physical properties of the quantum and the thermal Ising models, whereas for the more practical purpose of performing an optimisation, the typical strategy would be to set $\lambda$ to a constant value and implement a suitable anneal schedule $s(t)$. In the next section we shall consider both possibilities: indeed as we shall see the former provides a guide for how best  to perform the latter.
\FloatBarrier 

\section{\label{Sec:Ising} Solving the 2-D Ising Model}

\label{sec:solving} 

Given the similar Ising encodings used on quantum and thermal annealers, the first interesting analysis that can be made on these two particular systems  
is simply to probe the phase diagram of the basic 2-D Ising model in each case. This can be thought of as a way to understand and calibrate the behaviours of the 
 two systems, in order to guide the design of the anneal schedule for more complicated problems. However the 2-D Ising model is of course a well studied statistical mechanical  system in its own right, and in the thermal case it famously displays a characteristic phase transition (which partly explains its ability to solve Ising encoded problems). There has been surprisingly little equivalent analysis of the 2-D Ising model on the quantum annealer. In this section we perform such a study of the behaviour of the quantum Ising model, in order to compare and contrast it with the thermal one. 

Let us begin by defining the standard Ising 2-D model. This model is essentially the Hamiltonian in Eq.~(\ref{eq:ising}) with only adjacent couplings, which are degenerate and negative (i.e. ferromagnetic). We can implement this model by defining a 2-D $N\times N$ grid, with grid positions $(\hat{i} , \hat{j}) \in (1\ldots N,~1\ldots N)$ corresponding to $N^2$ qubits labelled $i\in (1\ldots N^2) $ as follows: 
$$i~=~ \hat{i} N+\hat{j}~. $$The negative coupling between adjacent qubits in the grid then corresponds to 
\begin{align} 
J_{\hat{i} N+\hat{j}, \hat{k} N+\hat{\ell}}~&=~- \lambda 
\left( \delta_{\hat{i},\hat{k}-1}\delta_{\hat{j},\hat{\ell}}\,+\,\delta_{\hat{i},\hat{k}+1}\delta_{\hat{j},\hat{\ell}} \right.\\
  & \qquad   \qquad\left. +\,\,\,\,  \delta_{\hat{i},\hat{k}} \delta_{\hat{j},\hat{\ell}-1}\,+\, \delta_{\hat{i},\hat{k}}\delta_{\hat{j},\hat{\ell}+1}\right) \, , \nonumber
\end{align}
where  $\delta_{\hat i\hat j}$ is the Kronecker-$\delta$.  (As each coupling provides a 
negative contribution if opposite spins are adjacent, it can also be thought of as an encoding that provides a solution to the following discrete and somewhat trivial problem: given two different possible colours, how can one fill an $N\times N$ grid so that there are as few differently coloured adjacent squares as possible.)
\begin{figure*}[tbh]
	\centering
	\begin{subfigure}{0.49\linewidth} \centering
		\includegraphics[width=\textwidth]{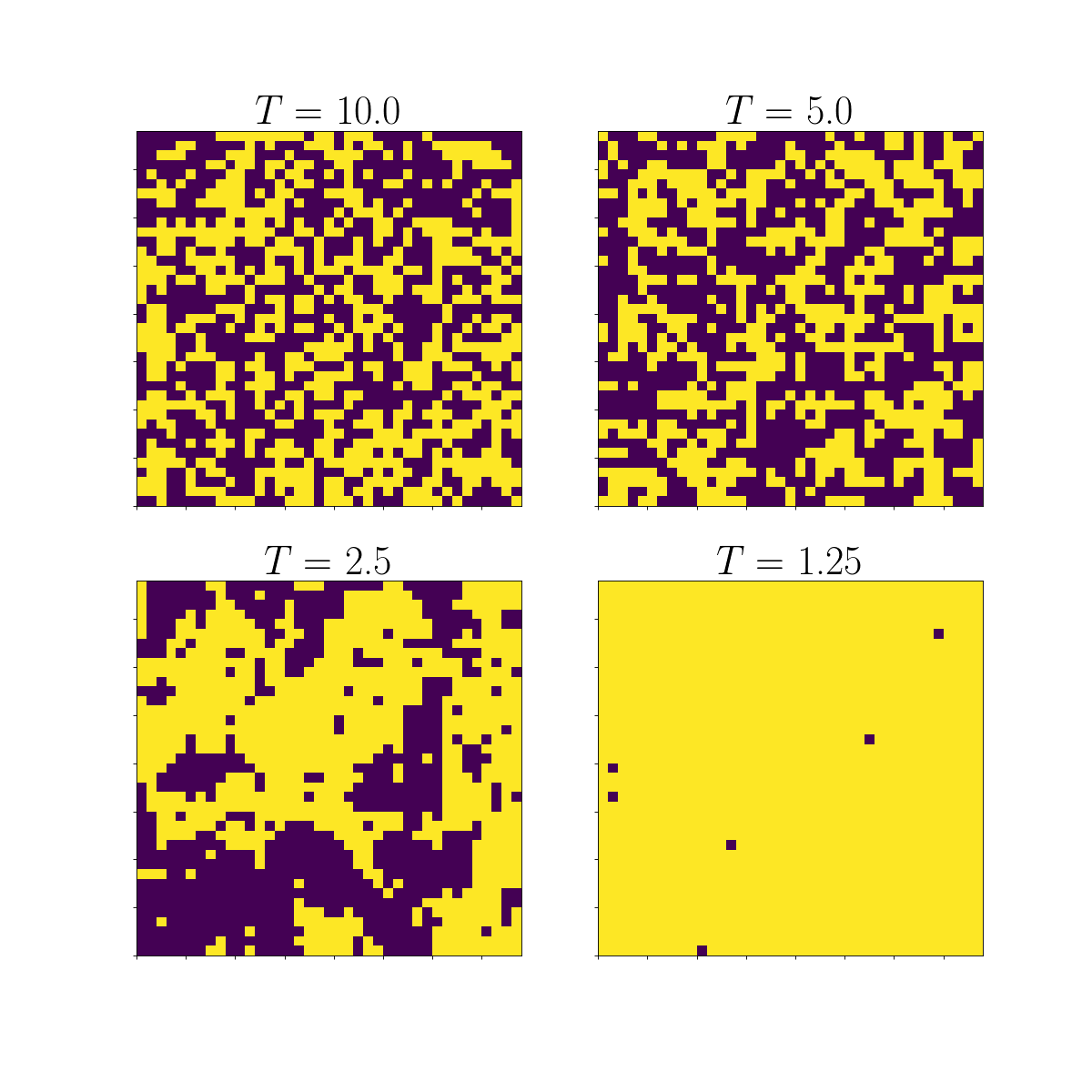}
		\caption{Ising model at different temperatures \label{fig:thermalising}}
	\end{subfigure}
	\begin{subfigure}{0.49\linewidth} \centering
		\includegraphics[width=\textwidth]{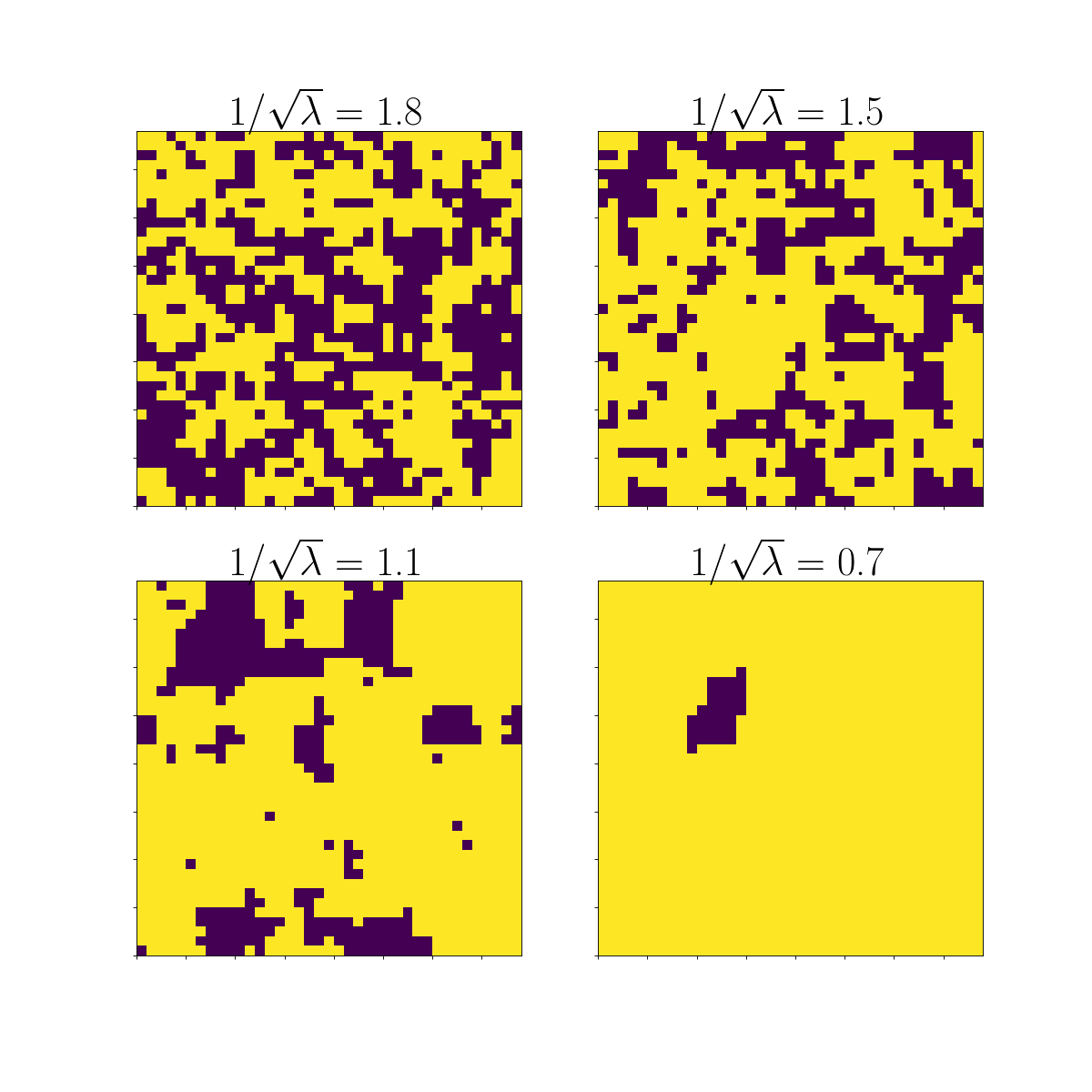}
		\caption{Quantum Ising model at different couplings \label{fig:Qthermalising}}
	\end{subfigure}
	\caption{Comparison of quantum versus classical annealers. Here we take $s=0.2$ for all the quantum Ising cases, with an anneal time of  $100\mu {\rm s}$. In order to produce these plots the coupling is kept constant but the anneal schedule is set to turn on the $s=0.2$ quantumness and then ramp it down as rapidly as possible (i.e. in $0.8\mu {\rm s}$), in order to take a snapshot of the system.}
	\label{fig:Qsched}
\end{figure*}

We begin by simply observing how the two systems behave, in Fig.~\ref{fig:Qsched}. As explained in the introduction the parameter in the quantum annealer that corresponds to temperature $T$ of the thermal annealer is $1/\sqrt{\lambda}$. By eye one can see that the behaviours of the two systems on varying these parameters are remarkably similar. 

We of course wish to quantify this by as usual studying and comparing how the energy $\eta$  and magnetisation ${\cal M}$  of the Ising models varies with $T$ and $1/\sqrt{\lambda}$. As $\lambda$ is our variable for the quantum system, we must for comparison remove it from the Hamiltonian by defining these quantities as follows:
\begin{align}
\eta \,&=~ \frac{1}{\lambda N^2} \sum_{i,j}^{N^2} J_{ij} \sigma_i^z\sigma_j^z \nonumber \\
{\cal M} \,&=~  N^{-2}\left\langle \left| \sum_{i=1}^{N^2} \sigma_i ^z \right|  \right\rangle  ~,
\end{align}
where they are defined as a density normalised per qubit (c.f. volume). Note that with these normalisations the energy is $0\leq \eta \leq-1$, with $-1$ corresponding to the perfect solution (i.e. with all spins aligned), while the magnetisation is correspondingly $0\leq {\cal M} \leq 1$, with ${\cal M}=1$  for perfectly aligned spins.

\begin{figure*}[htb]
	\centering
	\begin{subfigure}{0.4\linewidth} 
		\includegraphics[width=\textwidth]{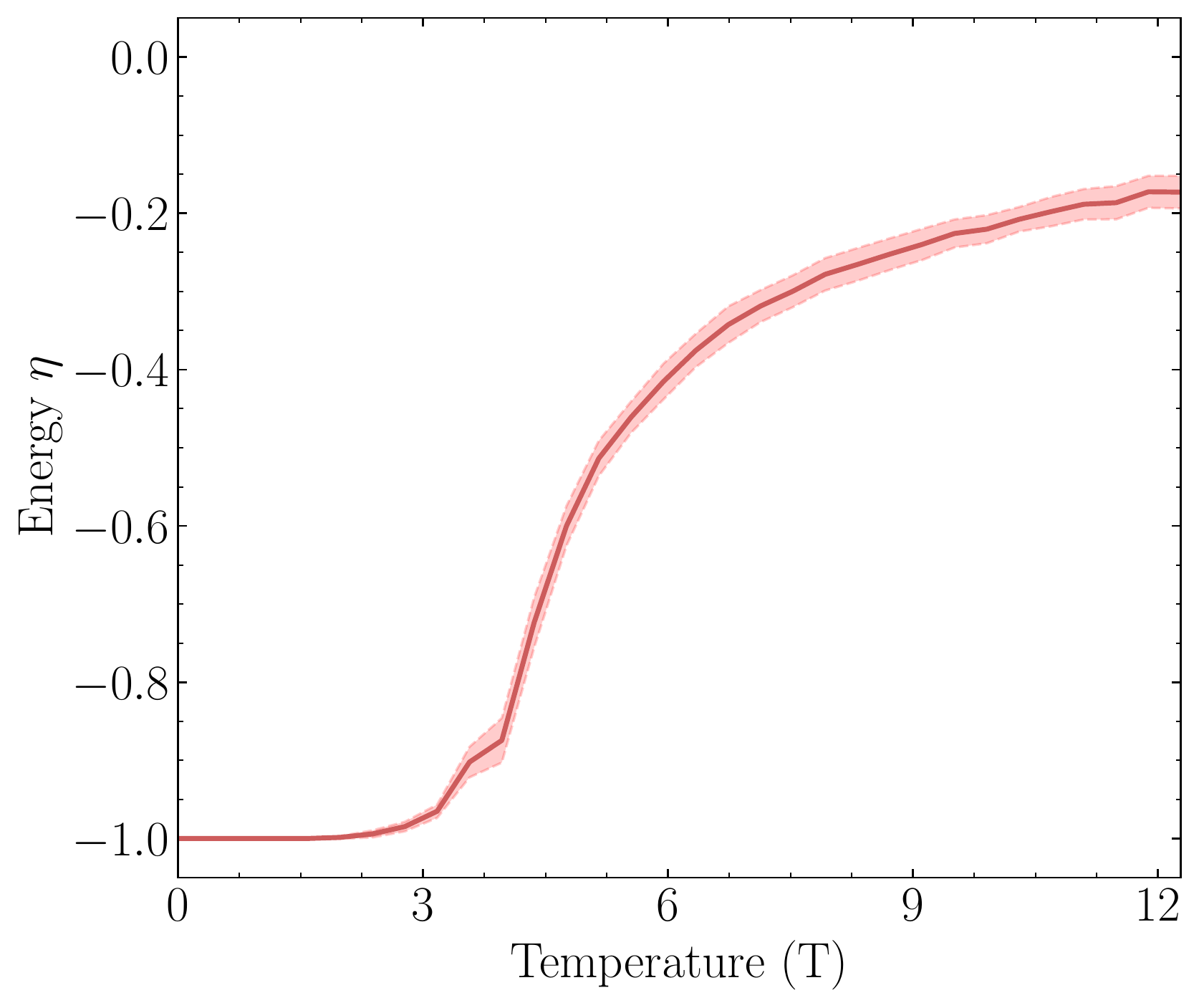} \\
		\includegraphics[width=\textwidth]{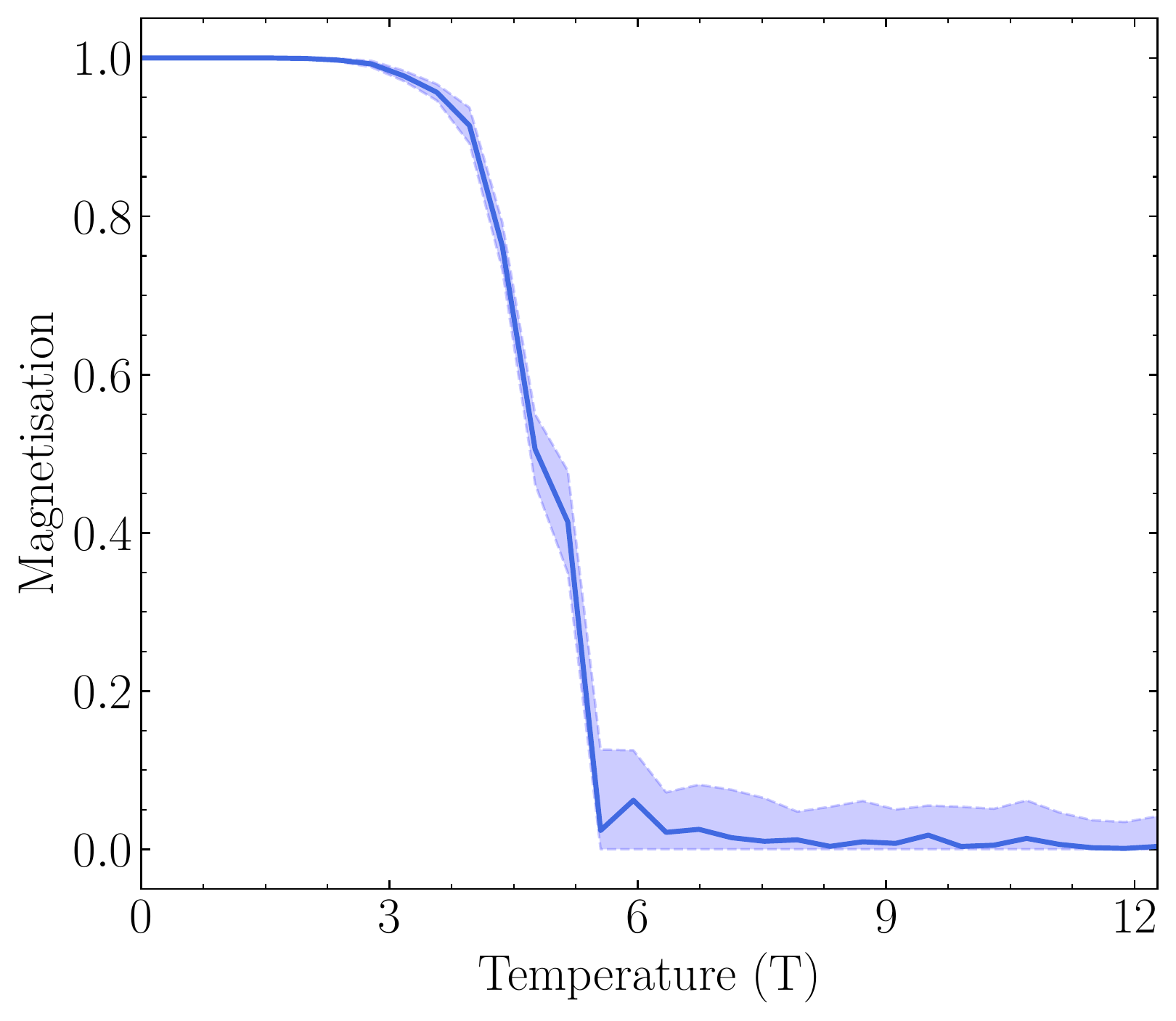}
		\caption{Thermal annealing}
		\label{fig:ising_class_T}
	\end{subfigure}\qquad\qquad
	\begin{subfigure}{0.4\linewidth} 
		\includegraphics[width=\textwidth]{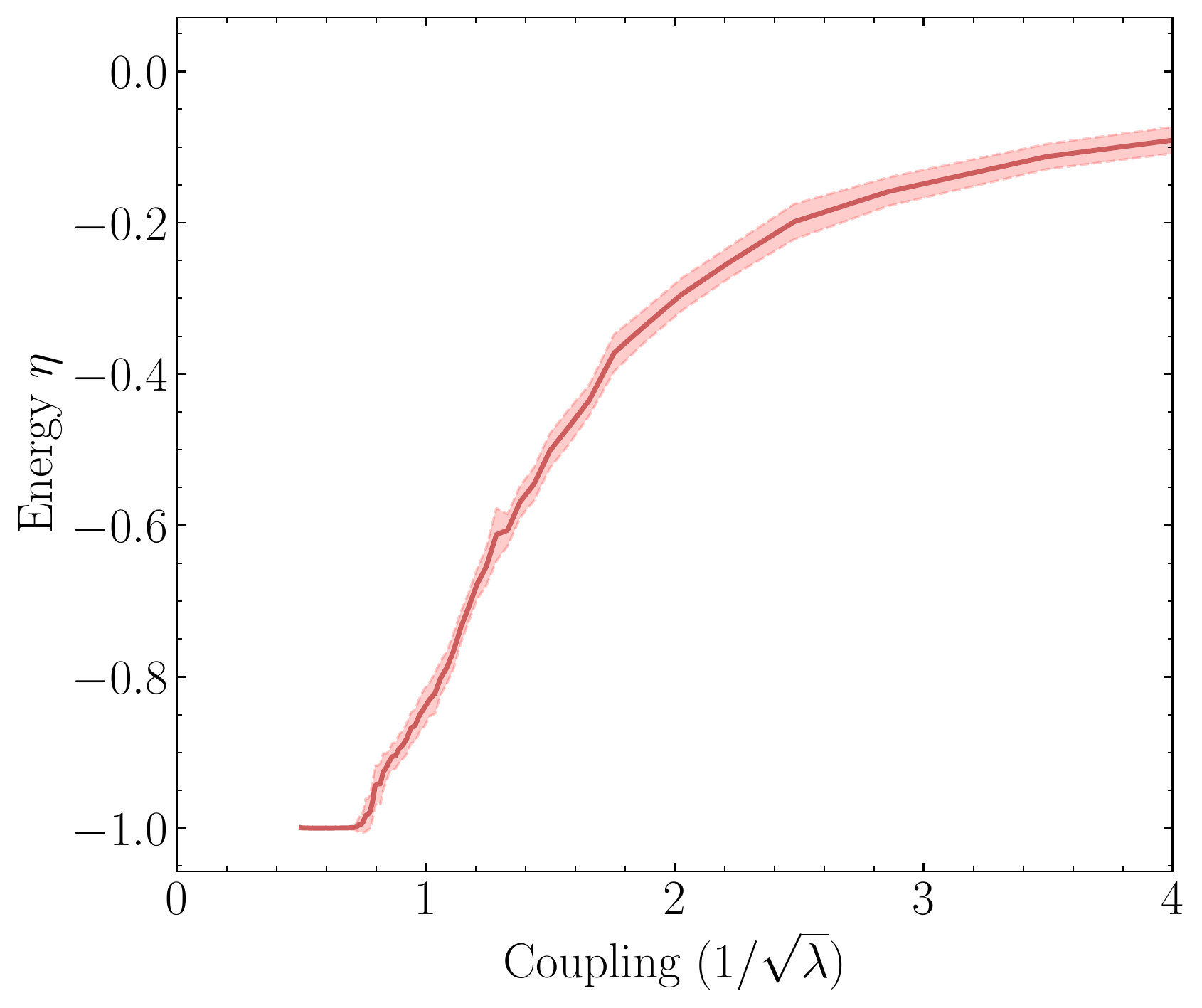} \\
		\includegraphics[width=\textwidth]{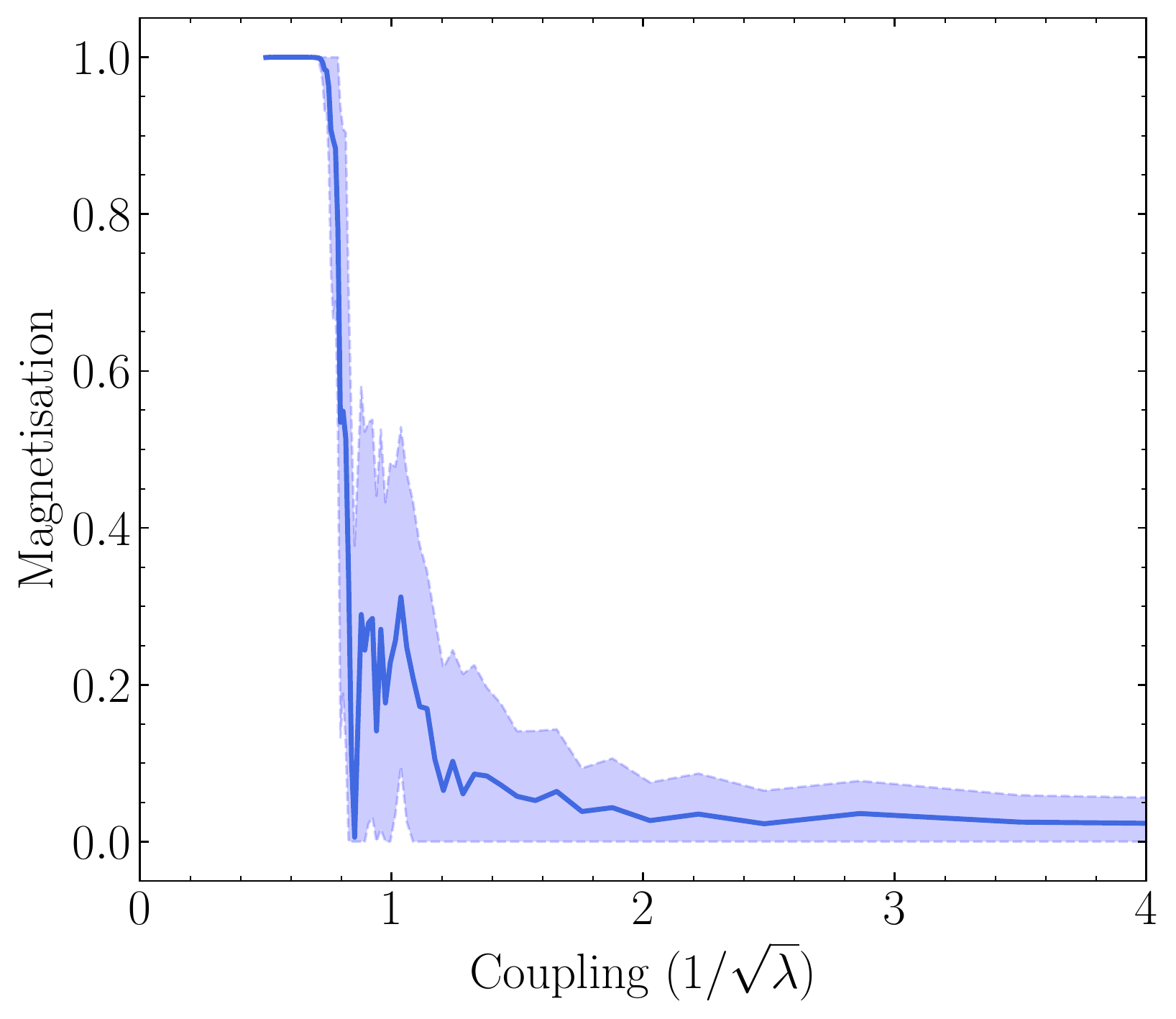}
		\caption{Quantum annealing}
		\label{fig:ising_quant}
	\end{subfigure}
	\caption{ Comparison of the calculation of the energy and the magnetisation using either thermal (Fig.~\ref{fig:ising_class_T}) or quantum annealing (Figs.~\ref{fig:ising_quant} and \ref{fig:quantum_S_search}) quantum annealing. Fig.~\ref{fig:ising_class_T} shows the energy $\eta$ and magnetisation ${\cal M}$ for the 2-D Ising model at different temperatures and with $N=40$. Fig.~\ref{fig:ising_quant} shows the result for quantum annealing ($N=40$, $s=0.3$, $t=100 \mu {\rm s}$) when Similarly for thermal and quantum annealing, the uncertainty bands are obtained by running the same setup 100 times.}
	\label{fig:ising_TvsQ}
\end{figure*}
 
Beginning with the thermal system, the Ising model was placed on a grid with $N=40$. For each temperature, we allowed the thermal annealing process to iterate 100 times to find the equilibrium, before iterating another 100 times to find the energy of the system. From this latter 100 iterations the mean energy was found, and standard deviation error bars  drawn. The same initial starting point was used in all runs, initially chosen by randomly assigning the spins. 
Figure \ref{fig:ising_class_T} shows how the energy and magnetisation of the system then depends on temperature for a suitable choice of parameters, and shows the expected Ising phase transition behaviour.

Figure \ref{fig:ising_quant} shows the equivalent set of plots for the quantum annealer, with the temperature replaced by $1/\sqrt{\lambda}$ and where  we have again set the number of qubits in the system to be $40\times 40$. Here and throughout we perform {\it{reverse}} annealing, that is the anneal schedule starts with the annealer at $s=1$, the classical scenario, then it decreases $s$ to a set value where it is held for a ``tunnelling period'', before then returning to $s=1$. Unless otherwise specified, when we refer to a given value of $s$ in a run we are referring to the value during this tunnnelling period.  
 The ramp up and ramp down times in the schedule are limited 
 by the physical constraints of the annealer, such that the minimum ramp up or down time is $(1-s)\mu {\rm s}$. 
 
 For the plots in  Fig. \ref{fig:ising_quant}  we took $s=0.3$ and a relatively short quantum tunnelling period of $100\mu {\rm s}$. 
 This optimises the number of  reads we can perform (i.e. 100 in practice) for each value of $\lambda$ to find the mean and error bars. Despite the short time period for the runs, it was sufficiently long for the system to reach its equilibrium for  $s=0.3$ as we shall discuss shortly.  
  Of course in order to ``freeze'' the results and read them the annealer must be brought back to the classical state $s=1$ as quickly as possible, namely in $0.7 \mu {\rm s}$ for this choice of parameters.
From these figures it is clear that the quantum 2-D Ising system exhibits a remarkably similar kind of phase transition to the thermal system.

%================================== STUDY OF ANNEALING 

We can now probe this characteristic behaviour of the respective Ising models in order  to devise a set of annealing procedures for both systems to carry forward to the 
study of optimisation of more generic problems. For the 
thermal systems this will consist of a suitable temperature/time profile. For the quantum system the procedure consists of  a suitable choice of overall typical 
coupling $\lambda$, together with a time profile for the parameter $s(t)$. We are also interested in the dependence on the lattice size $N$, as clearly this will limit the 
achievable precision. 

\begin{figure*}[!htbp]
	\centering
	\begin{subfigure}{0.4\linewidth} \centering
		\includegraphics[width=\textwidth]{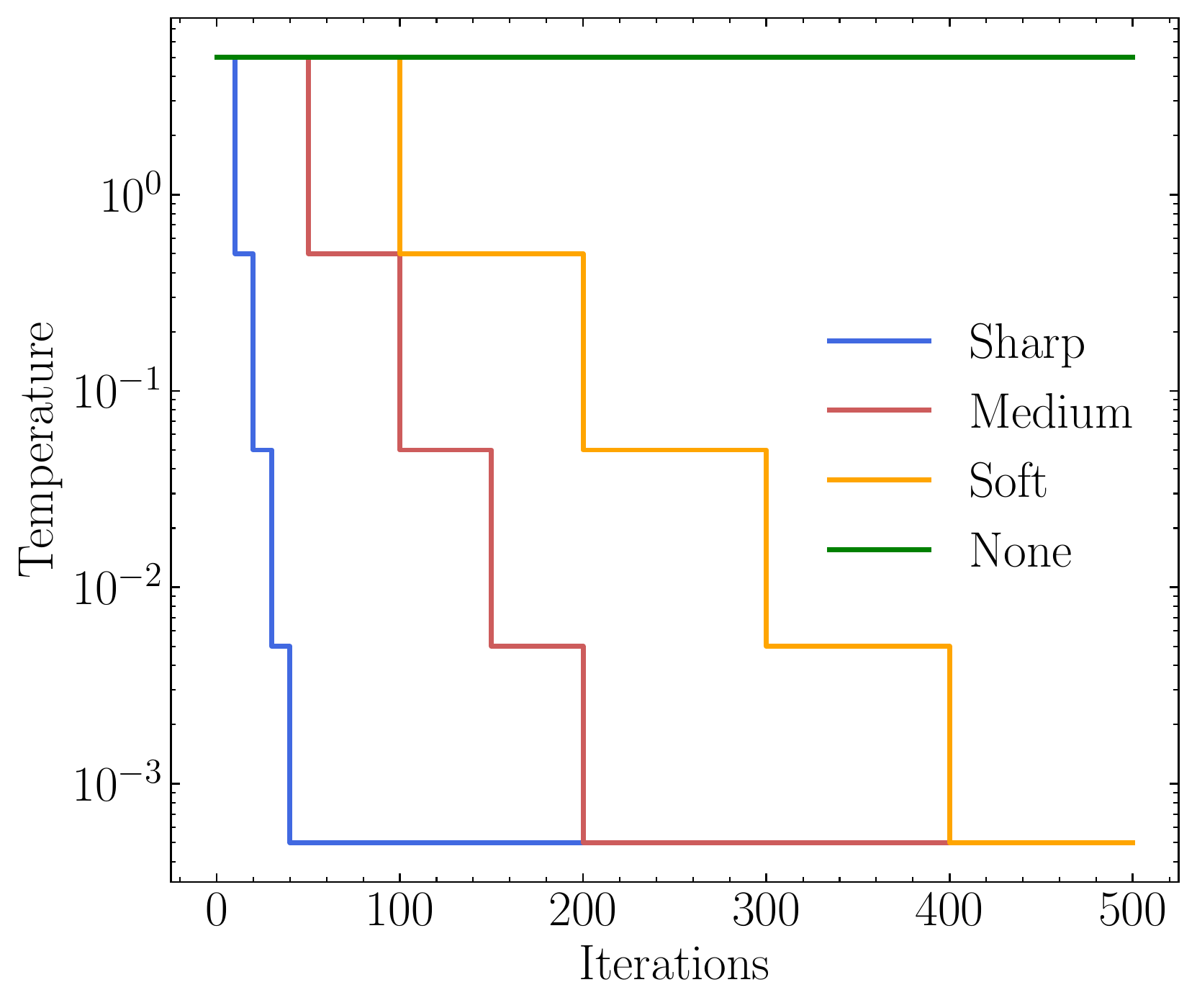}
		\caption{Schedule}
	\end{subfigure}\qquad\qquad
	\begin{subfigure}{0.4\linewidth} \centering
		\includegraphics[width=\textwidth]{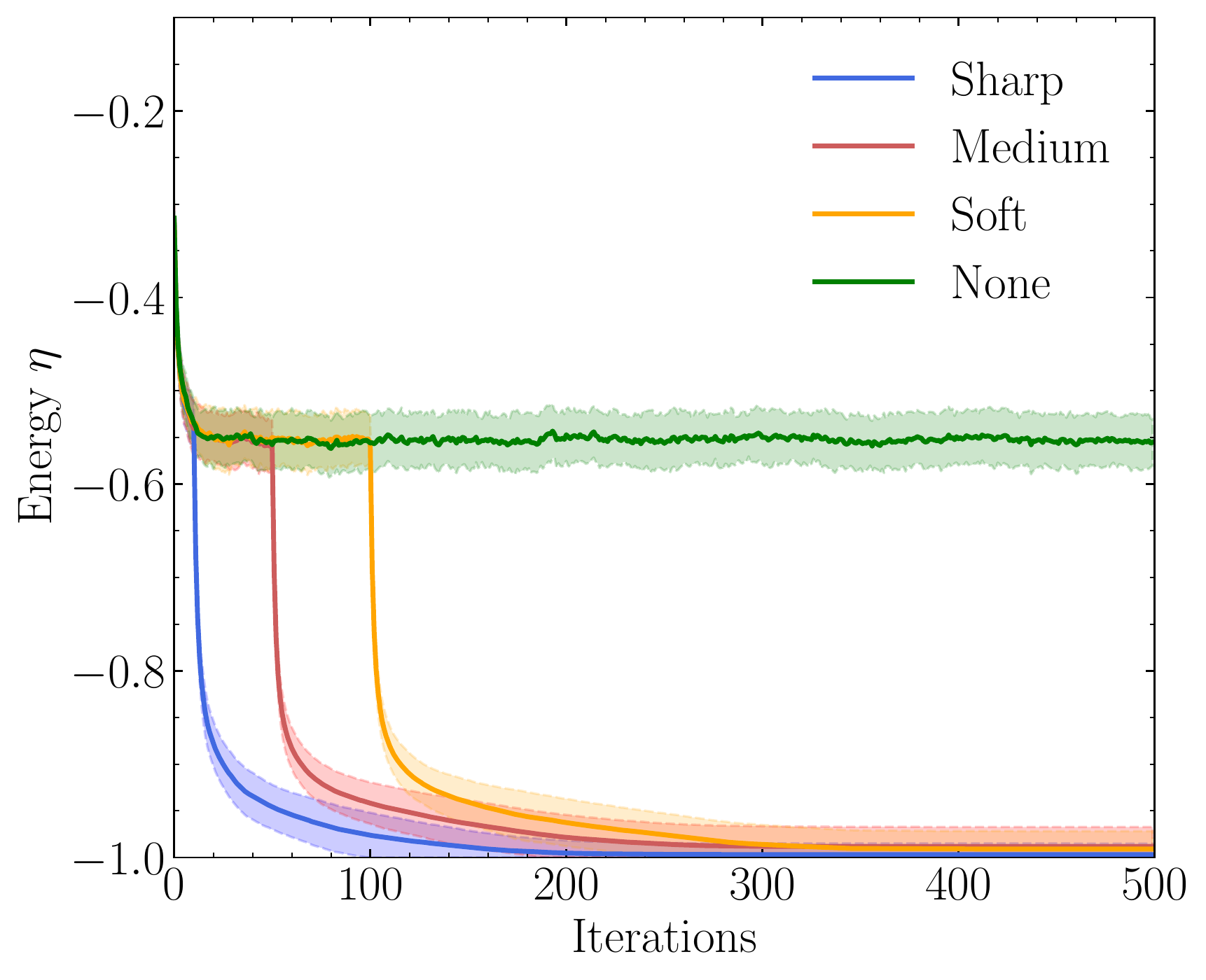}
		\caption{Energy}
	\end{subfigure}
	\caption{A comparison of four thermal annealing schedules. In (a) we show our choice of different schedules, ranging from a very sharp drop in temperature to a slow drop. In (b) the results of performing thermal annealing with these schedules are shown. We choose a grid of $N=40$, and run each setup 100 times to generate the error bars.}
	\label{fig:thermalsched}
\end{figure*}

Beginning with the thermal system, the equilibration is as we said mainly determined by the anneal schedule, which is a profile of decreasing temperature. For comparison we set four schedules: a constant schedule, one where the temperature falls off dramatically, a gentle decay, and one where there is a slow decay. The schedules, and the corresponding results (which are displayed in terms of the energy) are shown in Fig.~\ref{fig:thermalsched}. To draw the standard deviation bands we ran each model 100 times. Each schedule scenario begins from its own randomised initial set of spins, and is repeated 100 times.

%==================================

\begin{figure*}[htb]
	\centering
	\begin{subfigure}{0.4\linewidth} \centering
		\includegraphics[width=\textwidth]{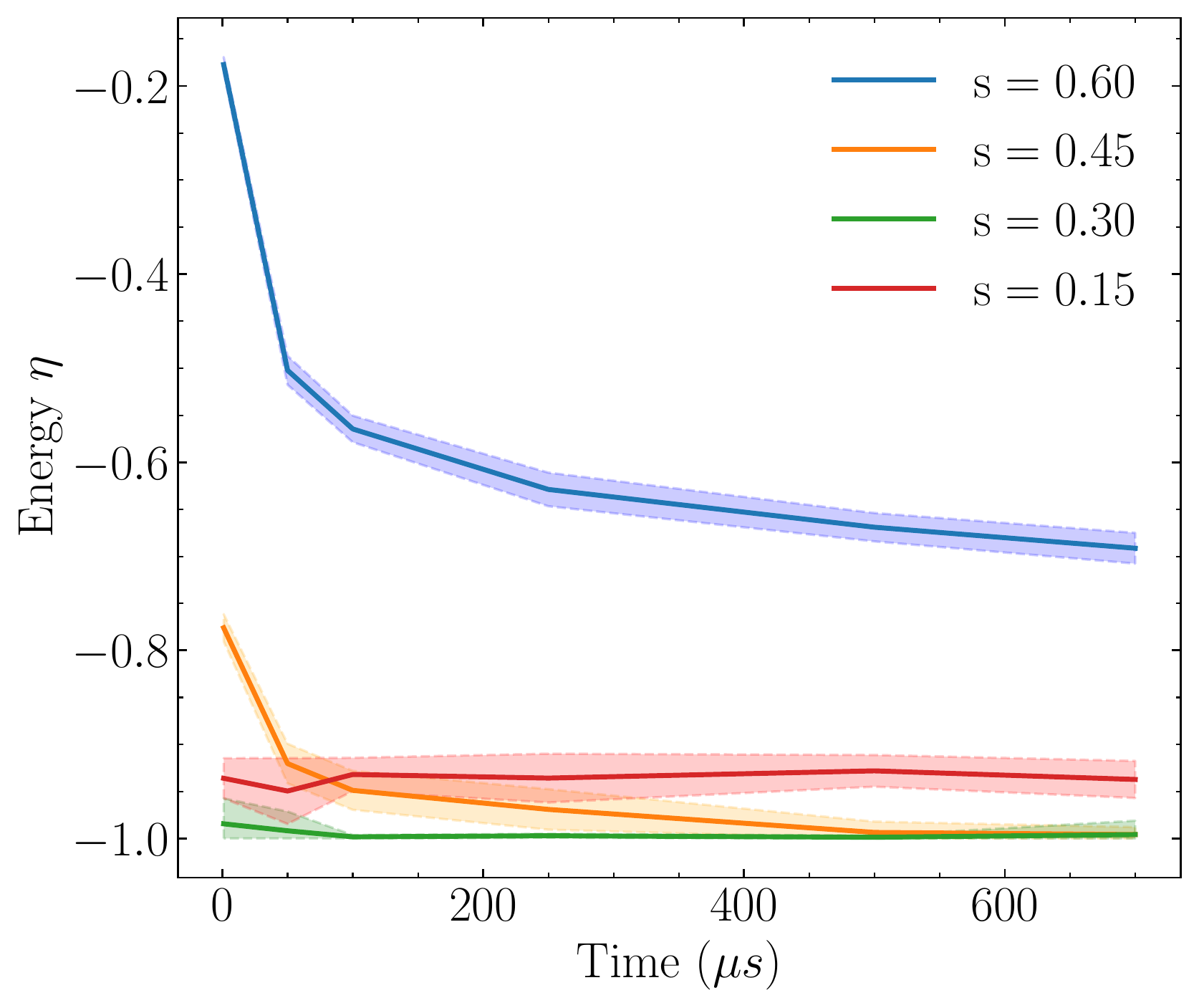}
		\caption{Schedule}
	\end{subfigure}\qquad\qquad
	\begin{subfigure}{0.4\linewidth} \centering
		\includegraphics[width=\textwidth]{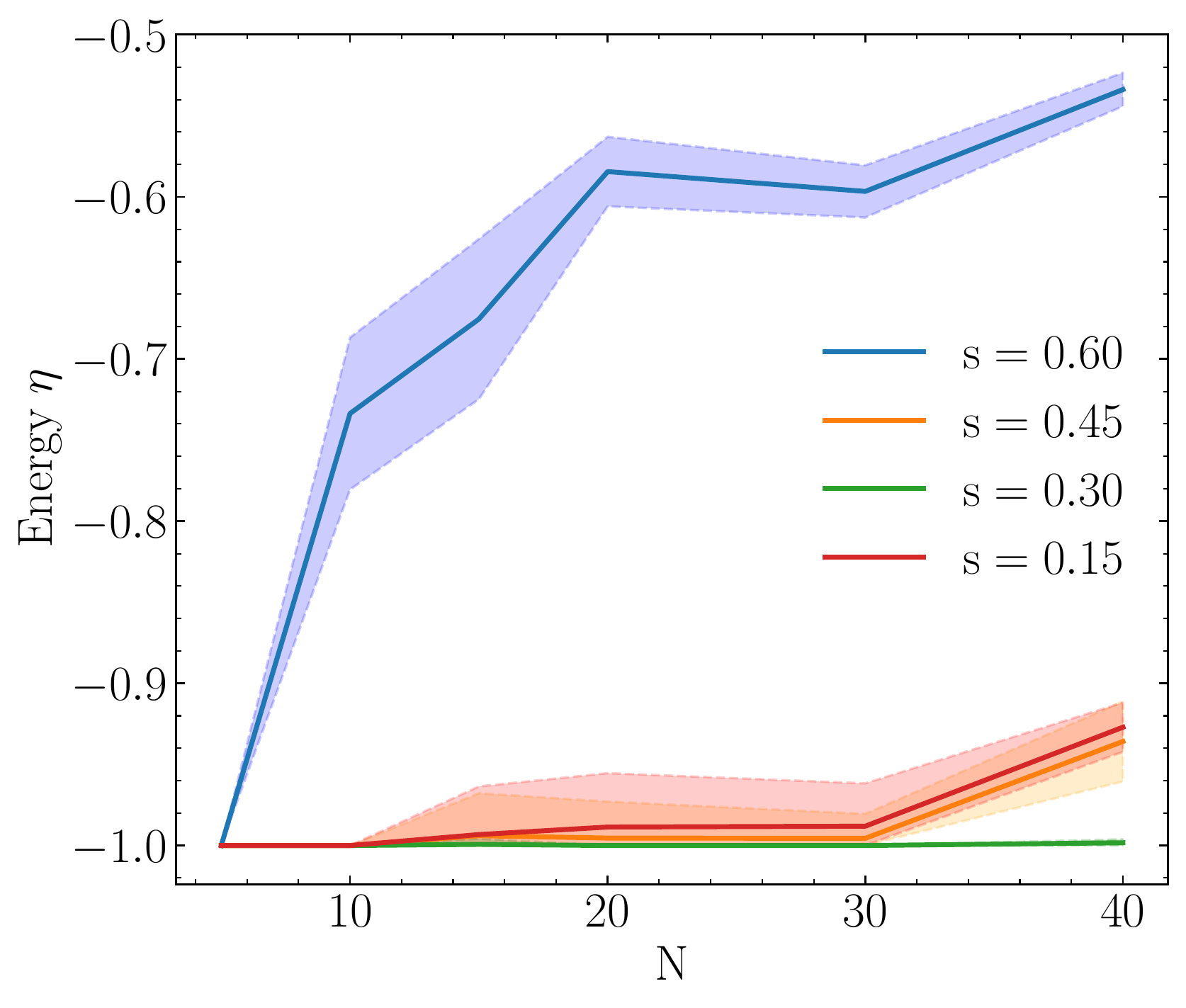}
		\caption{Energy}
	\end{subfigure}
	\caption{Effect of changing the anneal time and $N$ on the average energy, for several values of $s$.}
	\label{fig:quantum_S_search}
\end{figure*}

Turning to the quantum annealer, the  important parameter to gain control over is $s$. Fig.~\ref{fig:quantum_S_search} shows the effect of changing this value. We see the quantum annealer's ability to optimise is heavily dependent on our choice. However with a value of $s=0.3$ we consistently optimise the system even for very short times. To determine the mean and standard error bars for these plots a run for each choice of $s$ was performed 25 times. We note that the performance of the quantum annealer to find the ground state of the Ising model is not monotonic in $s$. Instead, we find that for $s=0.3$ the quantum annealer finds the global minimum with a tunnelling time of $t\simeq 100 \mu {\rm s}$ very efficiently, whereas smaller $s$ tends to lead to randomised fluctuations that simply prevent the system from ever settling in the correct minimum. Indeed we can appreciate from Fig.~\ref{fig:Qthermalising} that the optimal choice of $s$ (and also $\lambda$) is in this sense (much like the maximum temperature in the thermal system) a compromise between focussing on deeper minima and enabling quantum tunnelling.

In Fig.~\ref{fig:quantum_S_search}b we show how changing the lattice size $N$ affects the performance. Here, each model was again run for 25 times. Unsurprisingly, for larger values of $N$ it becomes more difficult to sample the configuration space within the tunnelling time of $t\simeq 100 \mu {\rm s}$. However, while the performance of the quantum annealer for $s \gg 0.3$ or $s \ll 0.3$ clearly deteriorates with growing $N$, for $s=0.3$ the quantum annealer finds the global minimum very reliably even for the largest $N$ that can physically be placed on the annealer. 

\begin{figure*}[htbp!]
	\centering
	\begin{subfigure}{0.4\linewidth} \centering
		\includegraphics[width=\textwidth]{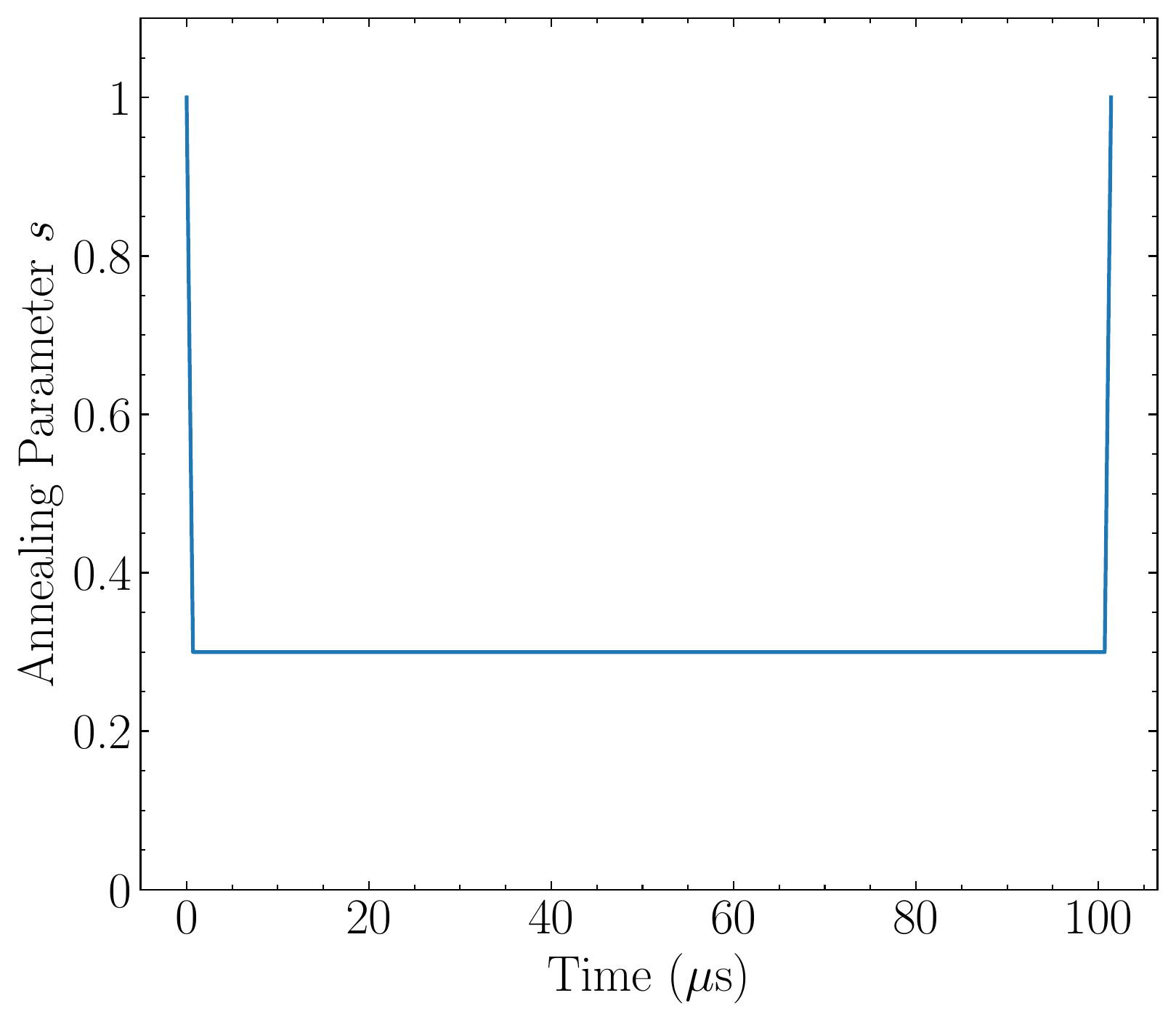}
		\caption{Anneal schedule used for comparative study}
	\end{subfigure}
	\begin{subfigure}{0.4\linewidth} \centering
		\includegraphics[width=\textwidth]{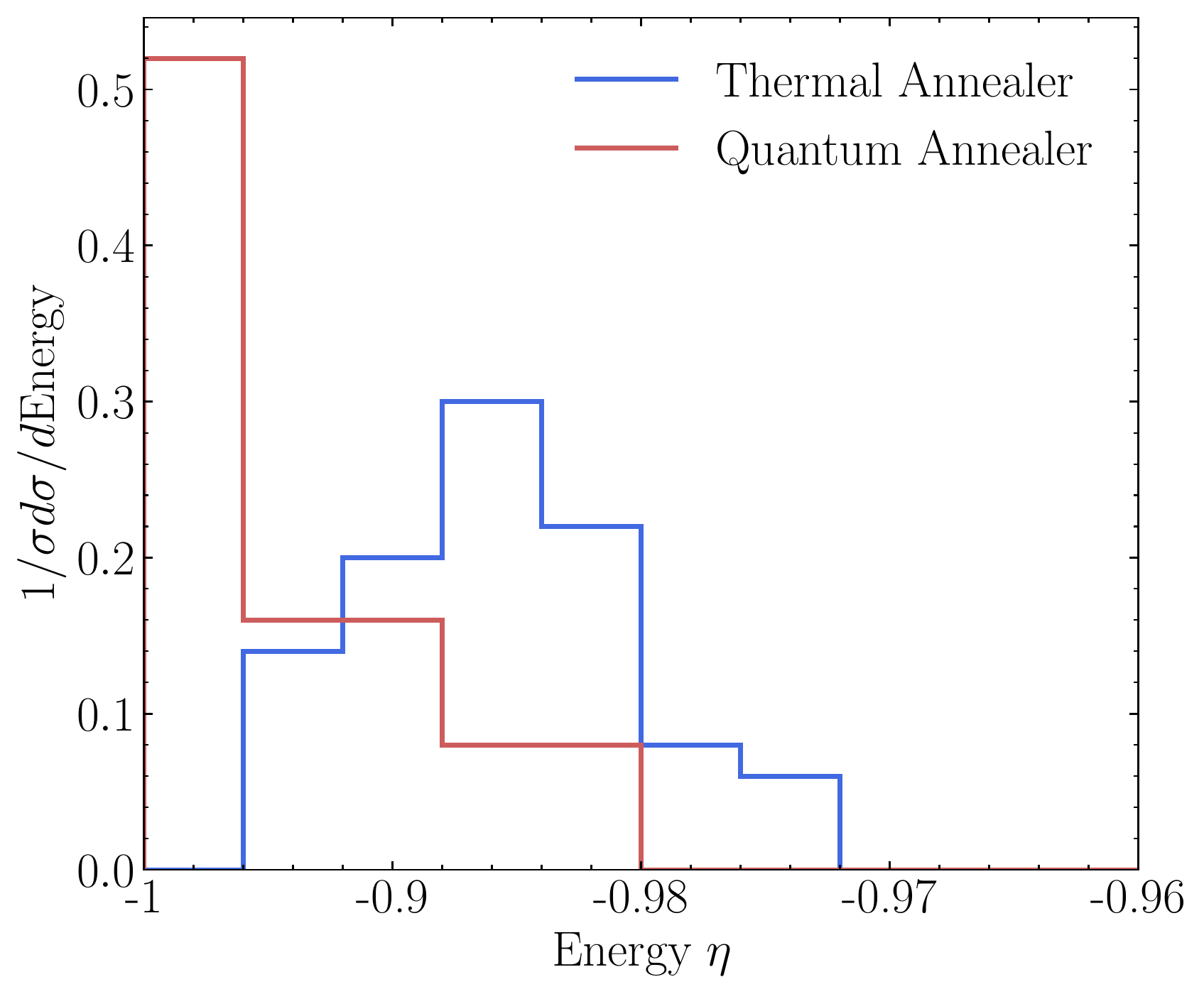}
		\caption{Distribution of minimum energy}
	\end{subfigure}
	\caption{Comparision of the distribution of energies of the Ising model found using thermal and quantum annealers, starting from randomised initial states. The thermal annealer uses the medium schedule from Fig.~\ref{fig:thermalsched}a. For the quantum annealer, we chose $s=0.3$ with a quantum anneal period of 100 $\mu {\rm s}$ and the very simple reverse annealing schedule (shown for completeness) in Fig.~\ref{fig:Energy_Comparison}a. Both setups are applied to grids of $N=40$. We use 50 and 25 randomised initial states for the thermal and quantum anneal, respectively. 
	}
	\label{fig:Energy_Comparison}
\end{figure*}

It is also worth mentioning at this point that of course the physical limitations of the D-wave annealer itself limits $N$. In particular the annealer does not provide the most
general Ising model as in Eq.\eqref{eq:ising}, but only provides a subset of the $J_{ij}$ couplings. Therefore in practice any desired model has to be ``embedded'' onto the annealer using 
locked spin-chains. Thus due to the physical constraints of the annealer (in particular the fact that the maximum number of available qubits is $\sim 5000$) large $N$ 
embeddings begin to become artificially frustrated for $N> 40$. 

The results are shown in Figure \ref{fig:Energy_Comparison}b. The quantum annealer clearly shows a higher probability of finding the correct ground state and with a smaller variance. It is also notable that the evaluation time is vastly longer for thermal annealing: although this is obviously a function of the available architecture, a thermal annealing run takes approximately $\mathcal{O}(10^6)$ times longer than a run on the quantum annealer which takes $\mathcal{O}(100\, \mu {\rm s})$. 
We conclude that the quantum annealer has a clear edge in speed and reliability over simulated thermal annealing in calculating the ground state of the 2-D Ising model, a discrete highly complex system.

We now wish to compare the thermal and quantum methods with {\it {each other}}, to see how efficient each method is at finding solutions and, conversely, how dependent the results are on the choice of initial conditions. To do this, we  ran both the quantum and thermal annealer for a number of randomised initial states with an optimal annealing configuration in each case. 

For the quantum annealer we selected a value of $s=0.3$ and a schedule with a quantum period of 100 $\mu {\rm s}$ with ramp up and down times of $(1-s)\,\mu$s as shown in Fig.~\ref{fig:Energy_Comparison}a. We then ran 25 initial states, 25 times each.  For the thermal annealing we chose the medium schedule from Fig.~\ref{fig:thermalsched}a.  For each initial state, we ran the thermal annealing process 50 times to find an average energy for 50 initial states. Both setups use $N=40$ and $\lambda = -2.0$.

\FloatBarrier 

\section{\label{Sec:Function} Function optimisation}

We now wish to extend our study to consider the solution of more practical problems that 
are encoded onto non-minimal Ising models. The problem we shall discuss is the very generic one of 
finding the minimum of a continuous function of two variables. The useful aspect of such a generic study is that the 
analysis can now be broadened to bring in for comparison other optimisation techniques that do not use Ising model encodings at all. In this paper we will 
as mentioned in the Introduction also consider the Gradient Descent Method  and the Nelder-Mead Method, described in appendices  \ref{appendix:gdm},  and \ref{appendix:nmm} respectively. However as a first step let us describe the Ising encoding of a continuous function which will be central to 
this part of the study. This is the subject of the next subsection. 

\subsection{Domain Wall Encoding of a Continuous Function}
\label{Sec:potential_1}

\begin{figure*}[tbh]
	\centering
	\begin{subfigure}{0.4\linewidth} \centering
		\includegraphics[width=\textwidth]{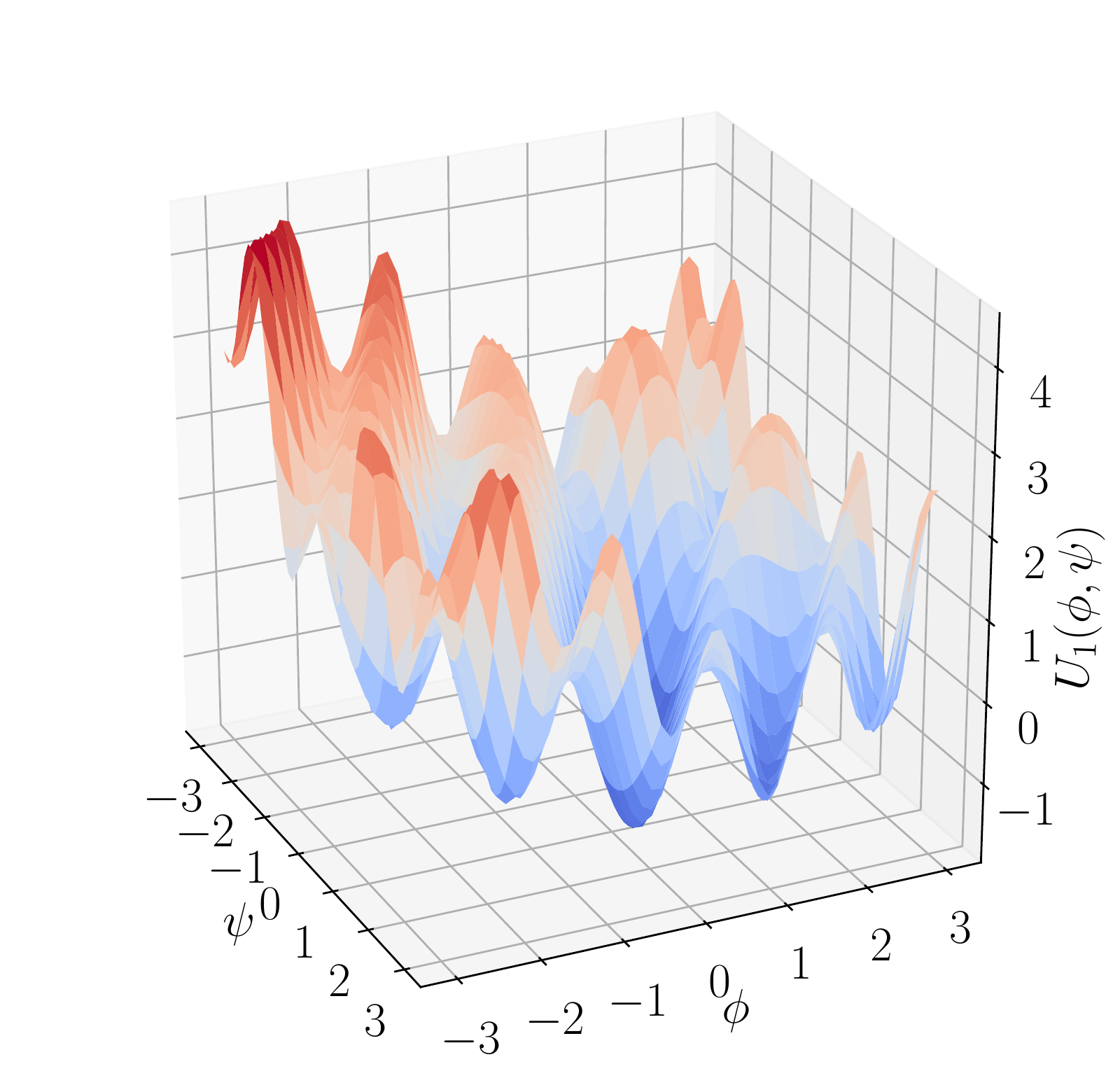}
		\caption{}
	\end{subfigure}
	\begin{subfigure}{0.4\linewidth} \centering
		\includegraphics[width=\textwidth]{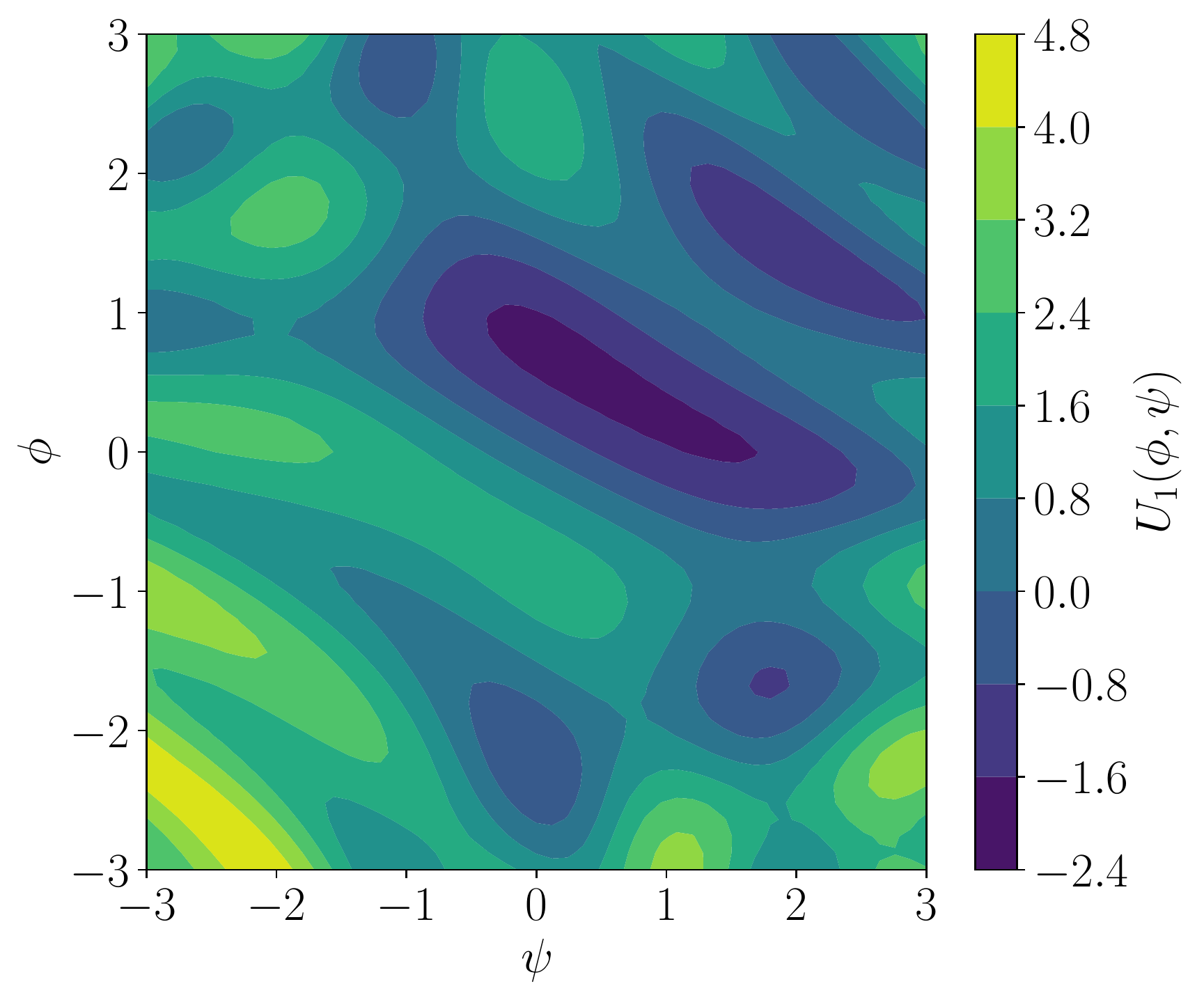}
		\caption{}
	\end{subfigure}
	\caption{The potential $U_1$ of Eq.~\eqref{eq:potential_1} rendered in 3D (a) and as a contour plot (b).}
	\label{fig:potentials}
\end{figure*}

In order to have a specific problem in mind we will consider minimising the following function of two variables $\phi,\psi$ shown in Fig.~\ref{fig:potentials}:
\begin{align}
\label{eq:potential_1}	
	U_1(\phi, \psi) ~=~ -\lambda &\left[ \phi(1-\phi) + \psi(1 - \psi) \right.  \\
	& \left.~~ +~ 12 \cos(\phi\psi) \sin(\psi + 2\phi) \right]~. \nonumber 
\end{align}
This ``corrugated''  function has of course been chosen because it is particularly unpleasant to minimise due to its many local minima. Indeed even in the 
domain shown in  Fig.~\ref{fig:potentials}, namely $\phi,\psi \in [-3,3]$, we can identify by eye 6 minima. 

Naturally the minimisation of such a function will be implemented by adopting it directly as the Hamiltonian (i.e. we regard $U_1$ 
as a potential in other words). Therefore as per the discussion in the Introduction and the previous section we have introduced an overall scaling of the couplings $\lambda$ that we can use to tune the efficiency of the annealers. Here Fig.~\ref{fig:potentials} shows the potential for  $\lambda=0.5$. 

The scalar values of the two variables $\phi,\psi$ can be represented using the ``domain-wall encoding''
introduced in Ref.~\cite{Chancellor19b}. 
The basic idea of the domain wall encoding is to discretize the continuous $\phi$ and $\psi$ domains into $N$ elements of size $\xi$ and $\zeta$ respectively. Then the value of each of the scalars is represented by the 
position of a ``frustrated'' spin on a spin chain where the spin flips sign from negative to positive, the so-called ``domain wall''. Thus the total number of 
qubits in a chain is $2N$, with the first $N$ qubits  
encoding $\phi$ and the second $N$ qubits encoding $\psi$. (For convenience of presentation we will use the same number of qubits for $\phi$ and $\psi$
but of course this is not mandatory.)
Clearly to approximate the continuous function $N$ should
be as large as possible (we will usually take either $N=40$ or $N=30$ depending on the difficulty of the problem we are considering). 
 Two scalar values are then faithfully described when there is a single ``frustrated''
position encoding $\phi$ in the lower block at $i=r_1\in [1,N]$, where the spin flips from negative for all lower indices 
to positive for higher indices up to $N$, and a second ``frustrated''
position encoding $\psi$ in the upper block at $i=r_2\in [N+1,2N]$. In other words a faithful encoding of two scalar field 
values is defined to have the following spin structure:
\begin{eqnarray}
&& \begin{tabular}{|l|}\hline
   $i$  \\ \hline
   $\sigma_i^z$  \\ \hline
   \end{tabular}
  \overbrace{
 \begin{tabular}{| c | c  | c | c | c }\hline
   $1$ & \ldots & $r_1$ & \ldots & N \\ \hline
   $-$ & $---$ & $+$ & $+++$ & $+$ \\ \hline
   \end{tabular}
   }^{\mbox{$\phi$}}\qquad\qquad  \label{eq:spinstruc}
   \\
  &&~~~ \qquad\qquad \underbrace{
 \begin{tabular}{ c | c  | c | c | c |}\hline
   $N+1$ & \ldots & $r_2$ & \ldots & 2N \\ \hline
   $-$ & $---$ & $+$ & $+++$ & $+$ \\ \hline
   \end{tabular}
   }_{\mbox{$\psi$}}\nonumber 
\end{eqnarray}
The scalar values are extracted from such a configuration by counting the negative spins, as follows:  
\begin{align}
\phi\,\,& =\,\,\phi_{0}+\xi (r_1-1)\,=\,\phi_{0}+\frac{\xi}{2}\sum_{i=1}^{N}(1-\sigma_{i}^z)\, ,\nonumber \\
\psi\,\,& =\,\,\psi_{0}+\zeta (r_2-1)\,=\,\psi_{0}+\frac{\zeta}{2}\sum_{i=N+1}^{2N}(1-\sigma^z_{i})\, ,
\end{align}
where $\phi_{0}$, $\psi_{0}$ are fiducial minimum values. 

This is the way that we wish the scalar values are to be represented, but how do we enforce only such faithful representations of $\phi$ and $\psi$, and not for example 
meaningless configurations in which the spins flips signs at more positions, or at no positions at all? This is done by adding two crucial components to the Hamiltonian, a 
domain wall enforcing component in the linear $h_i$ term of the Ising model in Eq.\eqref{eq:ising}, and a ferromagnetic coupling in the $J_{ij}$ terms that encourages spin alignment
between neighbouring sites:
\begin{align}
h_i^{(\text{chain})} & ~=~\Lambda'\,\,(\delta_{i,1}-\delta_{i,N}~+~\delta_{i,N+1}-\delta_{i,2N} )~,  \nonumber \\
J_{i<N,j<N}^{(\text{chain},\phi)}  & ~=~  -\frac{\Lambda}{2}\,(\delta_{i,j+1}+\delta_{i+1,j} )  ~,\nonumber \\
J_{i\geq N,j\geq N}^{(\text{chain},\psi)}  & ~=~  -\frac{\Lambda}{2}\,(\delta_{i,j+1}+\delta_{i+1,j} )  ~, \label{eq:Jhchain}
\end{align}
where  
$\Lambda,\Lambda'$ are parameters that are somewhat larger
than the largest energy scale in the problem. (For the best performance
they should not be very much larger.)

 The couplings $h_i^{\text{(chain)}}$
force the system to have negative spin $\sigma_{1}^{z}=-1$ at the bottom of each block, i.e. for $i= 1$ and $ N+1$, and
positive spin $\sigma_{i}^{z}=+1$ at the top of each block, i.e. for $i= N$ and $ 2N$. This is in accord with the desired spin structure in Eq.~\eqref{eq:spinstruc},
and it forces an odd number of spin flips between the two ends of a block. Meanwhile  $J_{ij}^{\text{(chain},\phi)}$ gives a (relative) penalty of $2\Lambda $ for 
each domain wall in the lower block, and separately $J_{ij}^{\text{(chain},\psi)}$ gives a (relative) penalty of $2\Lambda $ for 
 each domain wall in the upper block. This favours just a single wall in each block as required.
 
Having enforced only faithful representations of $\phi$ and $\psi$, it is then straightforward to see how one can 
now further encode the potential terms in  $U_1(\phi,\psi)$. 
To describe this, let us split the potential as
\be
U_1(\phi,\psi) ~=~ U_a(\phi) + U_b(\psi) + U_c(\psi,\phi) ~,
\ee
where $U_{a,b}$ are the terms that involve only one of the two variables, $\phi$ and $\psi$ respectively, while 
$U_c$ is the mixed term, in this particular case $$U_c~=~- 12\lambda  \cos(\phi\psi) \sin(\psi + 2\phi)~ .$$
Note that one has to be a little circumspect making this designation as we shall see.  

Consider first the simpler $U_{a}$ term. This piece can be encoded into either $h_i$ or $J_{ij}$. The encoding into 
$h_i$ can be performed by noting that the desired potential can be written 
\begin{equation}
U_a(\phi) ~=~ \frac{1}{2}\sum _{j=1}^N U_a(\phi_0 +  {\xi}{} j ) (\sigma ^z_{j+1}-\sigma ^z_{j})~,
\end{equation}
since there is a contribution only where $\sigma^z_j$ and $\sigma_{j+1}^z$ have different sign, namely at the position of a domain wall where $j=r_1-1$.  Assuming that $U_a$ is differentiable, and $\xi\ll 1$, this is 
equivalent to a linear coupling of the form 
\begin{equation}
\label{eq:h}
h_{j}^{{(U_a)}}~=~-\frac{\xi}{2}\partial_\phi U_a(\phi_{0}+\xi j)~.
\end{equation}
Alternatively the same coupling can be encoded into  $J_{ij}$  
by instead adding the couplings
\begin{equation}
\label{Ua:J}
J_{ij}^{(U_a)}~=~\frac{1}{2}U_{a}(\phi_{0}+\xi j)\left(\delta_{ij}-\delta_{(i-1)j}\right)~.
\end{equation}
It is easy to see that again this picks out a contribution from the location of the domain wall similarly to $h_i$ above. 
Indeed clearly there is no contribution from Eq.~\eqref{Ua:J} unless $i,j$ are equal or adjacent. If we consider then summing the $i$ past a given $j$,  
the above term reduces to  
\be 
\frac{1}{2}U_{a}(\phi_{0}+\xi j)\left(1 -\sigma_{j+1}^z\sigma_j^z \right)~,
\ee
which is then to be summed over $j$. Upon performing the $j$ sum there again remains a contribution only at the domain wall, where $j=r_1-1$. The equivalent encoding for $U_b(\psi) $ is trivially performed by summing instead over the upper block of indices $i\in [N+1,2N]$, swapping $\xi$ for $\zeta$, and obviously swapping $U_a$ for 
$U_b$. 

In the present context it does not matter if we decide to encode $U_a$ and $U_b$ into $h_i$ or $J_{ij}$. Where it does however make  difference is if one wishes to alter the 
potential at some point in the anneal in order to study some particular physical process. This was the case for example in Ref.~\cite{Abel:2020qzm} where the system was first allowed to settle into a minimum before the 
potential was changed by scaling $h_i$ (using the $h$-gain parameter) in order to develop a 
new lower ``true vacuum'' into which the system could then tunnel by barrier penetration. This possibility will not be used here.

 By contrast, the $U_c(\phi,\psi)$ terms in the potential that involve both variables can  only be encoded in $J_{ij}$. The logic for such terms follows that of the encoding of 
 a single variable term into $h_i$ in Eq.~\eqref{eq:h}. That is we add 
\begin{align}
\label{eq:Uc}
U_c(\phi,\psi) ~&\equiv ~  \sum_{i,j=1}^{N} U_c(\phi_0+i \xi ,\psi_0+j \zeta )\, \qquad  \\
&    \qquad \times {1\over 4} (  \sigma^z_{i+1}- \sigma^z_{i} ) 
(  \sigma^z_{N+j+1}- \sigma^z_{N+j} ) \nonumber \\
\,&\hspace{-1cm}= \, {\xi\zeta\over 4} \sum_{i,j=1}^{N} \partial_\phi \partial_\psi U_c(\phi_0+i \xi ,\psi_0+j \zeta )\, \sigma^z_{i} \sigma^z_{N+j} \, ,\nonumber
\end{align}
which upon symmetrising implies that 
\begin{align}
\label{eq:ju2}
J^{U_c}_{i,j+N}  & ~=~J^{U_c}_{N+j,i}  \\
& \qquad ~=~\, {\mu \over 8} \xi\zeta \,\partial_\phi \partial_\psi U_c(\phi_0+i \xi ,\psi_0+j \zeta )~.\nonumber 
\end{align}

We now return to our somewhat cryptic comment above about care being needed to decide which encoding is appropriate, 
which is something of a technicality but we discuss it for completeness: the issue is that Eq.~ \eqref{eq:Uc} 
may not be sufficient to encode the whole of $U_c$ because it is a double derivative. Suppose for example that the coupling $U_c$ 
contained a contribution $\phi\psi^2$. Then, written in discretised form, we see that there is a single-field term effectively given by the fiducial field values:
\begin{align}
\phi \psi^2 ~\supset~  \phi_0 \psi^2 ~+~ \psi^2_0 \phi ~+ \ldots ~.
\end{align}
Such terms could not be incorporated by the double derivatives implicit in $J_{ij}$ and Eq.~\eqref{eq:ju2}, and instead $h_i$ would have to be augmented to cancel spurious single 
field terms: in this particular case the required additional terms would be
\begin{align}
h_{j\,=\,1\ldots N}^{(U_c)} &~=~  -\,\frac{\xi}{4} \mu \psi^2_0 ~,\\
h_{N+j }^{(U_c)} &~=~  - \frac{ \zeta}{2}\, \mu \bar{\phi} \,(\psi_0 + j \zeta ) ~,
\label{eq:hu2}
\end{align}
where $\bar\phi $ is the average $\phi$ in the interval, i.e. $\bar\phi = \phi_0 + \frac{N\xi}{2}$.
In principle a similar consideration by trigonometric identities applies for the actual $U_c$ terms in the $U_1$ of  Eq.\eqref{eq:potential_1}. In practice 
we find that this issue is resolved by multiplying the potential by a factor that forces a constant value for $U_1$ at the boundaries, but
is unity everywhere inside the domain away from the boundaries.  For the functions we will consider we find that this issue does not make a significant 
difference, but it is something to be aware of for other studies.

%\FloatBarrier
\subsection{Results for the corrugated potential $U_1$}

To provide a baseline to compare the quantum annealer (QA) against, we will use three classical optimisation methods. As well as thermal annealing (TA), we will consider Nelder-Mead (NM) and Conjugate gradient descent (GD). The NM method uses a simplex to traverse the function space to find a minimum. The gradient descent method chosen is an extension to classical gradient descent which includes an adaptive step size. Both these methods, alongside thermal annealing, are discussed further in Appendix \ref{appendix:classical_opt}.

While the NM and GD methods use the continuous form of the potential, both the QA and TA method will use the discretised Ising encoding of the potential described in the previous subsection. Here, for the TA we chose to use $N=50$. Due to limitations imposed by the quantum annealer device we discretise the potential such that $N=20$ for most of the QA runs. This decision is kept for most of the paper. However later in Sec.~\ref{Sec:grid} we will discuss how the choice of $N$ might effect our findings, and will in fact conclude that it does not. In the near future when larger devices becomes available we believe all the conclusions we reach here will carry forward. 

Our choice of overall $\lambda$ is also influenced by the method. For the classical methods, larger $\lambda$ generally hinders the ability to find the minimum as it would be harder to escape local minima -- although of course  given the Boltzmann weighting, for the TA any change in $\lambda$ can be absorbed in a redefinition of $T$. This is not the case with quantum annealing, where as explained in Sections \ref{sec:solving} and \ref{sec:deeper}, scaling the potential will make the global minimum more ``visible''. Therefore, a slightly larger value for $\lambda$ is taken for the QA: here we take $\lambda=0.7$, instead of the $\lambda=0.5$ for the classical runs. 

Guided by the studies in Section \ref{sec:solving}, the thermal anneal schedule is set such that it runs for 4000 iterations, starting at a temperature of $T=1.1$. After 500 iterations the temperature is halved, then reduced to $5\%$ of its value every 500 iterations afterwards. The quantum annealer schedule is set such that the ramp-up period is $15\mu {\rm s}$. It then tunnels for $100\mu {\rm s}$ during which time $s$ is set to 0.1, before ramping down for $250\mu {\rm s}$. Note there is a somewhat different approach now: to find a global minimum we are free  to choose a slow ramping down to ensure we do not  ``excite'' the system.

\begin{figure*}[htb]
	\centering
	\begin{subfigure}{0.24\linewidth} 
		\includegraphics[width=\textwidth]{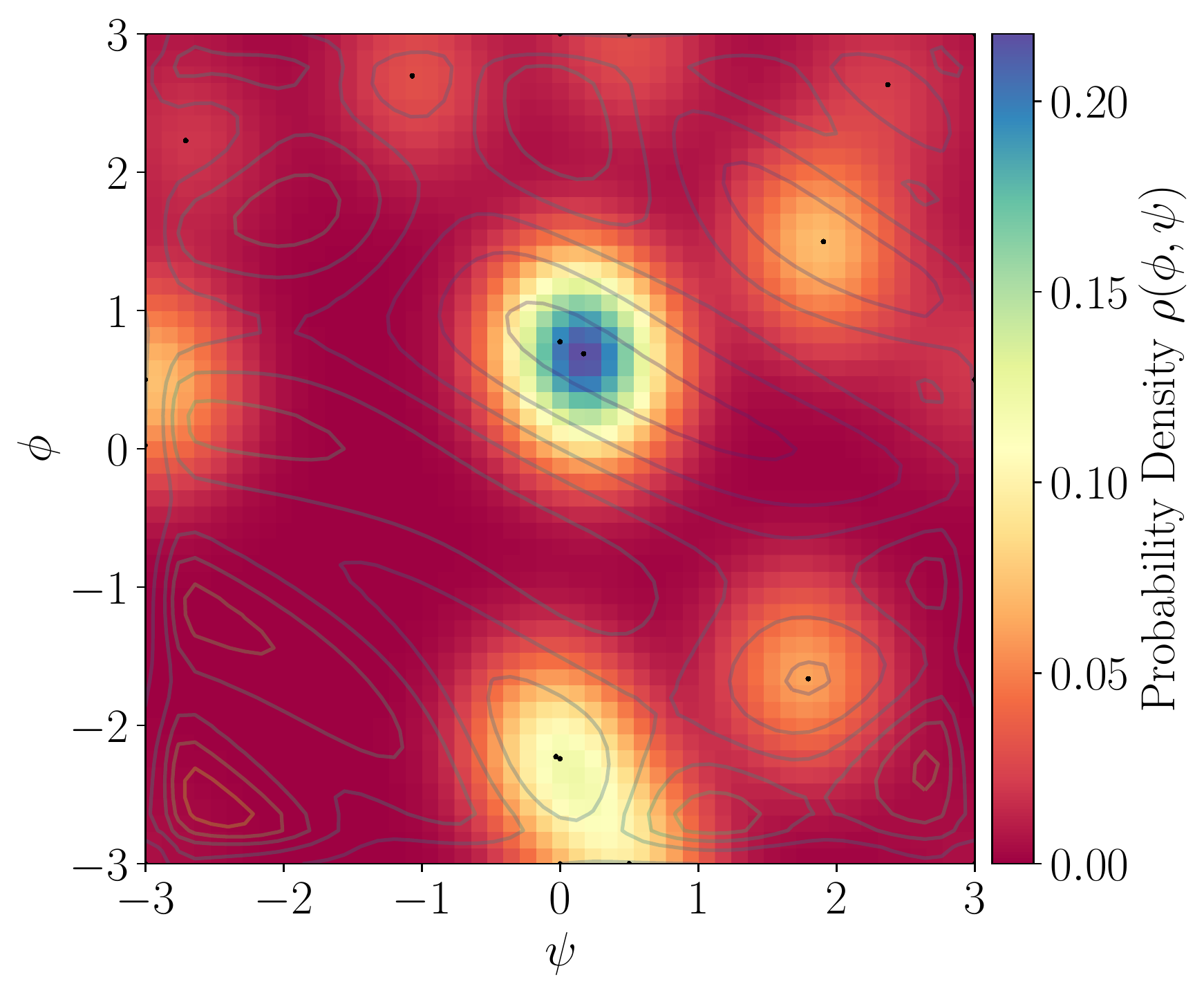}
		\includegraphics[width=\textwidth]{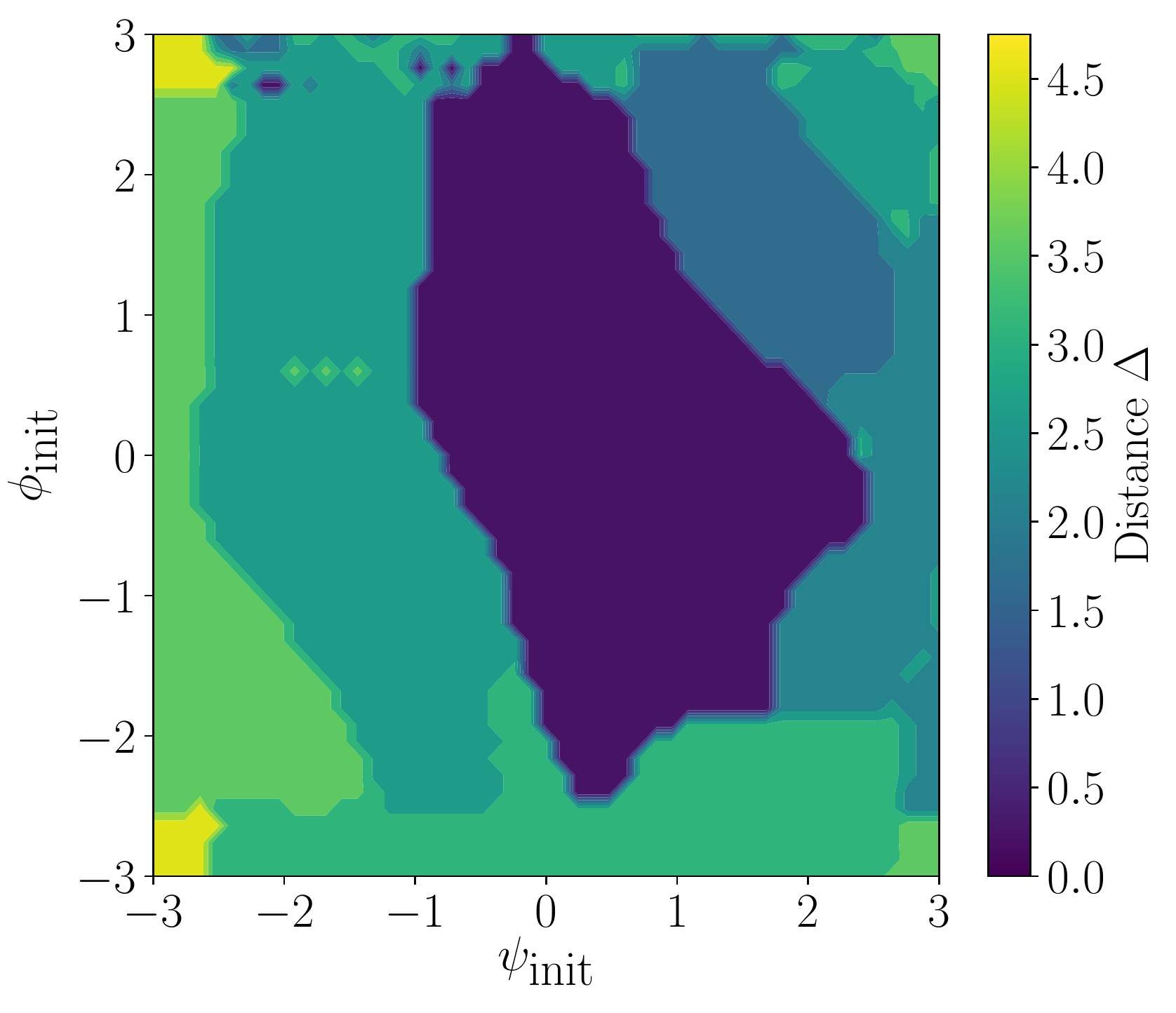}
	\caption{Nelder-Mead}
	\end{subfigure}
	\begin{subfigure}{0.24\linewidth} \centering	
		\includegraphics[width=\textwidth]{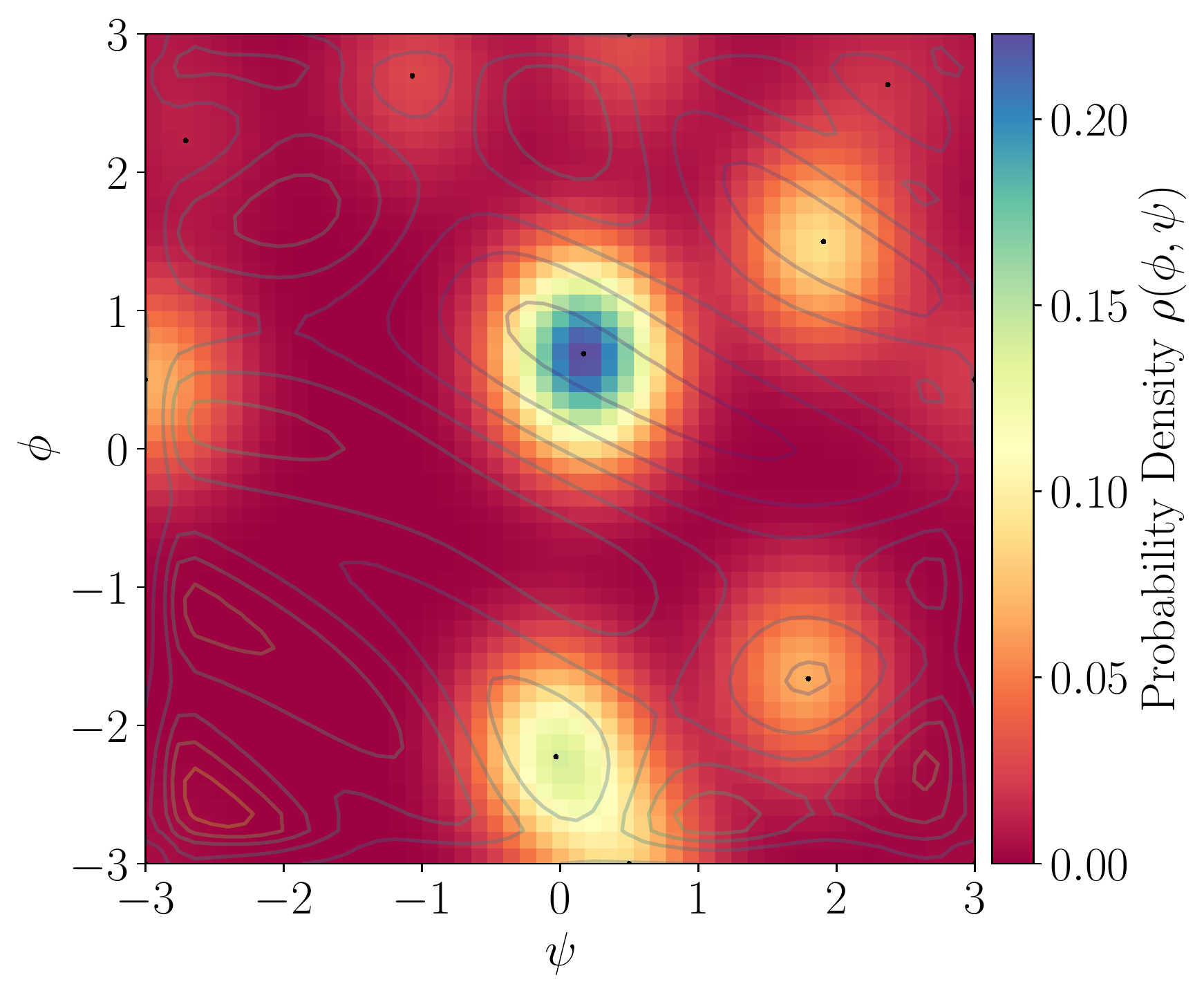} \\
		\includegraphics[width=\textwidth]{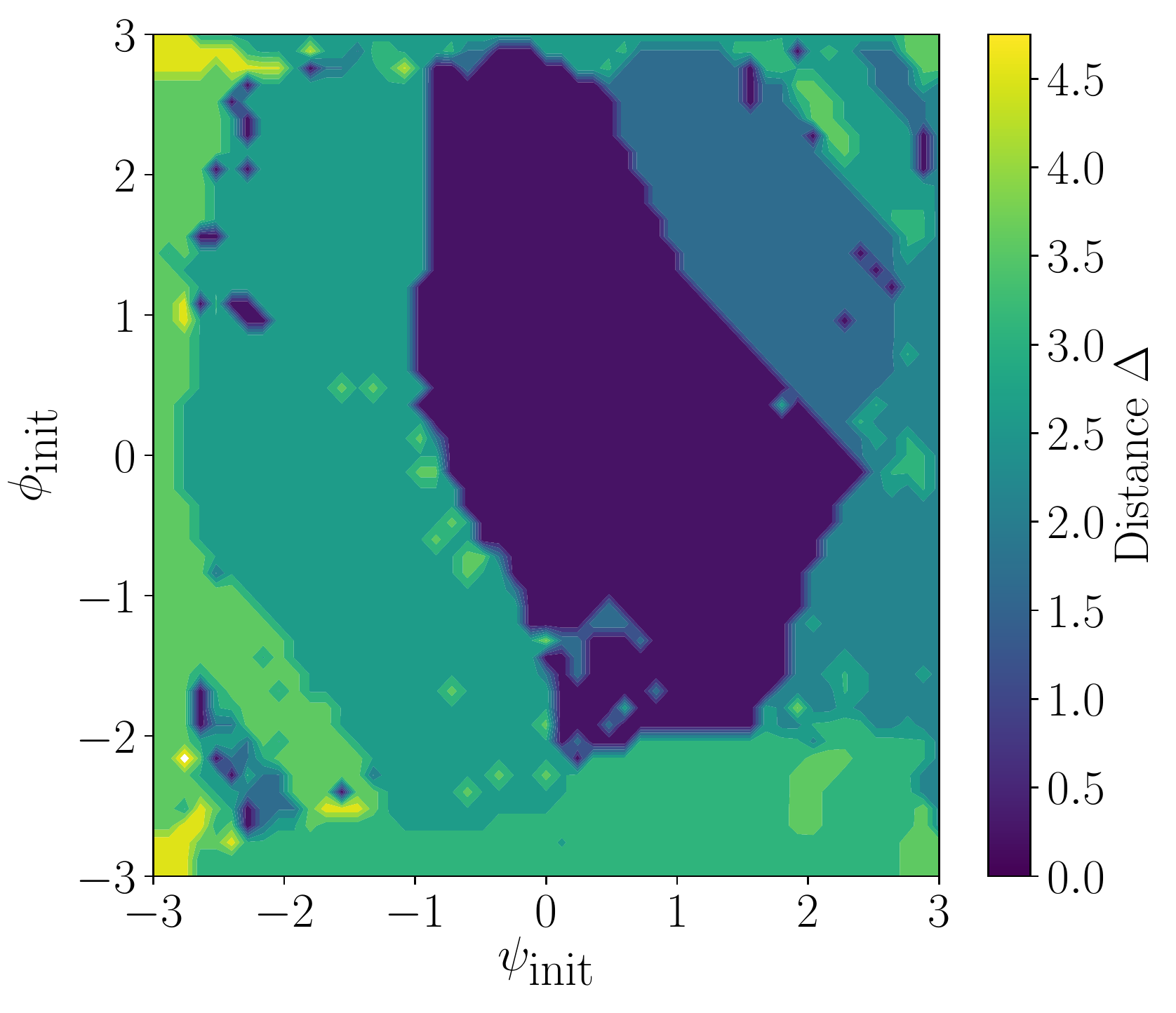}\\
	\caption{Gradient descent}
	\end{subfigure}
	\begin{subfigure}{0.24\linewidth} 	
		\includegraphics[width=\textwidth]{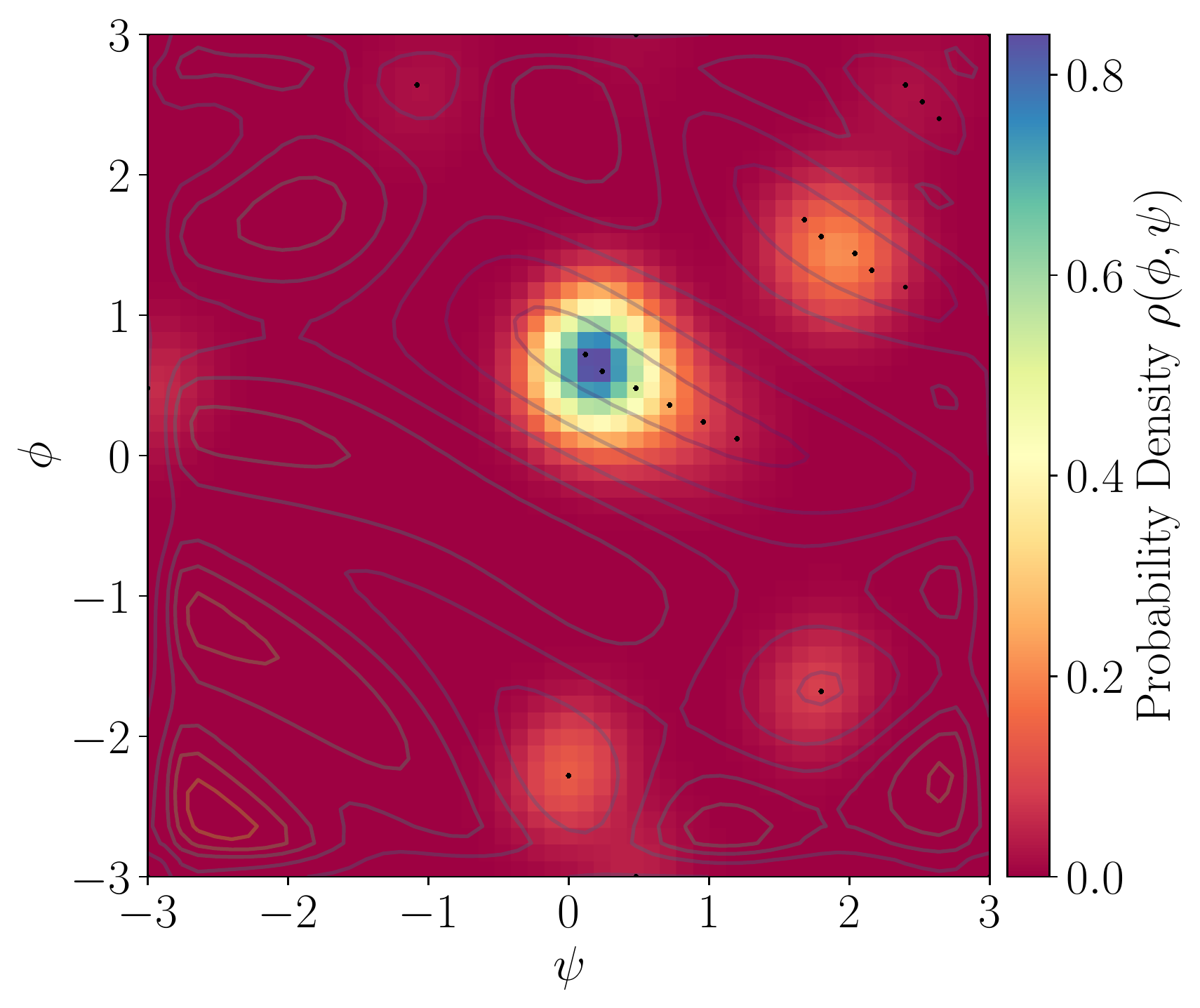}
		\includegraphics[width=\textwidth]{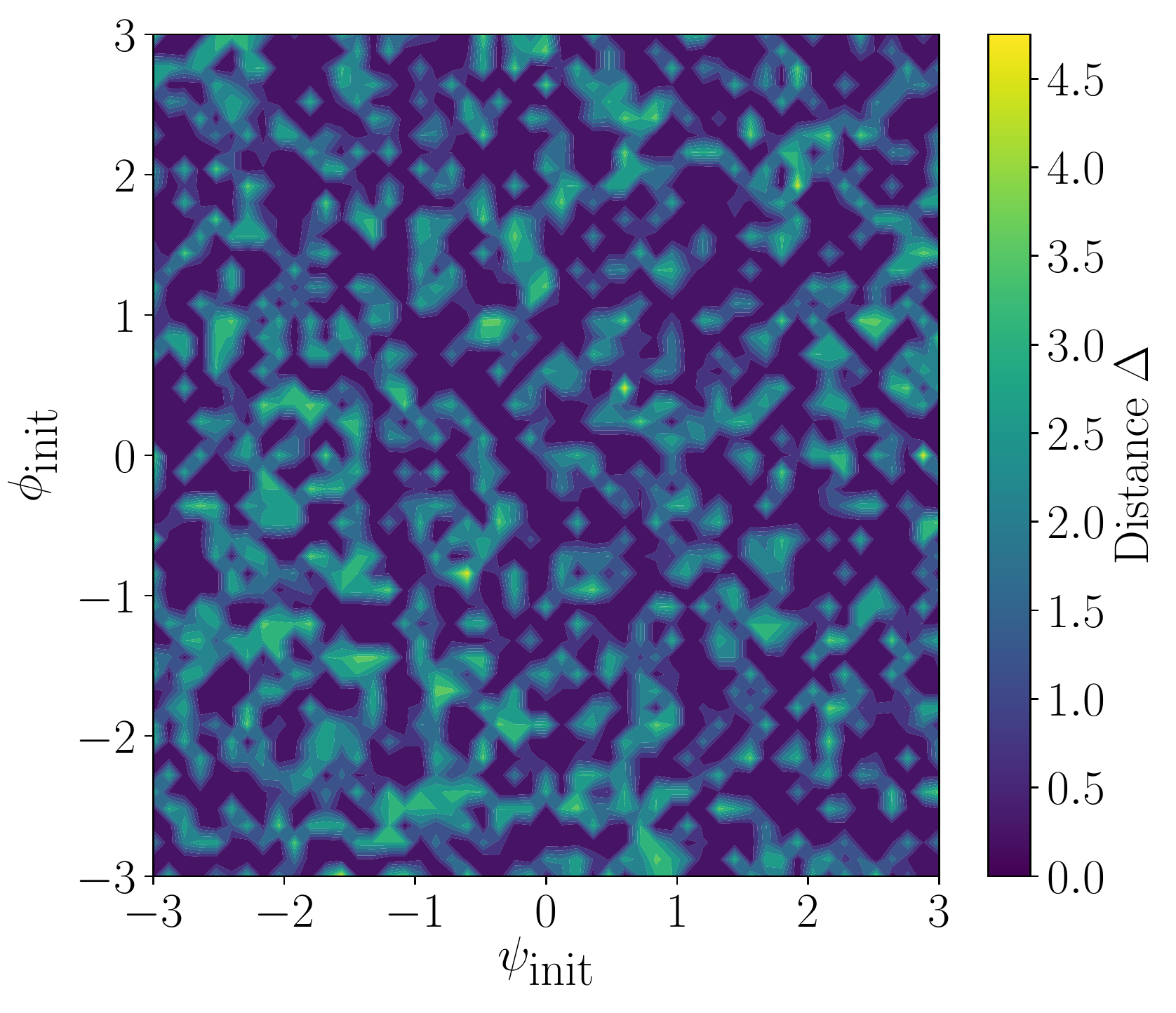}
		\caption{Thermal annealing}
	\end{subfigure}
	\begin{subfigure}{0.24\linewidth} 
		\includegraphics[width=\textwidth]{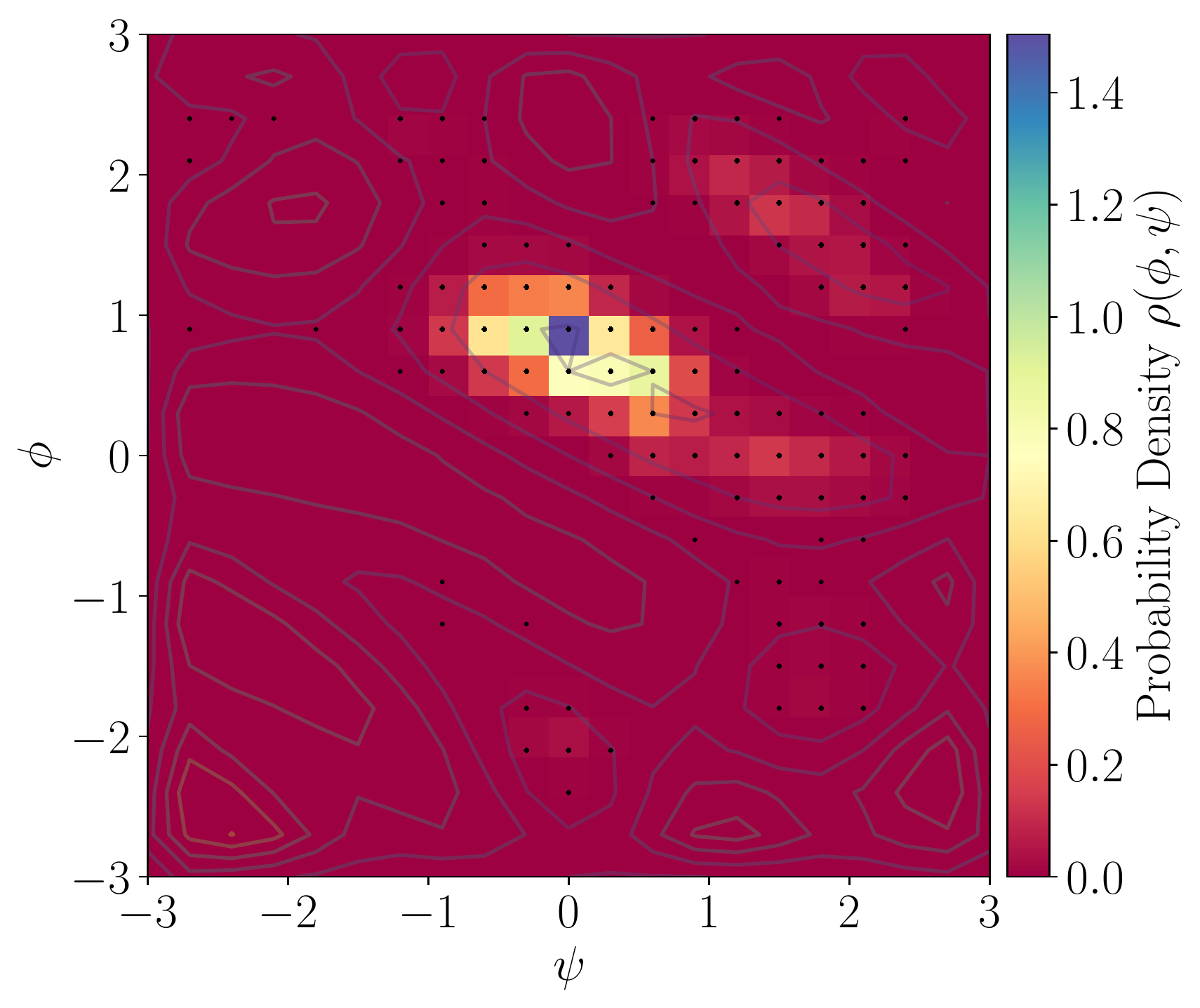}
		\includegraphics[width=\textwidth]{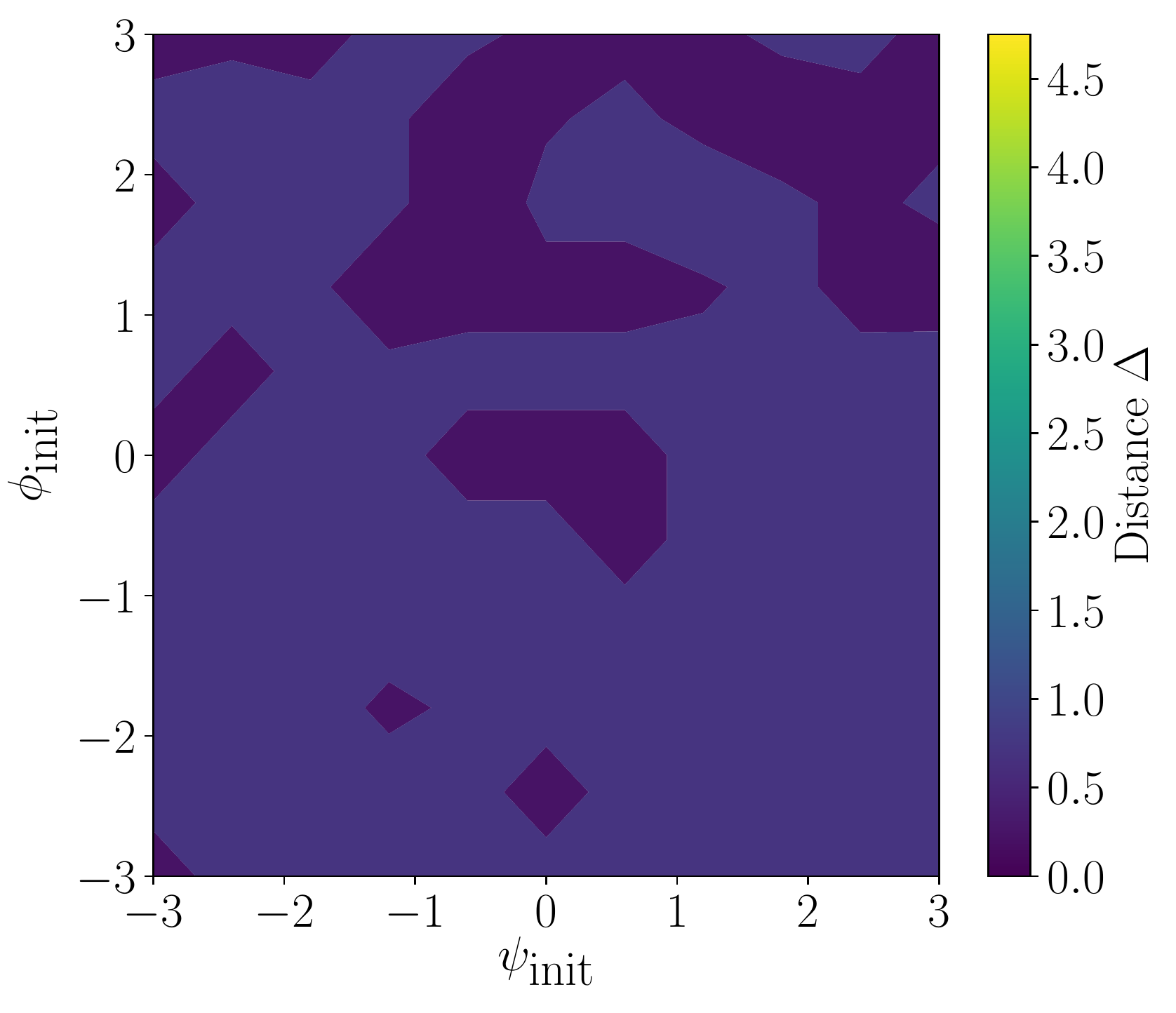}
		\caption{Quantum annealing}
	\end{subfigure}
	\caption{For a range of initial $\phi$ and $\psi$ values the distance between the predicted minimum and true minimum for corrugated potential $U_1$ is given. This is done for (a) Nelder-Mead, (b) gradient descent, (c) thermal annealing and (d) quantum annealing. For (a), (b) and (c) a grid size of $N=50$ and $\lambda=0.5$ is chosen with the models being initialised from 2500 different points. For the quantum annealing (d), we take $N=20$ and $\lambda=0.7$ and the model is initialised from $\sim100$ points. Each point is used to produce  100 reads, with the peak of the probability distribution being identified as the predicted minimum for that setup.
	}
	\label{fig:potential_1_results}
\end{figure*}

For each of the four optimisation methods we start the runs for a selection of initial starting points. This leads to Figure \ref{fig:potential_1_results}. Here, we show the distribution of results from the methods. For the classical methods we use 2500 initial conditions while for the quantum annealer, fewer initial conditions are used, around $100$ due to the limited time on the device. As is usual when performing quantum annealing, each initial position is run multiple times. Here, we choose the number of reads to be 100, with Figure \ref{fig:potential_1_results}d showing every successful point. (An unsuccessful point would be one for example that had ended up with more than two domain walls so that it does not faithfully describe the two variables).   We see that the three classical methods are more liable to fall into and get trapped in false minima than the quantum annealer. 

This effect is surprisingly marked, but in order to quantify it, it is instructive to plot the mean distance from the true minimum that the system ends up, for each given starting point. That is, for each run (i.e. for each particular $(\phi_{init}, \psi_{init})$) we take the result and calculate  the distance of the result to the correct true minimum,  $\Delta$. For the three classical methods this is straightforwardly run-by-run. However, for each set of initial conditions the quantum annealer has {\it already} been run multiple times when the results are collected. To select the predicted value from the set of 100 annealer runs corresponding to a given $(\phi_{init}, \psi_{init})$ we therefore take the peak of a fitted probability distribution. 
We justify this (namely that taking the peak of the distribution of 100 reads is not unfairly favouring the QA)  using the results  presented in Appendix \ref{appendix:starting}, where we show the distribution of $\Delta$ for a selection of initial starting points. We see that for the quantum annealer runs, the distribution is already very focussed around $\Delta=0$ for each of the 100 reads for a choice of $\phi_{init}$ and $\psi_{init}$. (For example taking instead the mean of the 100 reads would have produced very similar results.)

The lower panels of Fig.~\ref{fig:potential_1_results} show the resulting $\Delta$'s. Clearly NM and GD  perform very badly in such a potential, and we can  identify ``basins of attraction'' around the local minima from which they never escape. The TA performs noticeably better in the sense that it is not susceptible to just getting trapped in the closest minimum to its starting point, but on the other hand there is a remaining randomness to the minimum that it eventually {\it does} end up in. 
However we see again that the quantum annealer gets consistently much closer to the true minimum of the potential from virtually any starting point. Moreover this does not appear to depend on the size of the array (i.e. $N$) as we shall shortly demonstrate.

\subsection{\label{Sec:tanh} Multi well potential $U_2$}

\begin{figure*}[htb]
	\centering
	\begin{subfigure}{0.4\linewidth} \centering
		\includegraphics[width=\textwidth]{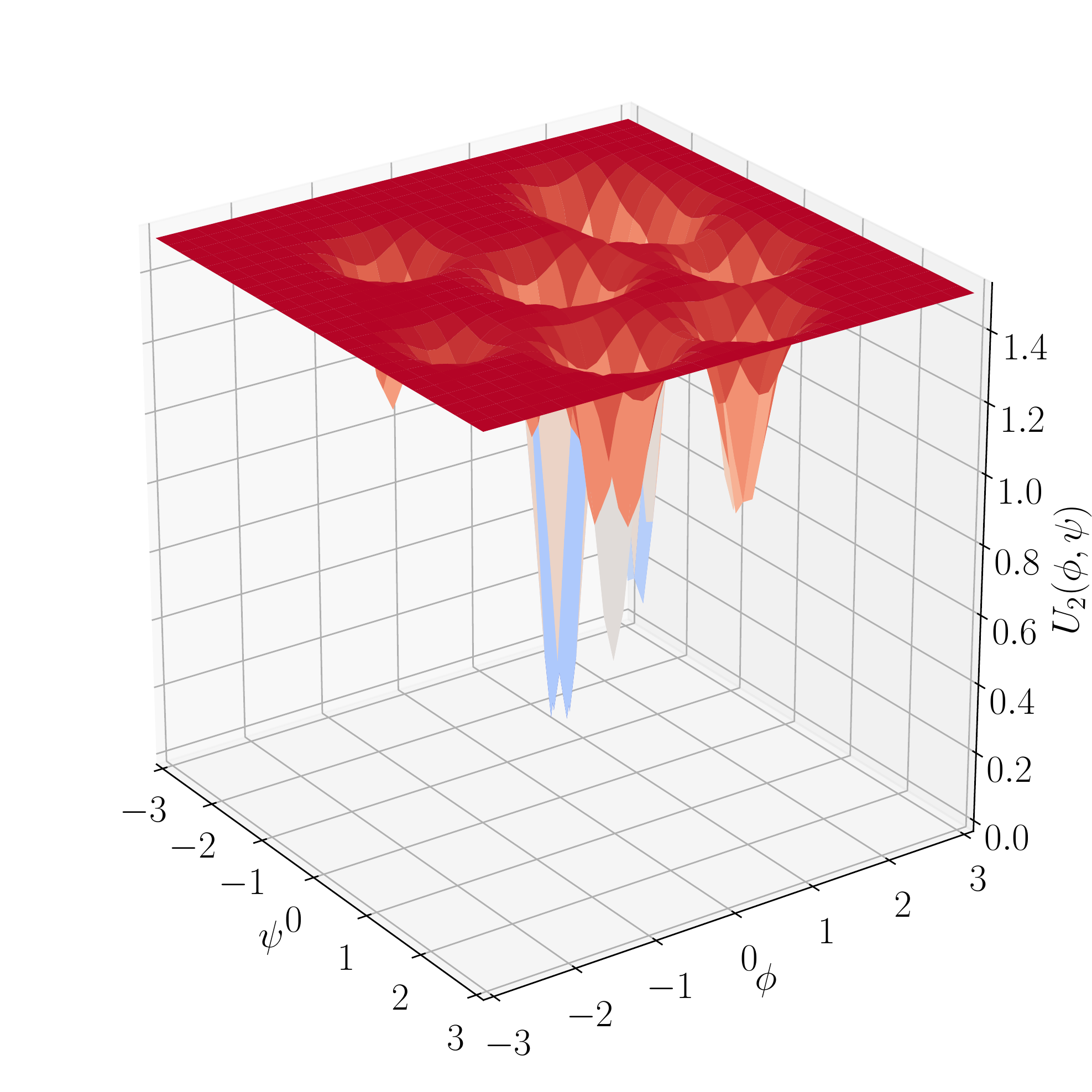}
		\caption{}
	\end{subfigure}
	\begin{subfigure}{0.4\linewidth} \centering
		\includegraphics[width=\textwidth]{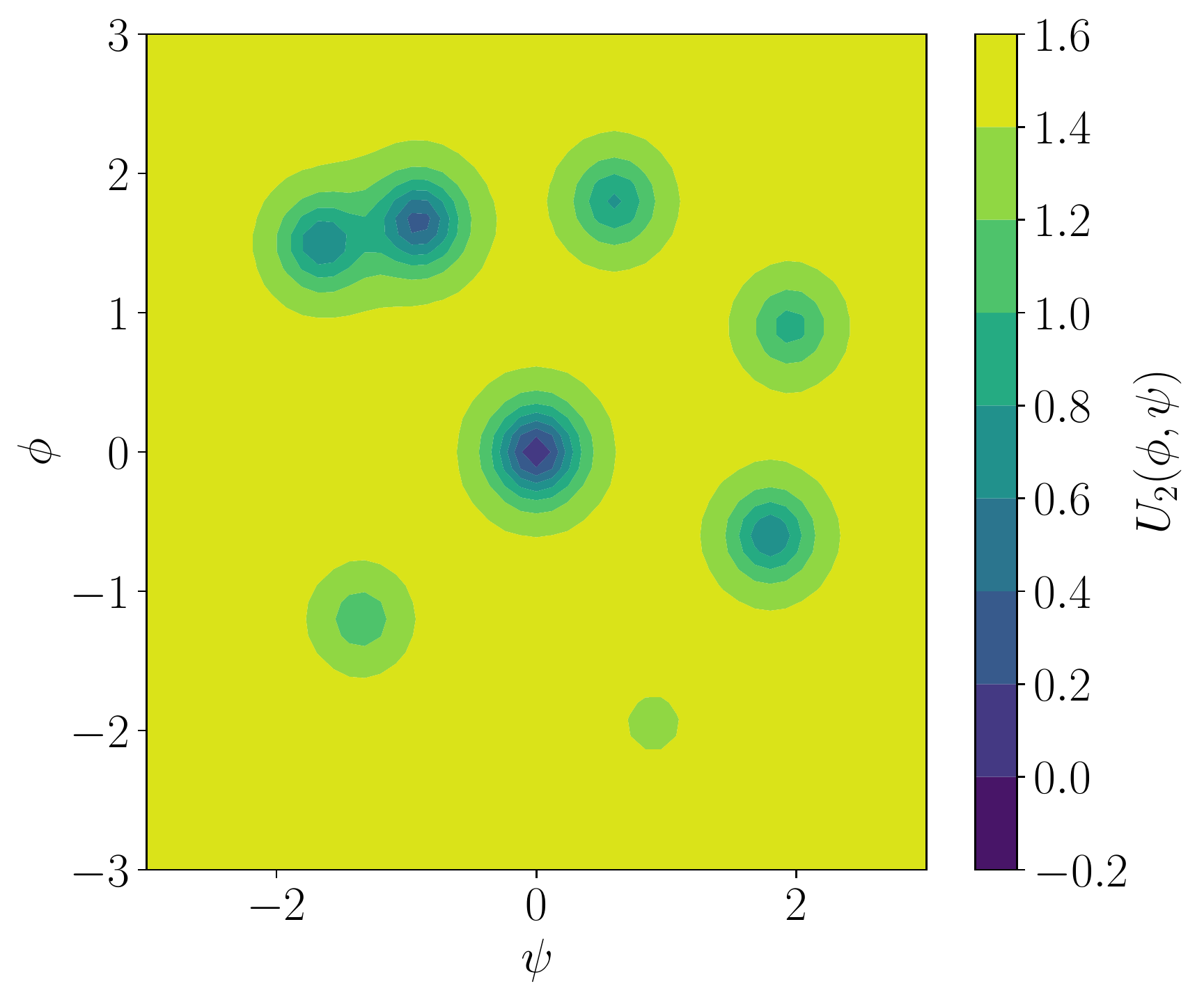}
		\caption{}
	\end{subfigure}
	\caption{The potential $U_2$ of Eq.~\eqref{eq:potential_2} rendered in 3D (a) and as a contour plot (b).}
	\label{fig:tanh_potentials}
\end{figure*}

To show the versatility of our method we now perform the same analysis on another potential that is even more difficult. 
This potential has the form of a flat plateau with multiple holes of varying depths, and is given by
\begin{align}
U_2(\vec{\phi}=(\phi, \psi))  &~=~\lambda\,\, \mbox{\Large (}  \, p_0 \,{\tanh}^2({ |\vec{\phi}|/\omega })
\label{eq:potential_2}	\\
&\qquad\,+\,\sum_a p_a \, \mbox{sech}^2  (| \vec{\phi}-\vec{v}_a|/\omega) \mbox{\Large )} ~,  \nonumber
\end{align}
where $(p_0,p_a)$ is a choice of minimum depths with $p_0> p_a$, where $\vec{v}_a$ is a choice of positions for each minimum, $\omega$ is the width of all the dips, and where $\lambda$ is again an overall scaling parameter. We select $\omega$ to be 0.3 and $\lambda$ to be 0.5 and 10 for the classical and quantum methods, respectively. The depths of the minima we took to be 
\begin{align}
(p_0,p_a)~=~(\, 3,0.9,0.3,1.2,1.8,1.5,1.8,2.4\,)~,
\end{align}
while the positions are given by
\begin{align}
	\vec{v}_a ~=~  [\, &(-1.2,-1.35),(-1.95,0.9), \nonumber \\
	&(0.9,1.95), (1.5,-1.65), \\\nonumber
	&(1.8,0.6), (-0.6,1.8), (1.65,-0.9)\, ]_a~ .
\end{align}
The resulting potential $U_2$ is shown in Fig.~\ref{fig:tanh_potentials}. 

We now carry out the same analysis as in Section \ref{Sec:potential_1}. During this analysis the thermal annealer is run on the same schedule as before. The quantum annealer is scheduled to ramp up for $15 \mu {\rm s}$  before tunnelling for $50 \mu {\rm s}$ and ramping down for 50 $\mu {\rm s}$. During the tunnel phase of the anneal $s=0.15$.

\begin{figure*}[htb!]
	\centering
	\begin{subfigure}{0.24\linewidth}
		\includegraphics[width=\textwidth]{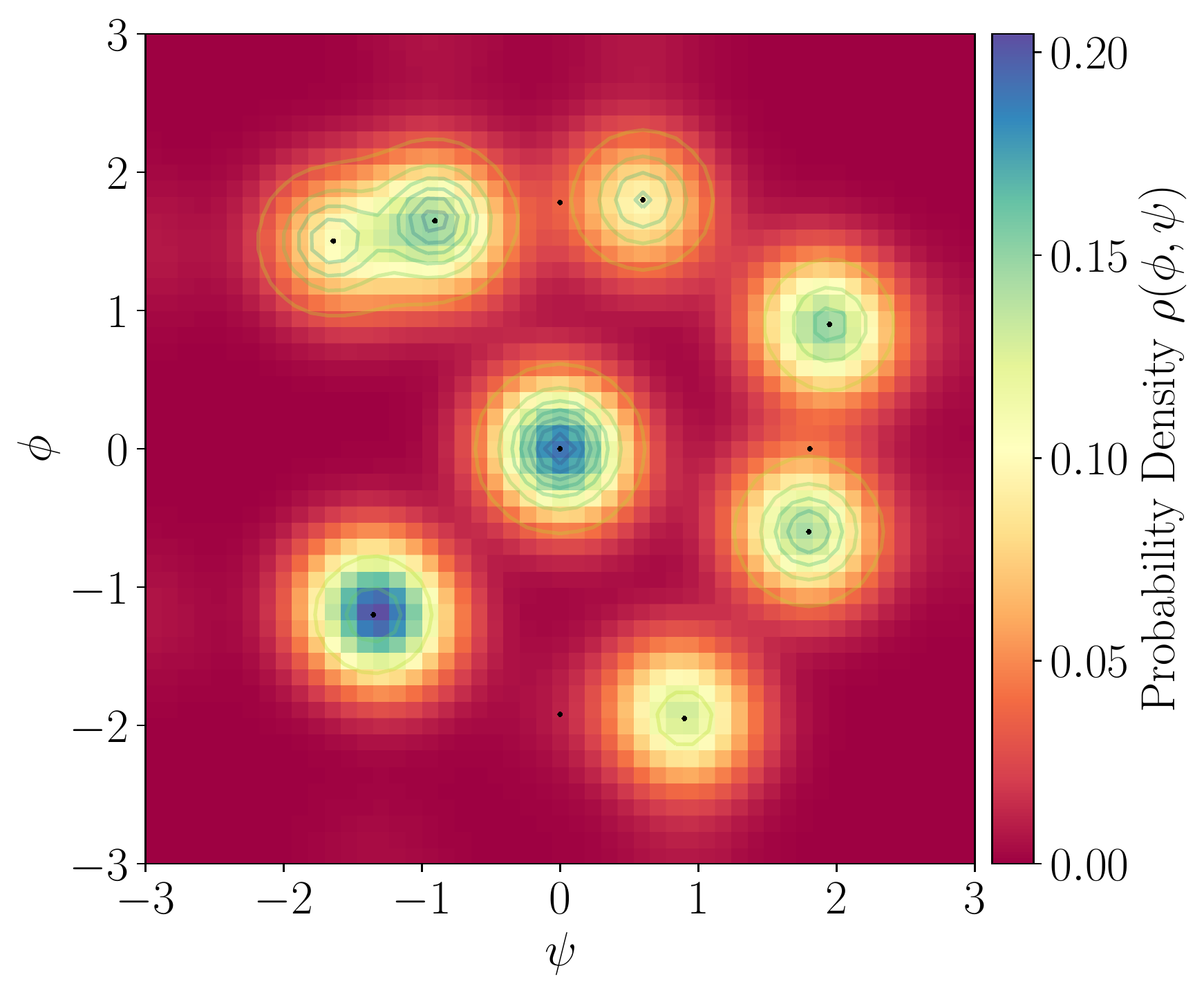}
		\includegraphics[width=\textwidth]{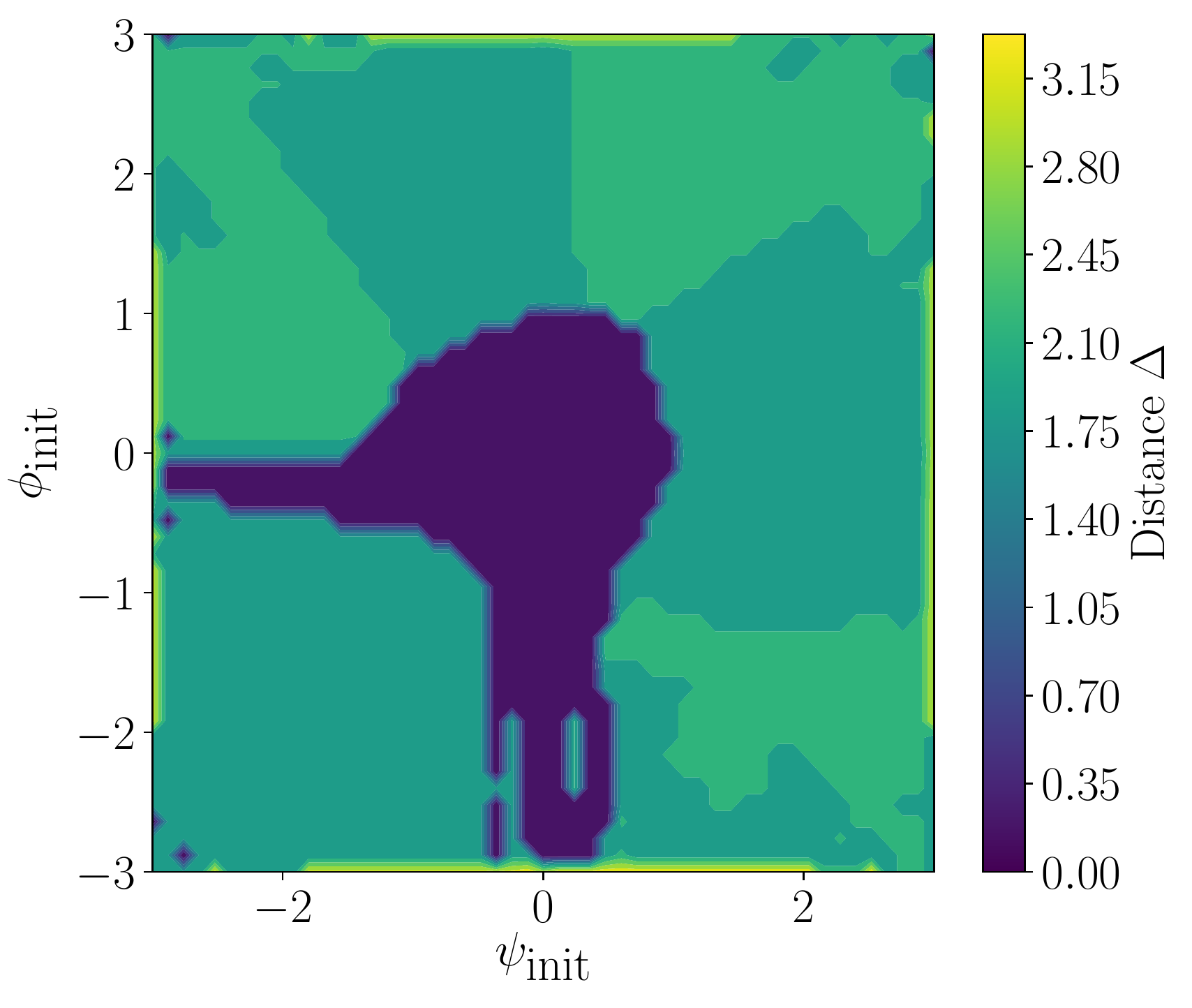}
		\caption{Nelder-Mead}
	\end{subfigure}
	\begin{subfigure}{0.24\linewidth} 
		\includegraphics[width=\textwidth]{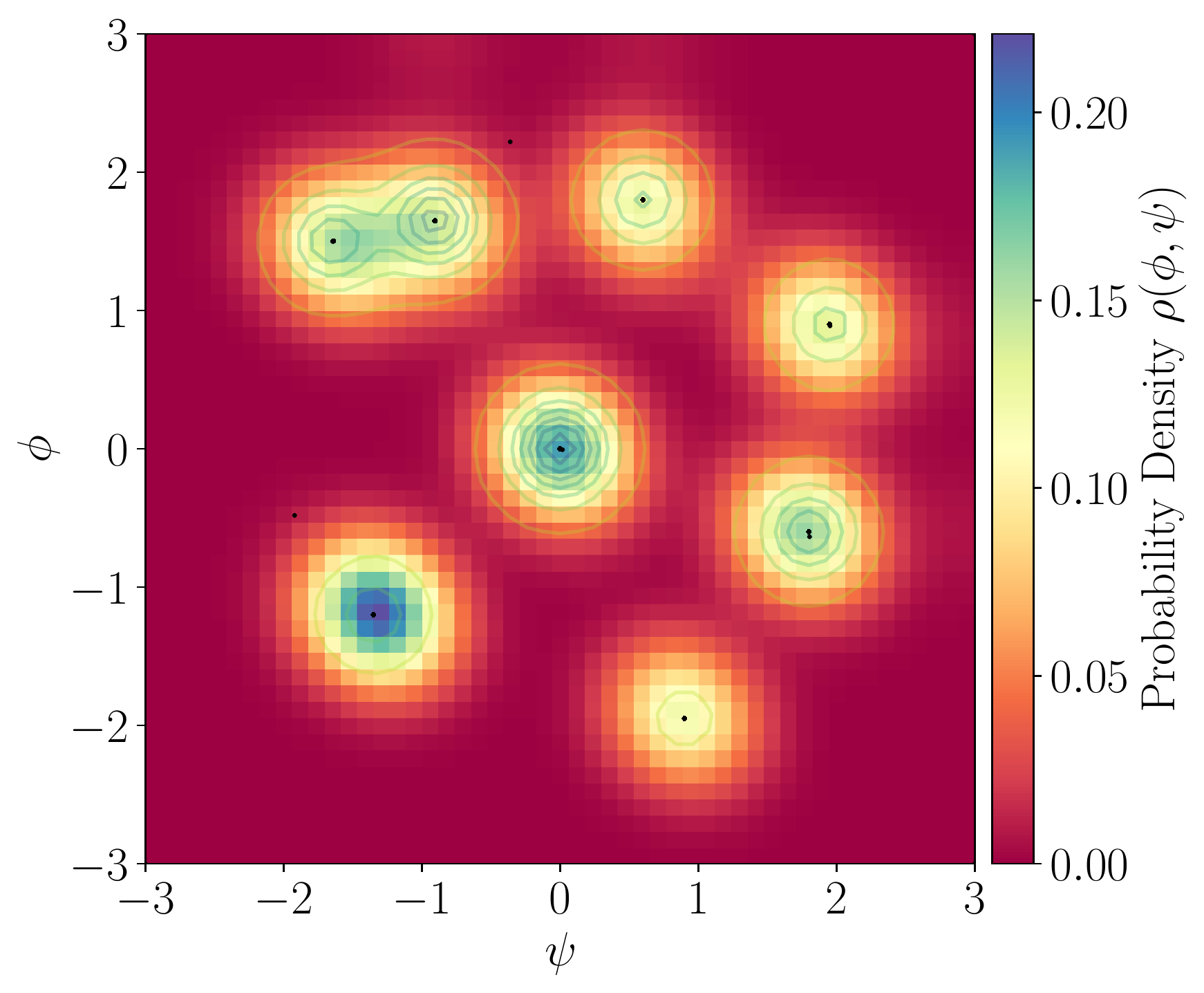}
		\includegraphics[width=\textwidth]{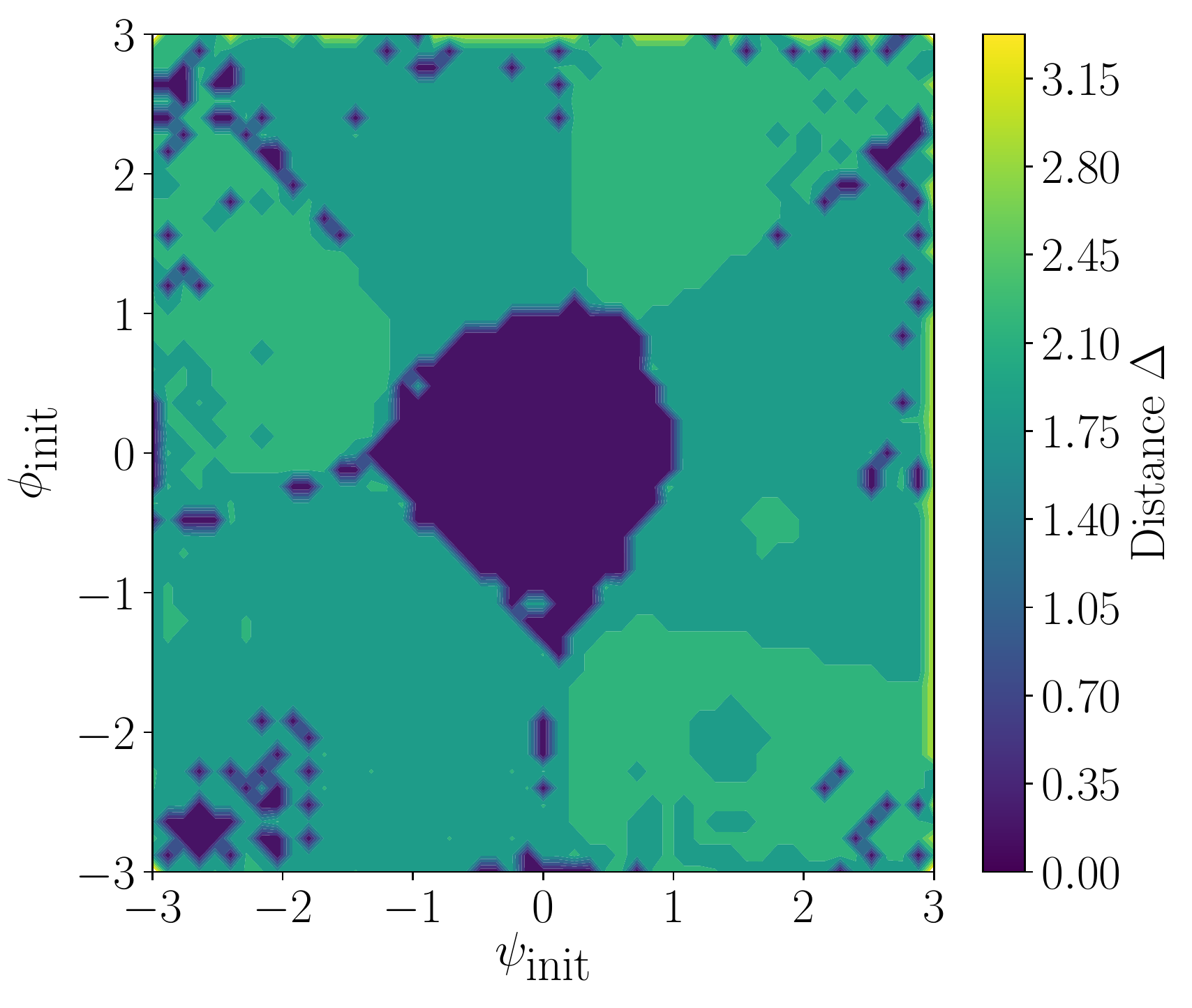}
		\caption{Gradient descent}
	\end{subfigure}
	\begin{subfigure}{0.24\linewidth} 
		\includegraphics[width=\textwidth]{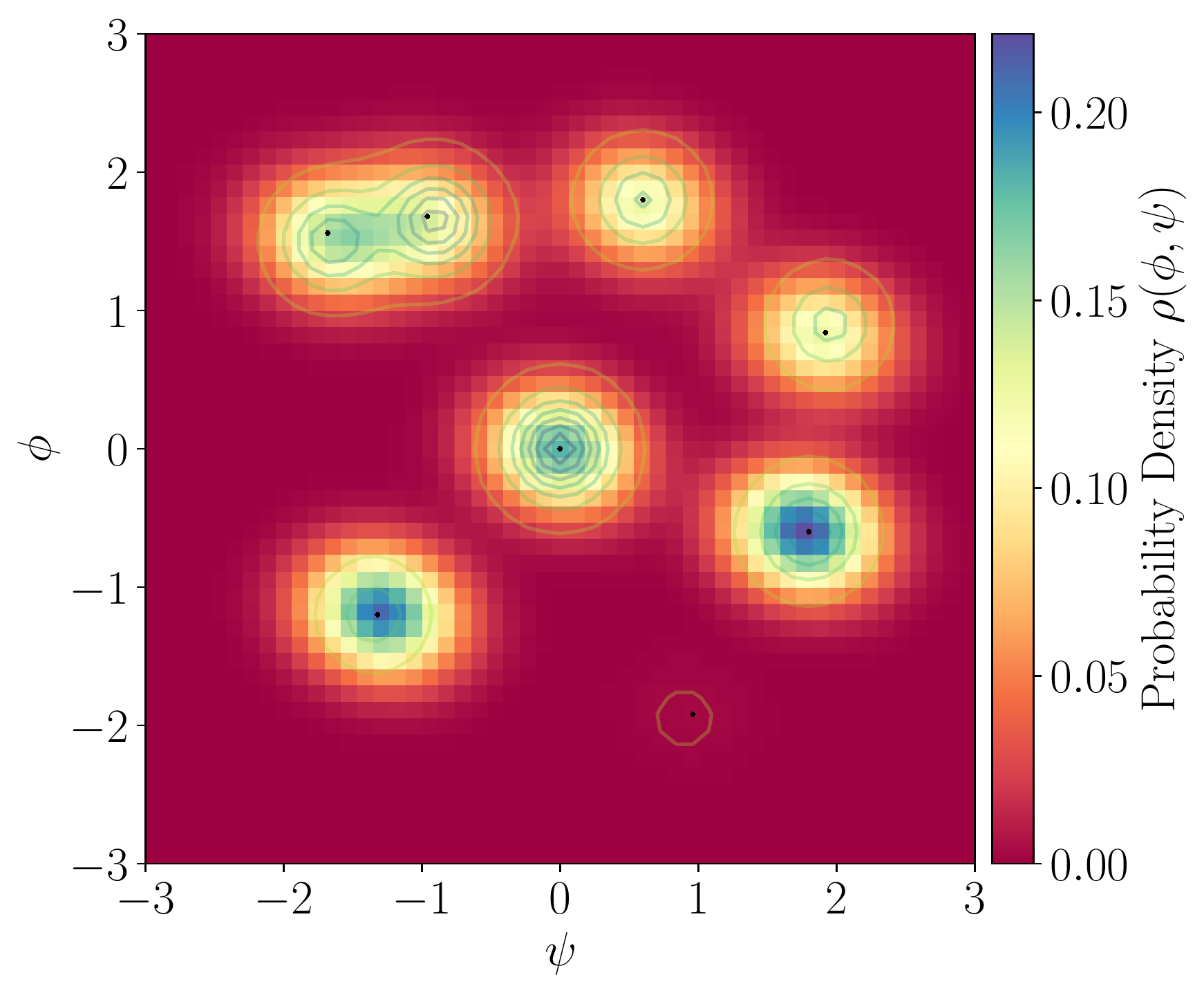}
		\includegraphics[width=\textwidth]{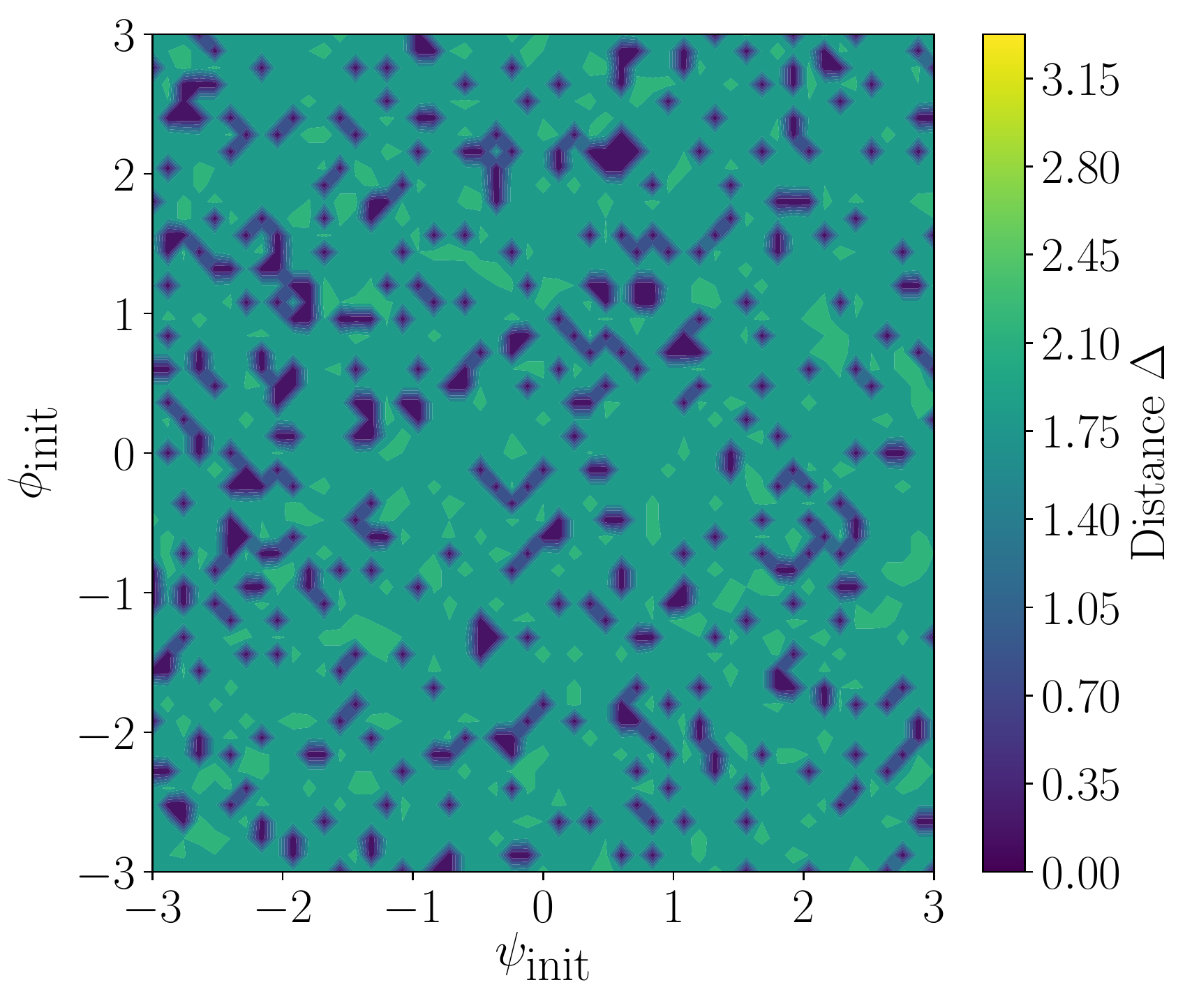}
		\caption{Thermal annealing}
	\end{subfigure}
	\begin{subfigure}{0.24\linewidth} 
		\includegraphics[width=\textwidth]{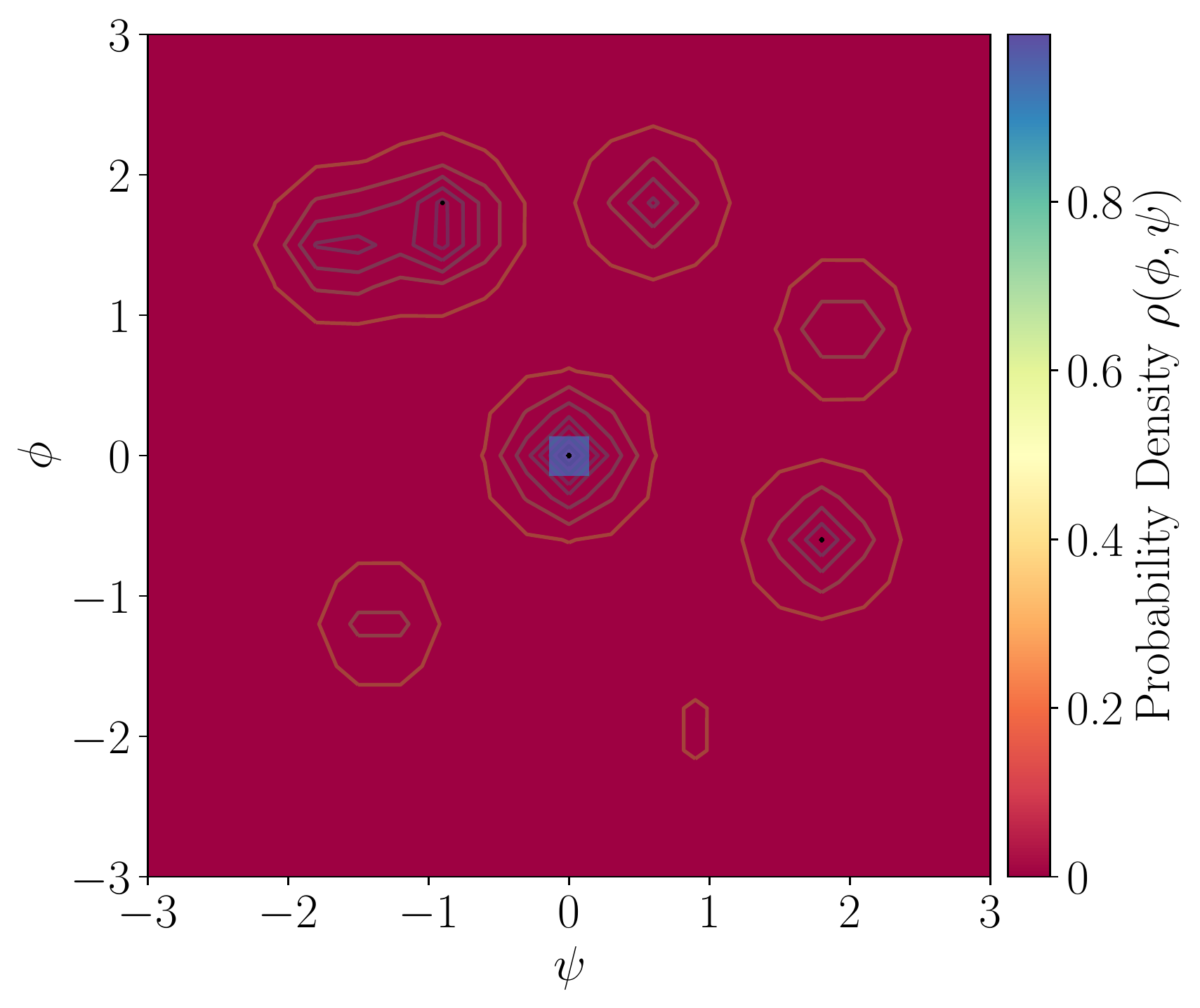}
		\includegraphics[width=\textwidth]{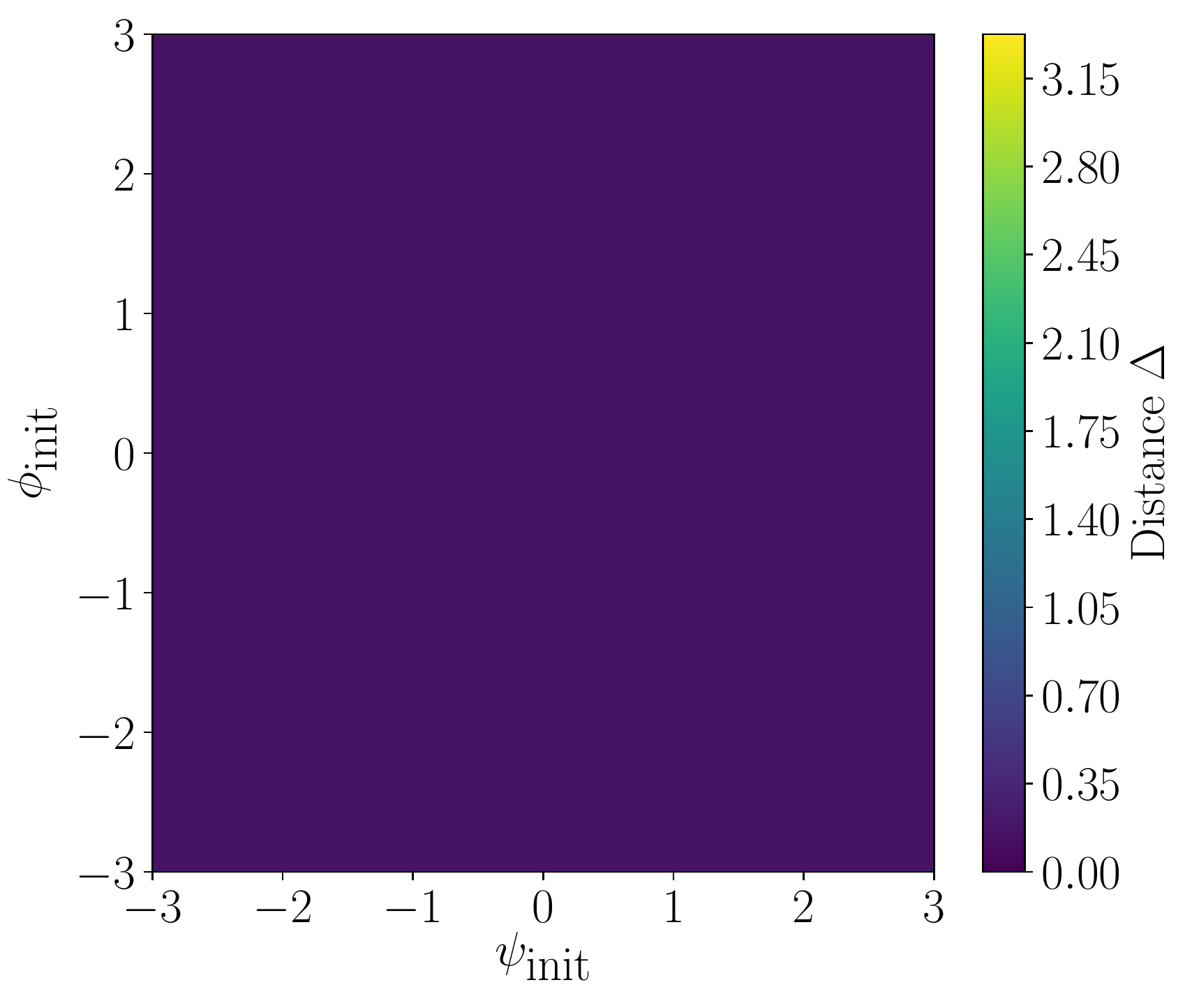}
		\caption{Quantum annealing}
	\end{subfigure}
	\caption{
		For a range of initial $\phi$ and $\psi$ values the distance between the predicted minimum and true minimum for multi-well potential $U_2$ is given. This is done for (a) Nelder-Mead, (b) gradient descent, (c) thermal annealing and (d) quantum annealing. For (a), (b) and (c) a grid size of $N=50$ and $\lambda=0.5$ is chosen with the models being initialised from 2500 different points. For the quantum annealing (d), we take $N=20$ and $\lambda=10$ and the model is initialised from $\sim100$ points. Each point is used to produce  100 reads, with the peak of the probability distribution being identified as the predicted minimum for that setup.}
	\label{fig:potential_2_results}
\end{figure*}

Running these optimisers, alongside the NM and CG methods, results in Figure \ref{fig:potential_2_results}. As was the case with the first potential, the quantum annealer performs more consistently than the classical methods. Instead of falling into one of the many local minima it is able to find the truth on almost every run. %Figure \ref{fig:initial_distances_tanh} shows the distribution of distances for four initial conditions, for the four methods. 

\begin{table}[]
	\centering
	\begin{adjustbox}{}
		\begin{tabular}[t]{lcccc}
			\hline
			\textbf{Method} & \textbf{Time/run  ($\mu$s)} \\ 
			\hline
			Nelder-Mead  &             4900             \\ 
			Gradient Descent    &           2900                \\ 
			Thermal Annealing   &       $5\times 10^5$          \\ 
			Quantum Annealing        &        115            \\
			\hline
		\end{tabular}
	\end{adjustbox}
	\caption{Timings for the four optimsation methods solving potential 2. For NM and CG $N=50$ while a value of $N=20$ is used to calculate both annealing methods.}
	\label{tab:timings}
\end{table}

So far, we have focused on the consistency of the quantum annealer. Another benefit is its speed. The quantum annealer's runtime is decided by its schedule. Here, that comes to $115 \mu {\rm s}$. The classical NM and GD methods are very dependent on the number of iterations each takes. By timing how long each optimisation takes that is used to create Figure \ref{fig:potential_2_results} we can find an average run-time for each method. Table \ref{tab:timings} shows the results. For the Nelder-Mead, an average optimisation takes 0.0049\,s, for the gradient descent method it takes 0.0029\,s. Thermal Annealing, by far, takes the longest. One thermal anneal run takes around half a second. For fair comparison, the thermal annealer time is calculated for an $N=20$ scenario.

%\FloatBarrier 

\subsubsection{Scaling the potential up or down}
\label{sec:deeper}
%\FloatBarrier

All three potentials chosen can be scaled up or shrunk down by adjusting the factor $\lambda$. Such a uniform operation does not change the position of the global minimum in the $(\phi,\psi)$ plane. Intuitively, it is clear what effect modifying this has on one of the classical methods. By increasing the depth of both local and global minima by the same amount, the probability for the classical algorithms to settle in a local minimum rather than the global remains unchanged to a good approximation. Classical algorithms that do not sample the entire configuration space are unaware of the depth of the extremum and whether they located the global one. However, the quantum annealer is instead aware of the depth of the potential well and by increasing the depth the quantum annealing process becomes more successful.

\begin{figure*}[htbp!]
	\centering
	\begin{subfigure}{0.32\linewidth} \centering
		\includegraphics[width=\textwidth]{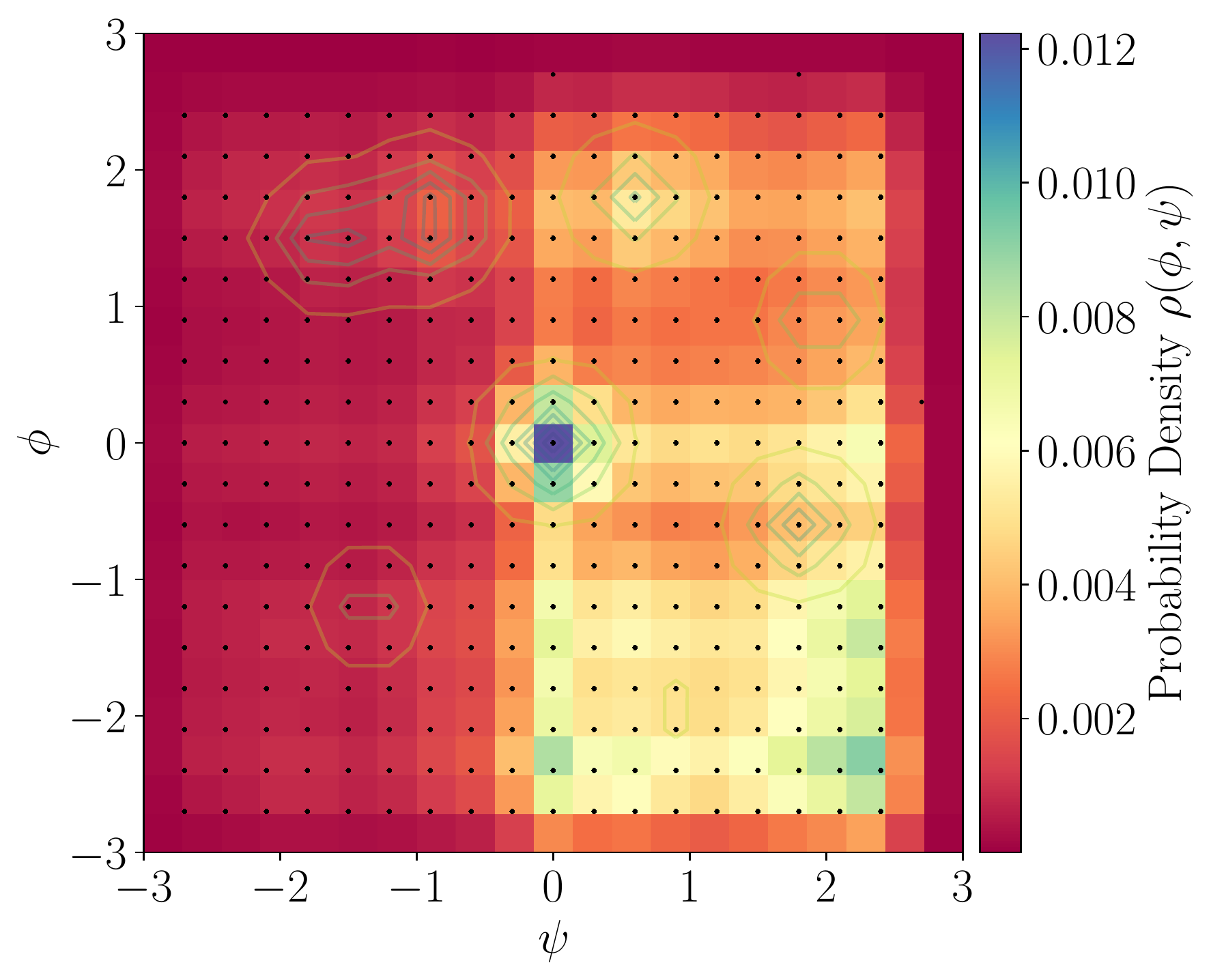}
		\caption{$\lambda=0.5$}
	\end{subfigure}
	\begin{subfigure}{0.32\linewidth} \centering
		\includegraphics[width=\textwidth]{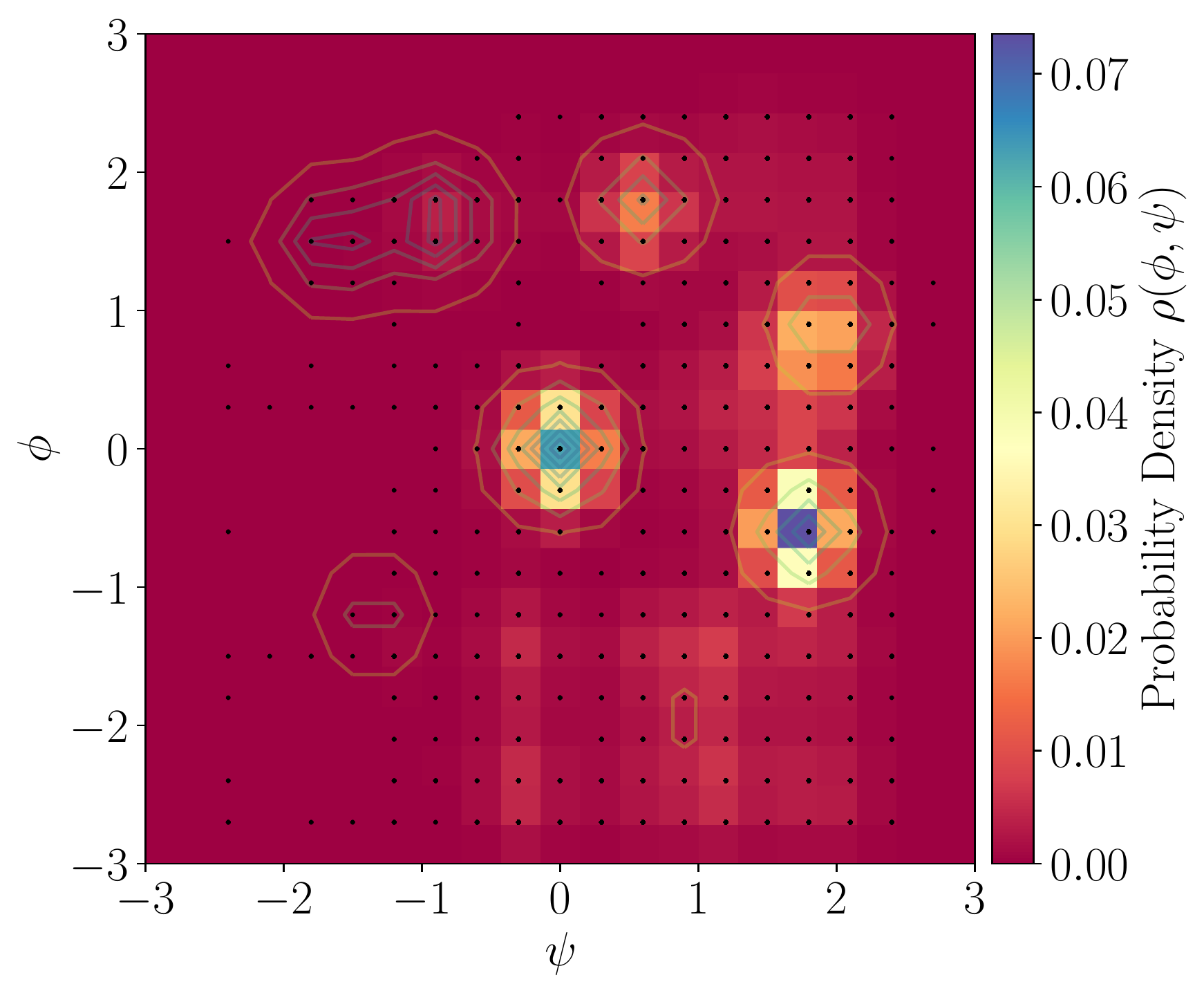}
		\caption{$\lambda=5$}
	\end{subfigure}
	\begin{subfigure}{0.32\linewidth} \centering
		\includegraphics[width=\textwidth]{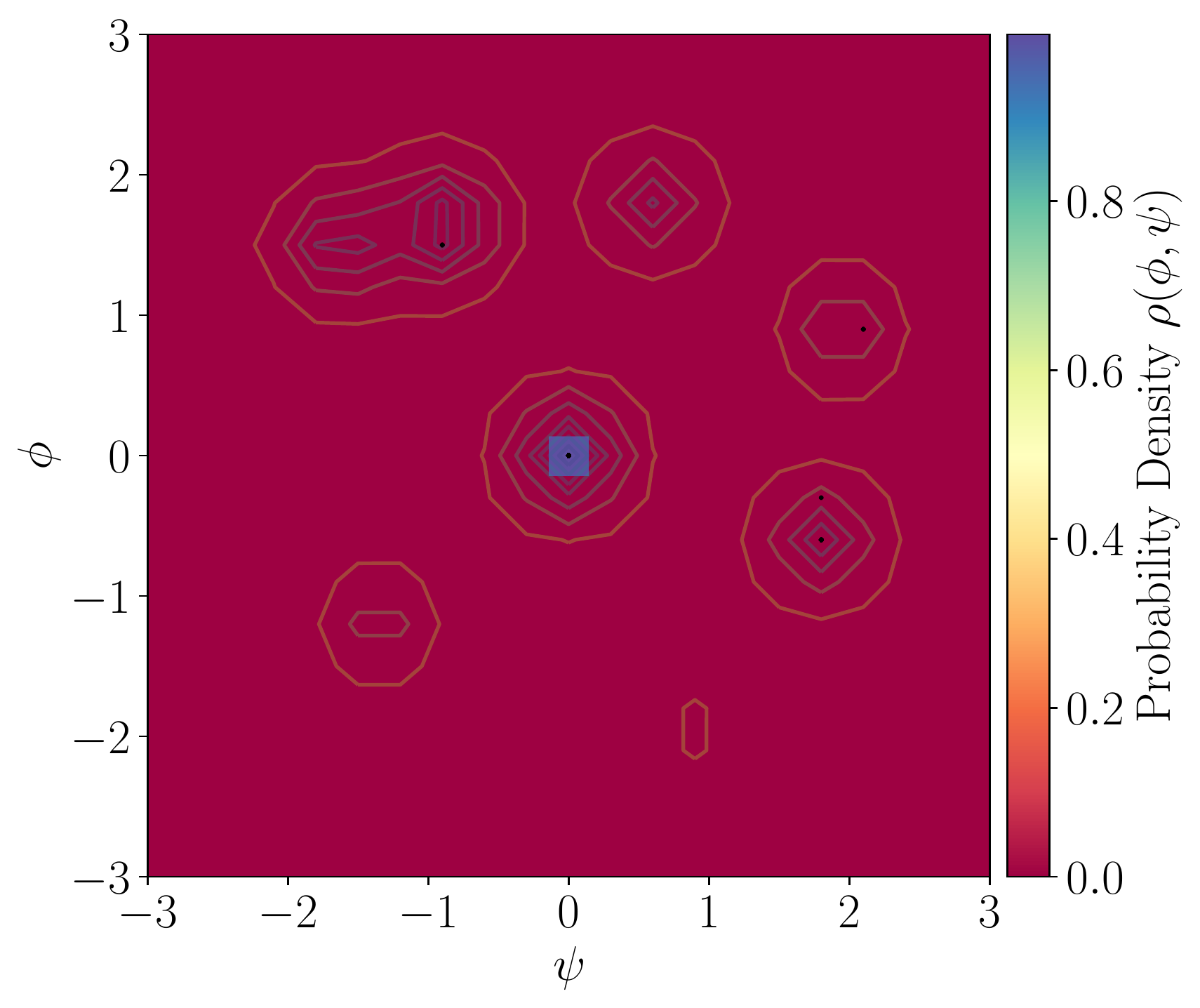}
		\caption{$\lambda=10$}
	\end{subfigure}
	\caption{The quantum annealer optimising the multi-well potential, $U_2$, for three values of the scaling parameter $\lambda$. All runs use $N=20$, a tunnelling and ramp down period that are both $50 \mu {\rm s}$, respectively and $s=0.15$.}
	\label{fig:Lambda}
\end{figure*}

We demonstrate this using the the multiwell potential example of Section \ref{Sec:tanh}. Fig.~\ref{fig:Lambda} shows the results of running the quantum annealer for a variety of values of $\lambda$. Starting from a low value of $\lambda$, the quantum annealer has difficulty finding the minima. As we increase the depth of the potential's minima from $\lambda=0.5$ to $\lambda=10$ (the choice we made in Section \ref{Sec:tanh}) the annealers ability greatly increases.

\subsubsection{\label{Sec:grid} Effect of varying the grid size}
\begin{figure*}[htbp!]
	\centering
	\begin{subfigure}{0.4\linewidth} \centering
		\includegraphics[width=\textwidth]{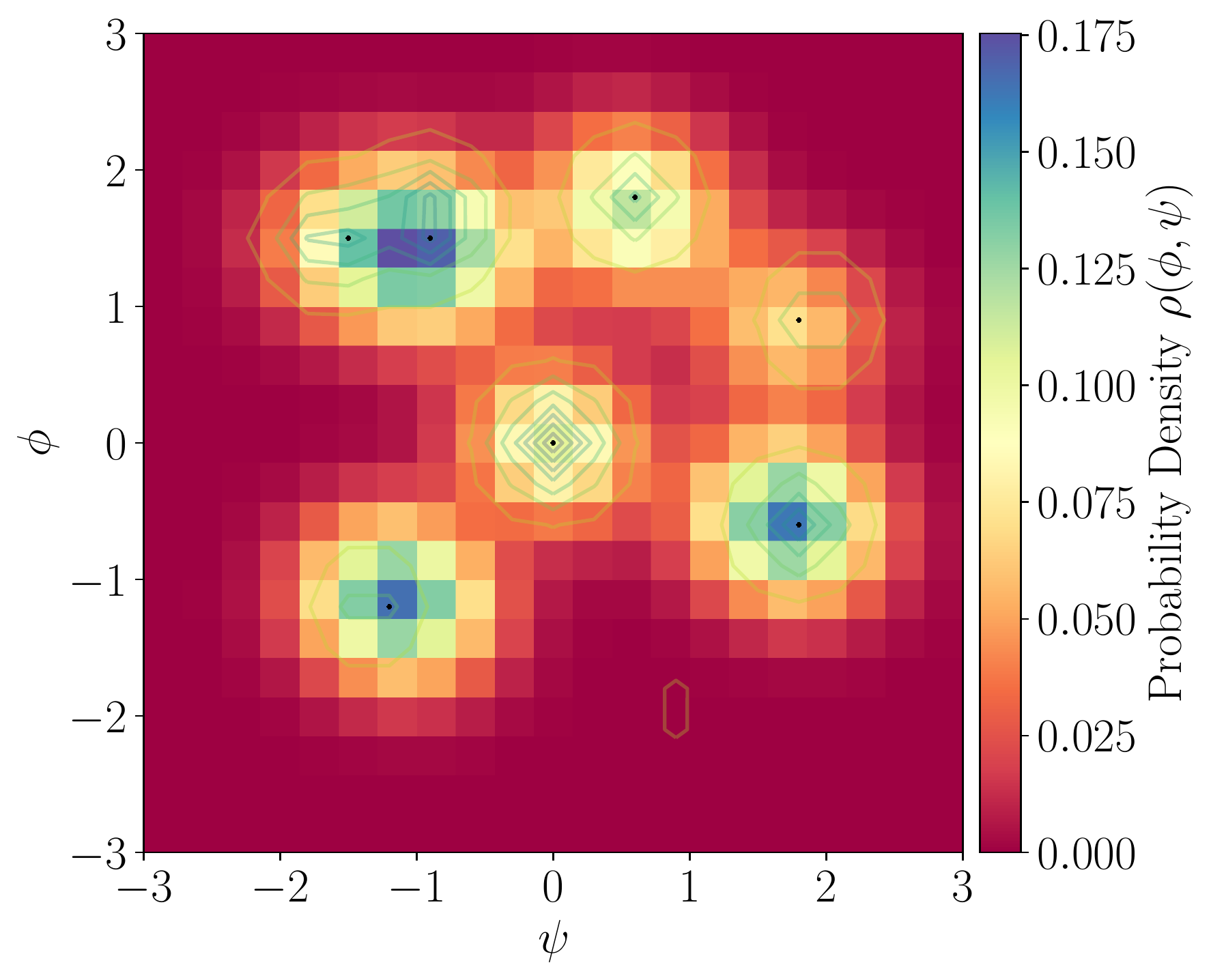}
		\caption{Thermal anneal on a $20\times 20$ grid}
	\end{subfigure}
	\begin{subfigure}{0.4\linewidth} \centering
		\includegraphics[width=\textwidth]{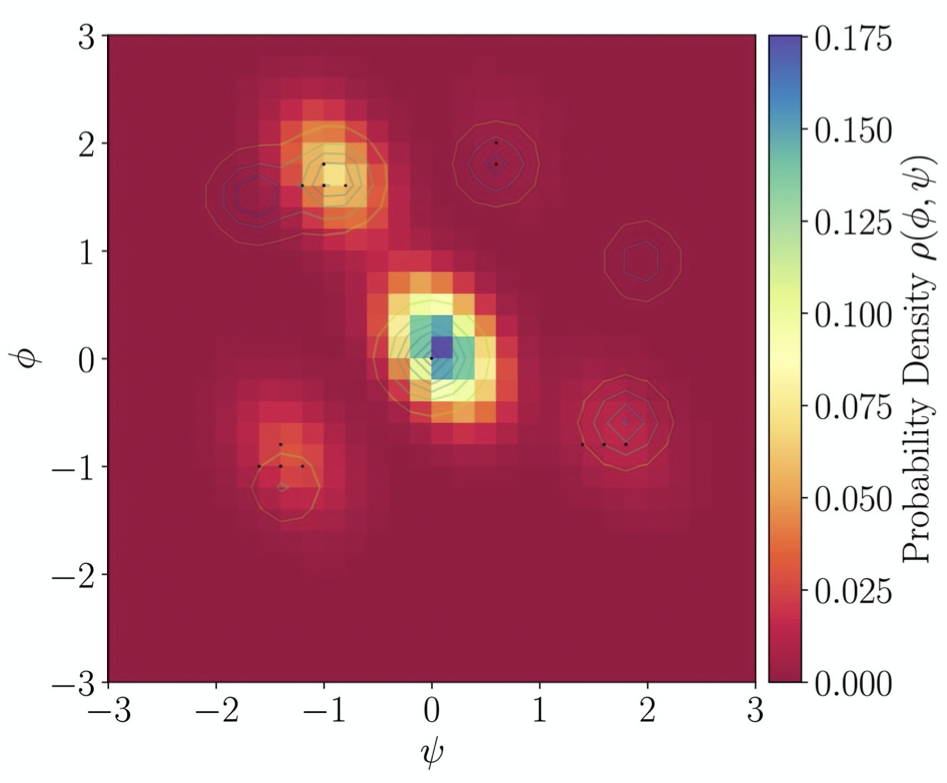}
		\caption{Quantum anneal with a $30\times 30$ grid}
	\end{subfigure}
	\caption{The results of thermal annealing on an $N=20$ grid (a) and quantum annealing on an $N=30$ grid (b) for the multi-well potential.}
	\label{fig:thermal2020}
\end{figure*}

It is worth briefly discussing our choice of $N$ in all the examples we have discussed. In particular an obvious remaining question is whether the different choices of $N$ and the lower possible $N$-values for the quantum annealer may somehow have favoured it. A direct comparison shows that this possibility can be dismissed. In Fig.~\ref{fig:thermal2020}a we perform a thermal anneal taking $N=20$. This shows us that the thermal annealing processes are not greatly 
influenced by the fineness of the discretisation of the potential.  We can see it is still unable consistently to find the true minimum. Likewise for the quantum annealer, the result is also independent of $N$. In Fig.~\ref{fig:thermal2020}b we present an $N=30$ run, showing that the same good results obtain, and that the superior performance of the quantum annealer persists even when it has a finer lattice. The grid size does not effect the qualitative features of the annealer's ability to find the minimum. Indeed the performance of both kinds of annealer is relatively independent 
of $N$ as would be expected on physical grounds. As mentioned our choice of $N=20$ for most of the QA studies shown here is mostly a technical limitation, not a conceptual one. 

\subsection{\label{Sec:potential_three} Volcano crater potential $U_3$}

\begin{figure*}[htbp!]
	\centering
	\begin{subfigure}{0.4\linewidth} \centering
		\includegraphics[width=\textwidth]{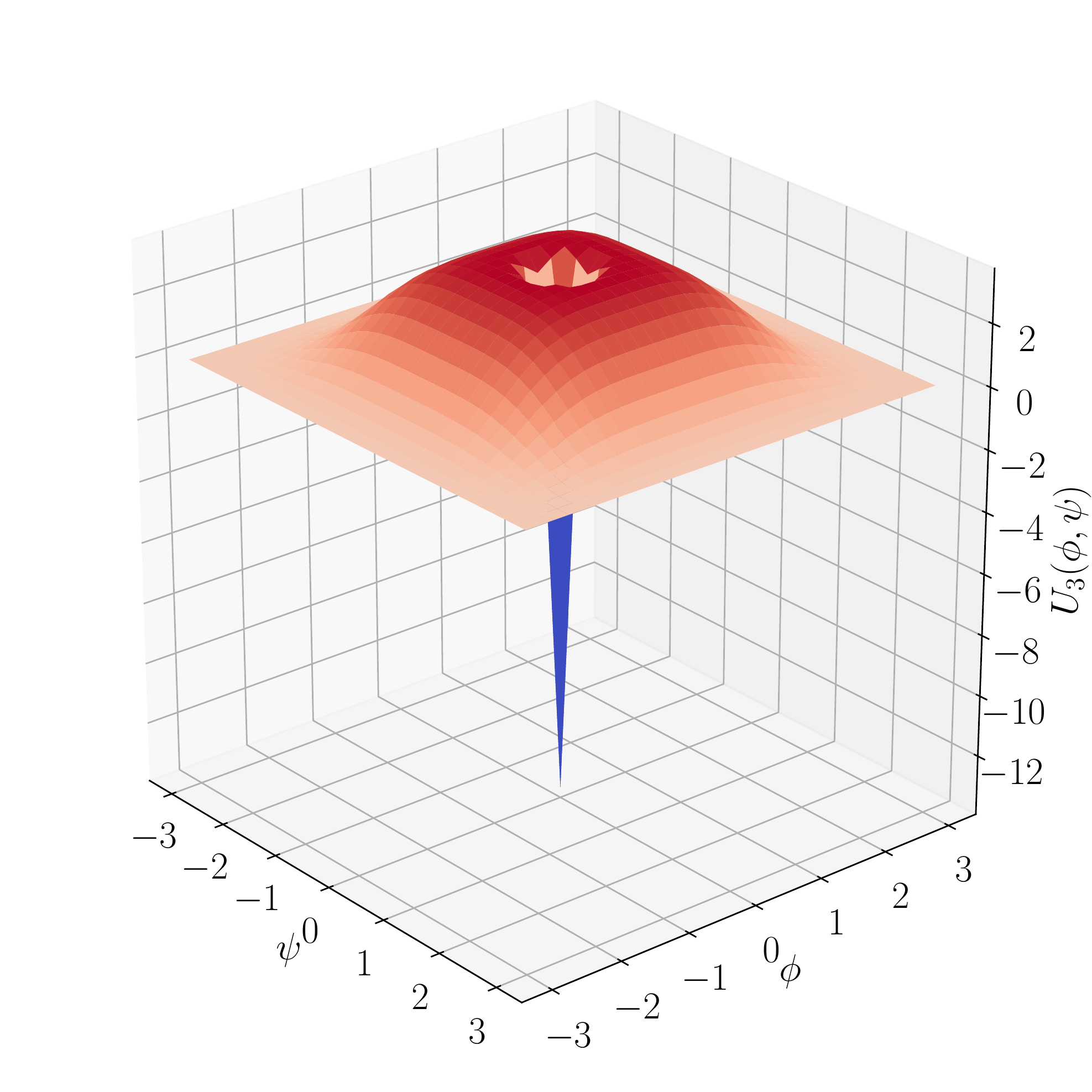}
		\caption{}
	\end{subfigure}
	\begin{subfigure}{0.4\linewidth} \centering
		\includegraphics[width=\textwidth]{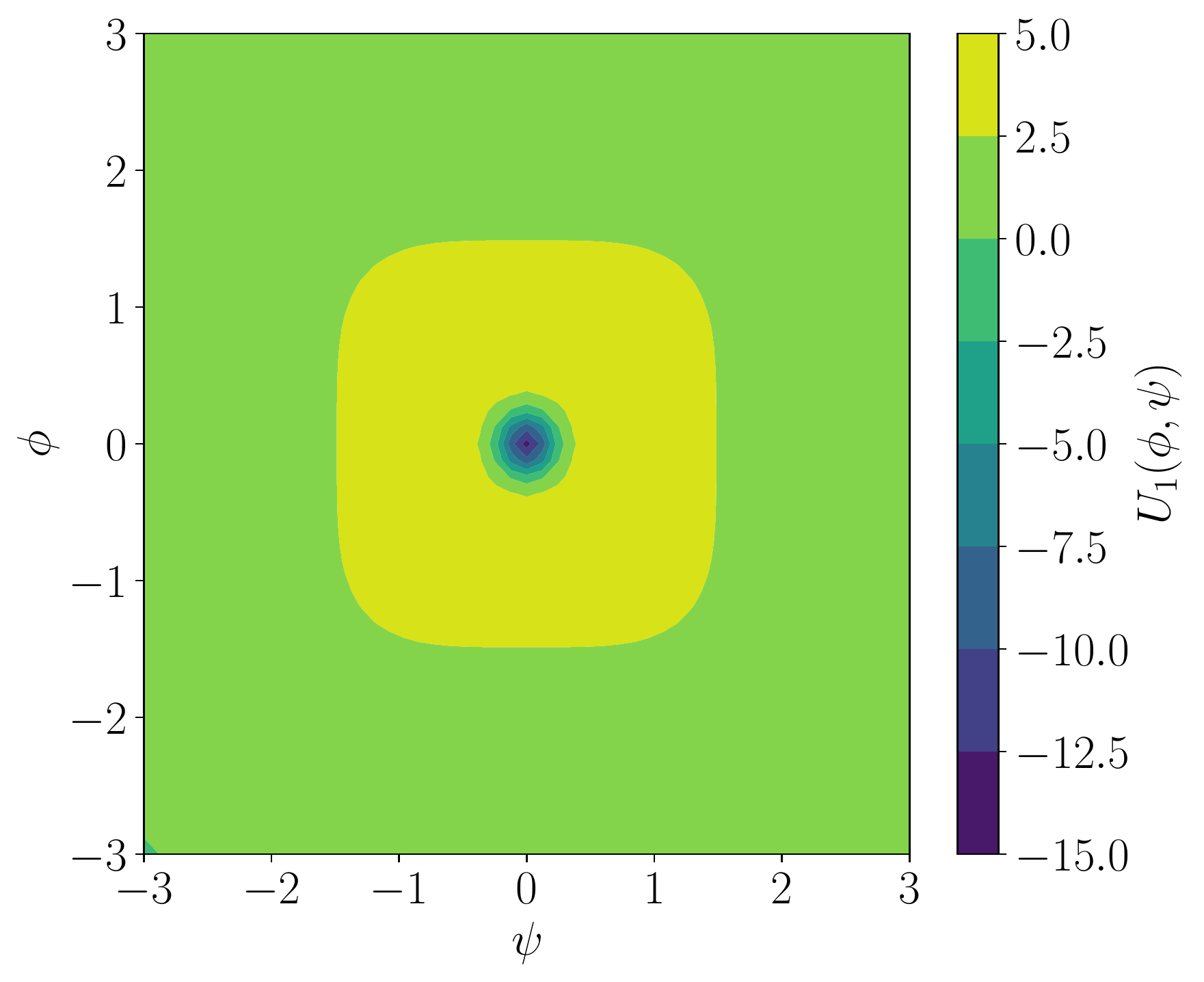}
		\caption{}
	\end{subfigure}
	\caption{A 3D rendering of the volcano potential of Eq.~\eqref{eq:potential_3} in (a), and contour plot in (b)}
	\label{fig:three_potentials}
\end{figure*}

Finally we consider a third potential that takes the following form
\begin{align}
\label{eq:potential_3}
U_3(\phi, \psi)  &~=~ \lambda\,\, \mbox{\Large (}  
2~e^{-(\phi^4+\psi^4 )/2}  \\
&\qquad - 10\, e^{ -20(\phi^2+\psi^2) }
% \nonumber \\ & \times~
\cos^2(\psi)\cos^2(\phi) \mbox{\Large )} ~,  \nonumber
\end{align}
where $\lambda$ is again a scaling factor, which is here chosen to be 1.7 for all the cases.
This particular ``volcano crater'' potential is actually adapted directly from Ref.~\cite{Balazs:2021uhg}. As can be seen from Fig.~\ref{fig:three_potentials} 
it is perhaps the single most difficult potential of all to minimise as it has a very narrow global minimum, surrounded by a repulsive potential that 
in principle rejects all attempts to try and find it.  We perform the same analysis as before, but this time the thermal annealer schedule is modified, allowing it to run for 1000 iterations before the temperature decay begins. This ensures that the result is not dependent on the initial starting condition. The quantum schedule ramps up for $15 \mu {\rm s}$, then the annealer tunnels for $100 \mu {\rm s}$ at $s=0.01$ before ramping down for  $250 \mu {\rm s}$.

\begin{figure*}[htbp!]
	\centering
	\begin{subfigure}{0.24\linewidth} \centering
		\includegraphics[width=\textwidth]{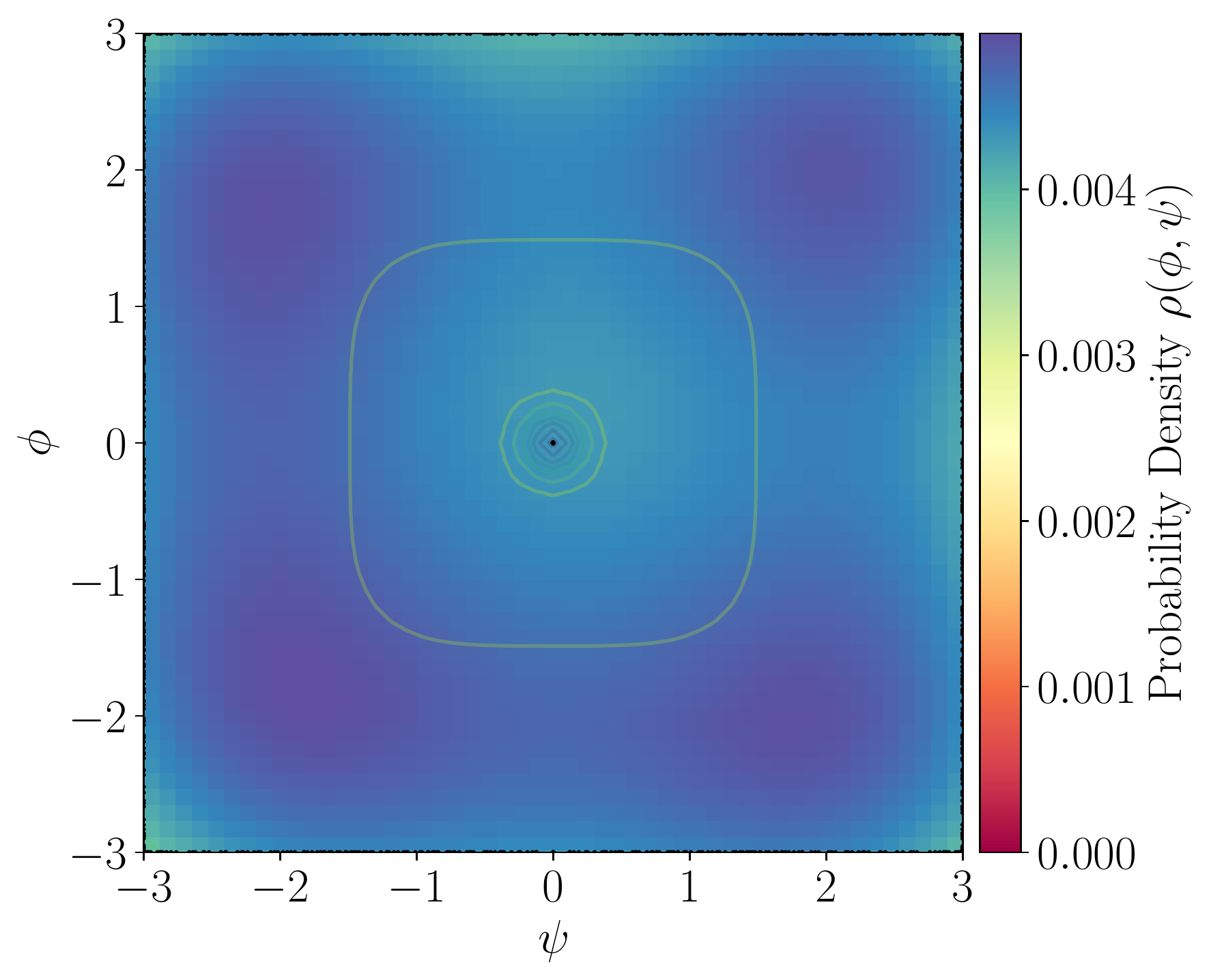}
		\includegraphics[width=\textwidth]{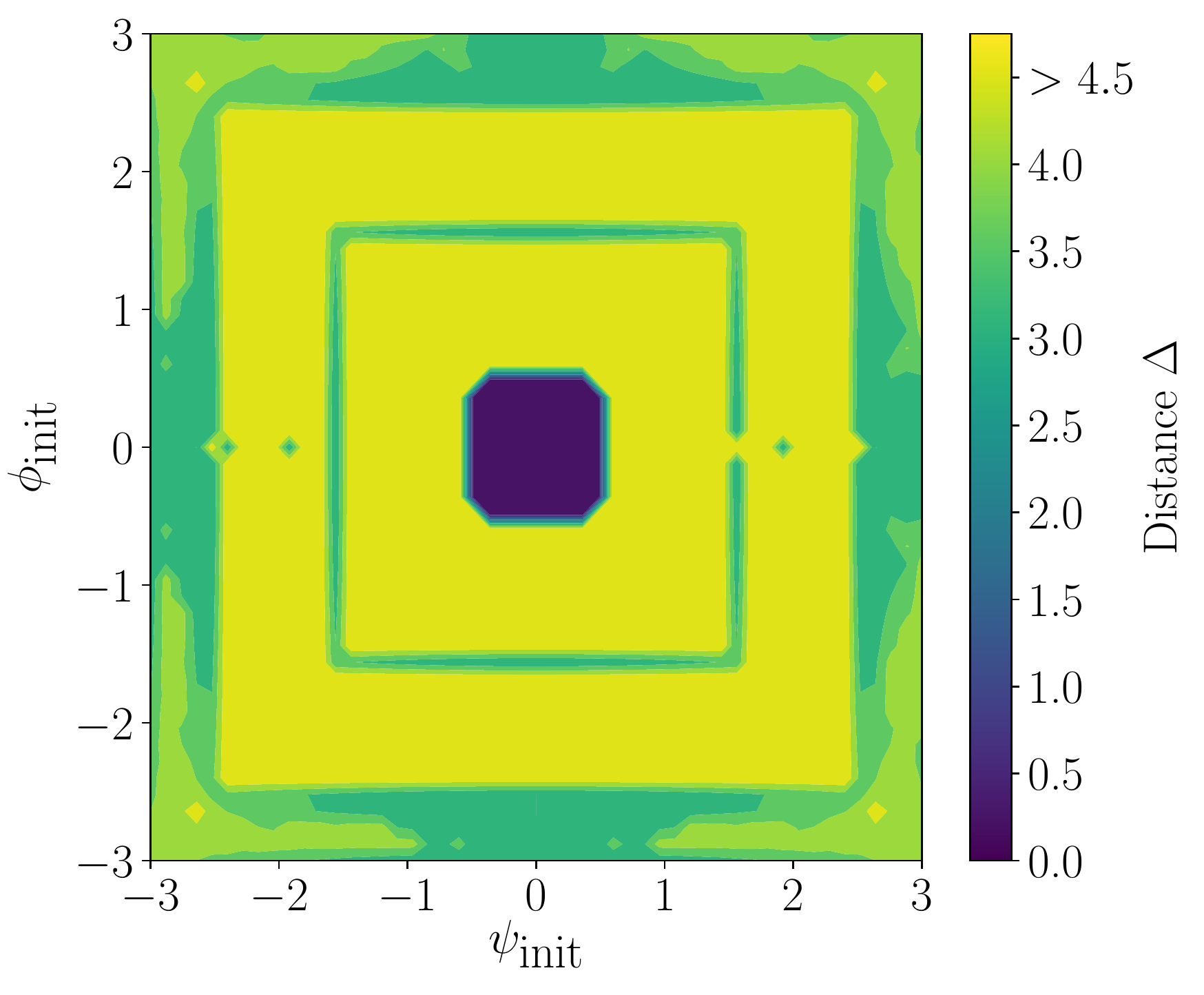}
		\caption{Nelder-Mead}
	\end{subfigure}
	\begin{subfigure}{0.24\linewidth} \centering
		\includegraphics[width=\textwidth]{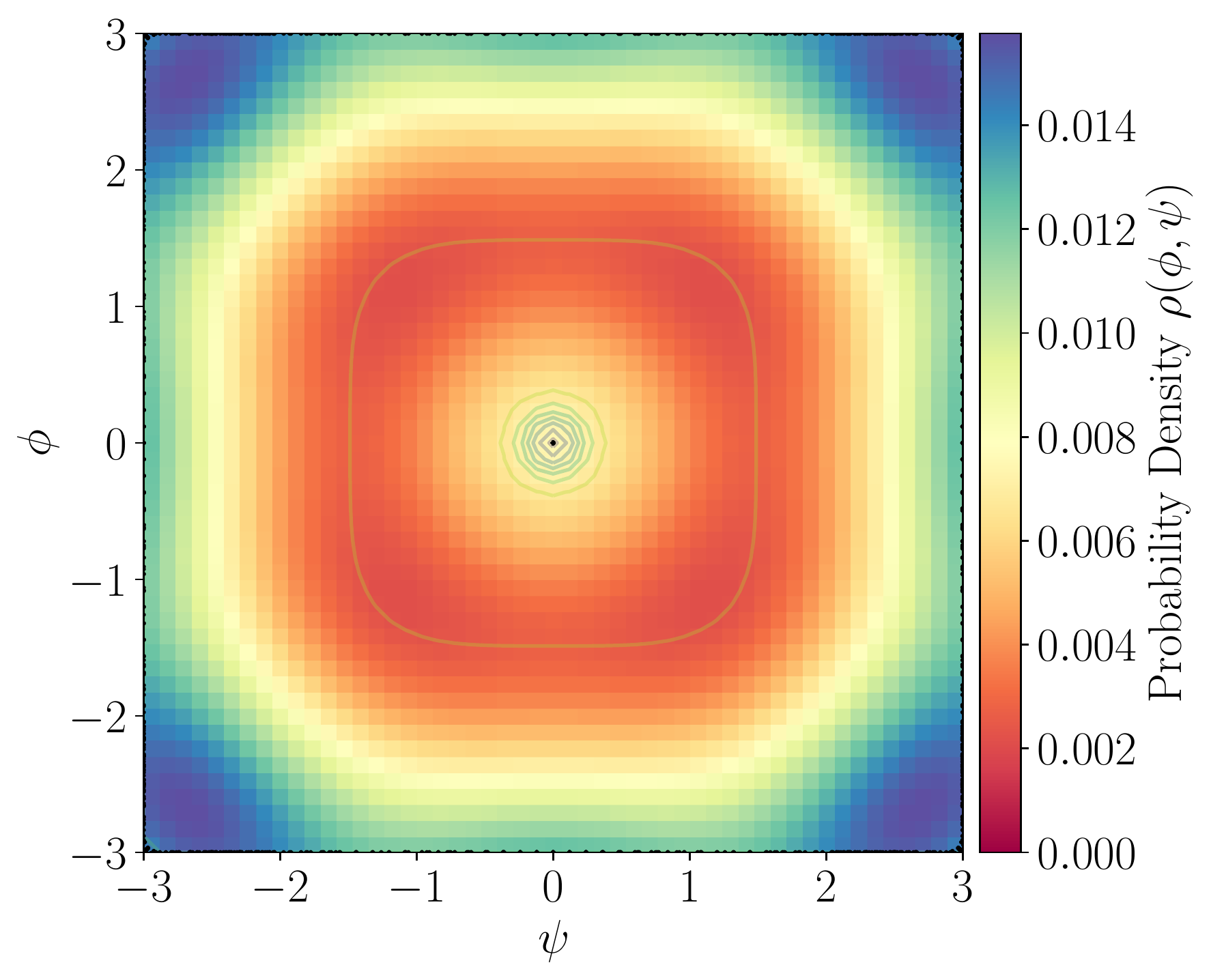}
		\includegraphics[width=\textwidth]{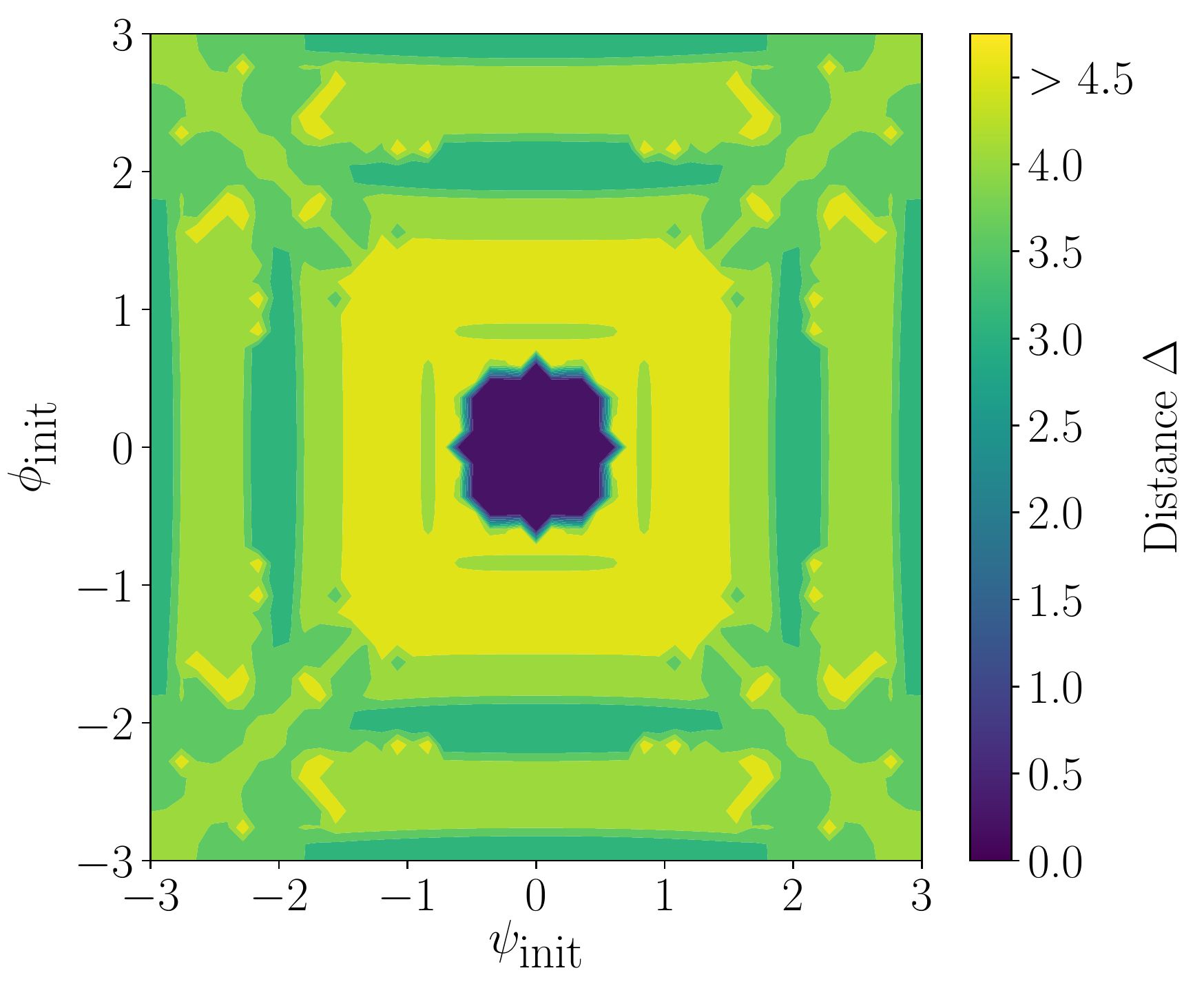}
		\caption{Gradient descent}
	\end{subfigure}
	\begin{subfigure}{0.24\linewidth} \centering
		\includegraphics[width=\textwidth]{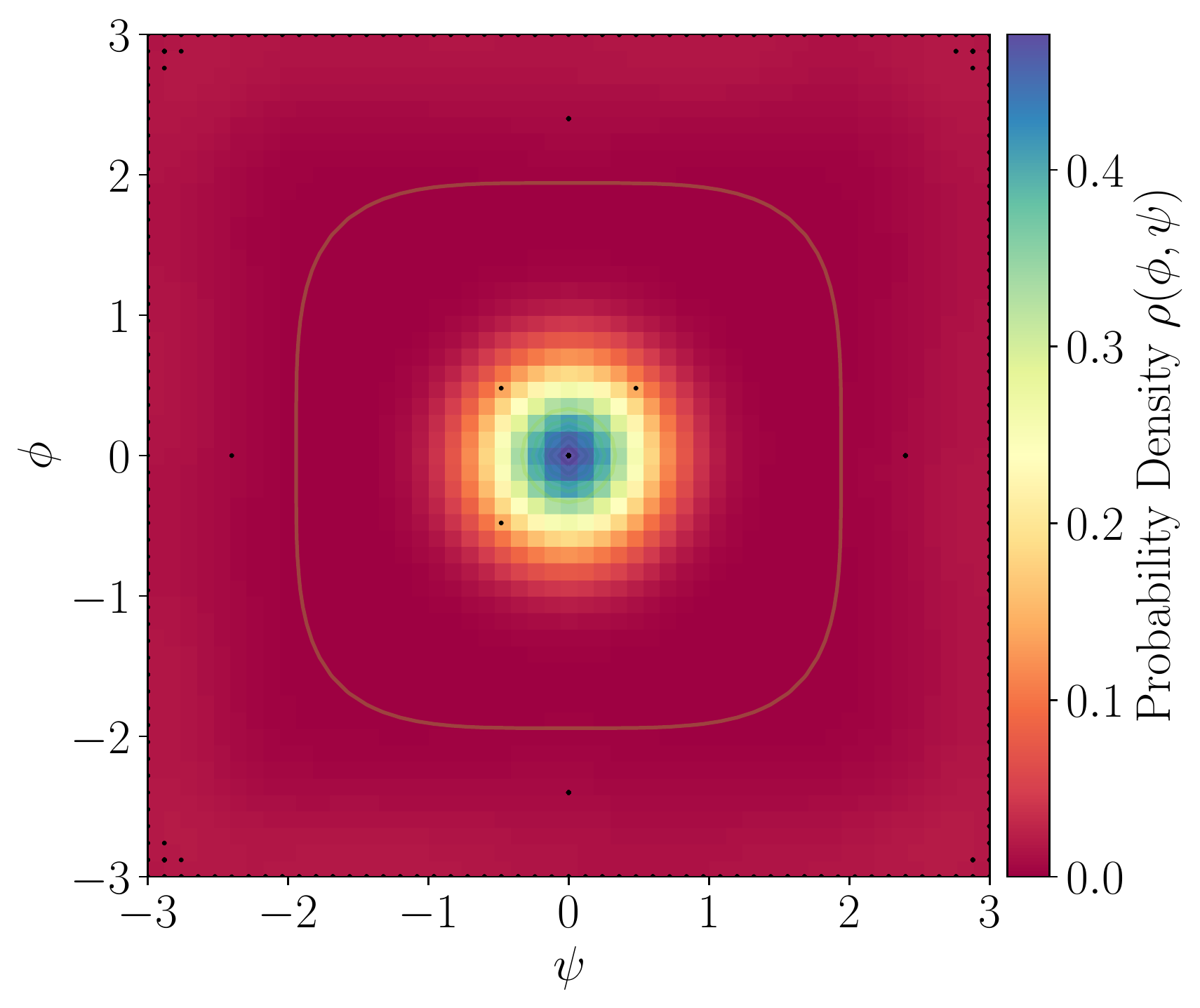}
		\includegraphics[width=\textwidth]{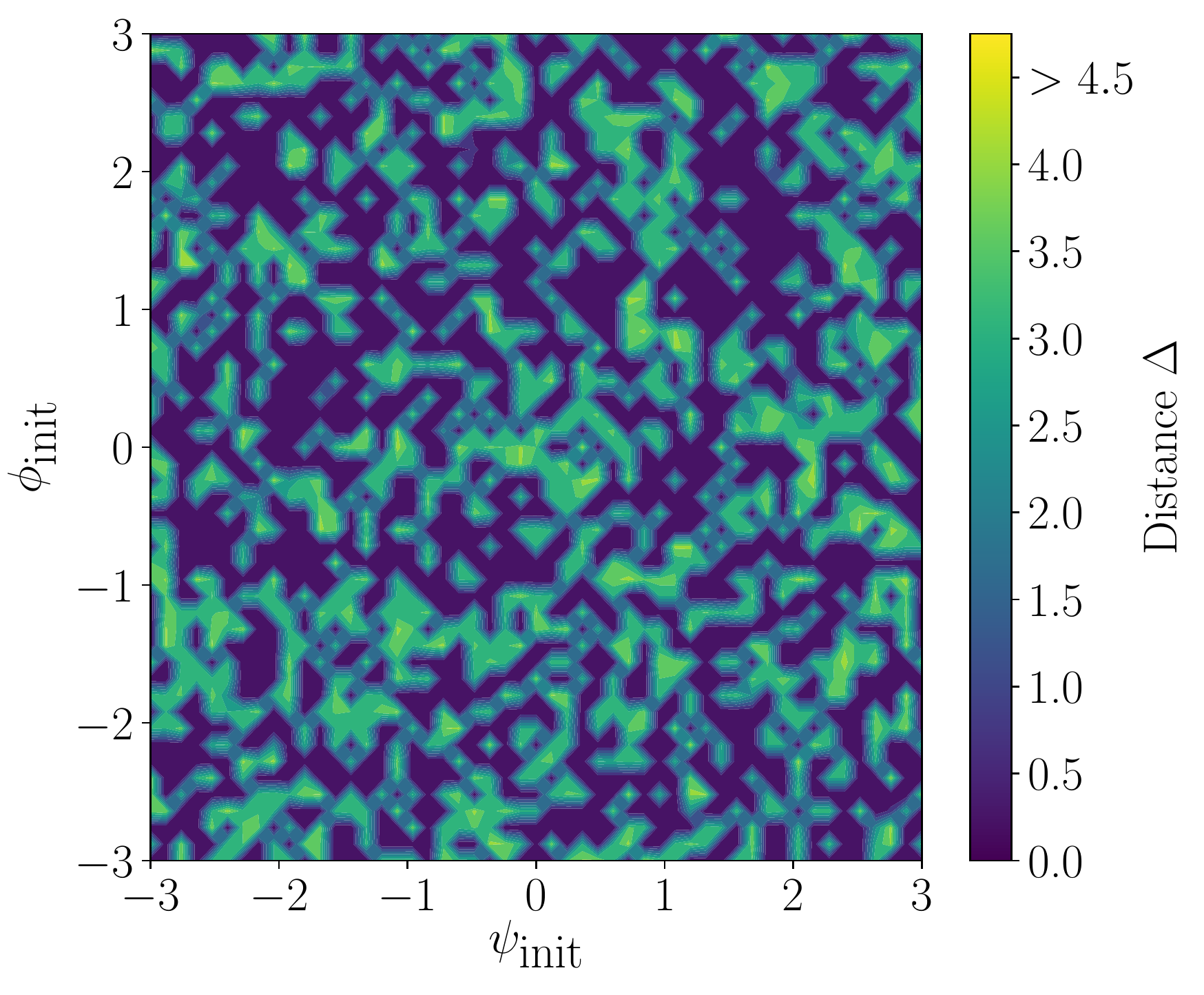}
		\caption{Thermal annealing}
	\end{subfigure}
	\begin{subfigure}{0.24\linewidth} \centering
		\includegraphics[width=\textwidth]{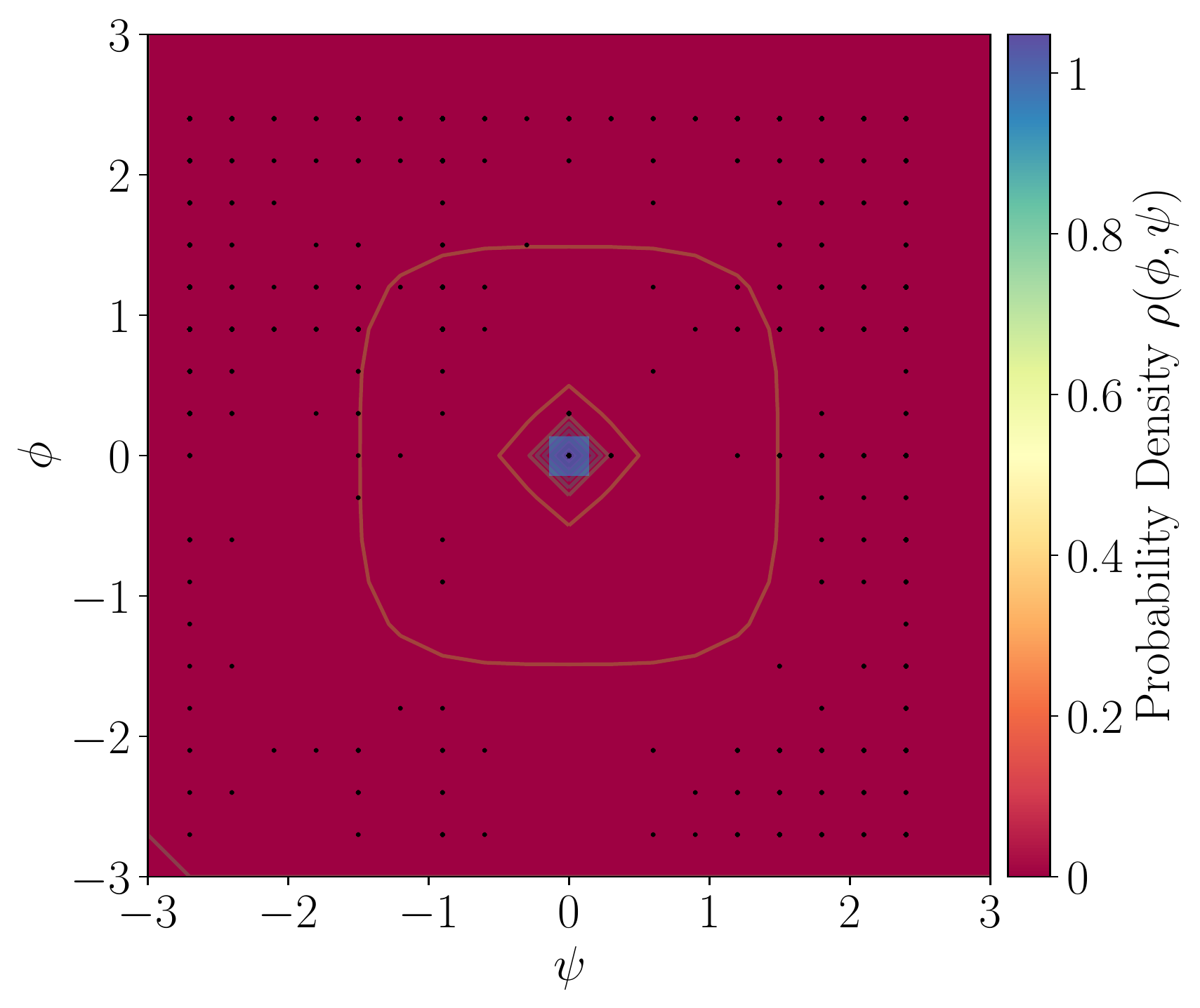}
		\includegraphics[width=\textwidth]{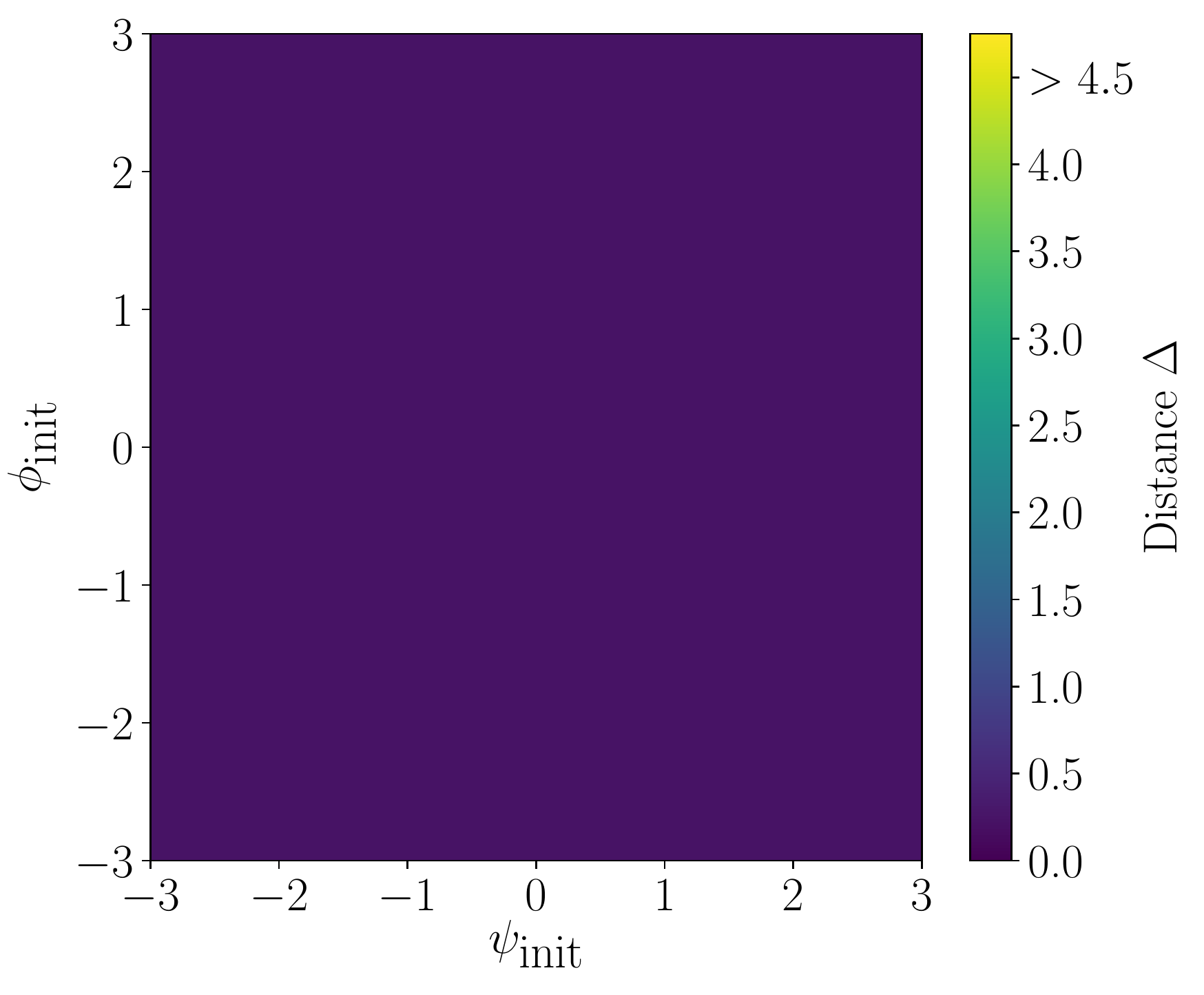}
		\caption{Quantum annealing}
	\end{subfigure}
	\caption{For a range of initial $\phi$ and $\psi$ values the distance between the predicted minimum and true minimum for volcano potential $U_3$ is given. This is done for (a) Nelder-Mead, (b) gradient descent, (c) thermal annealing and (d) quantum annealing. For (a), (b) and (c) a grid size of $N=50$ and $\lambda=1.7$ is chosen with the models being initialised from 2500 different points. For the quantum annealing (d), we take $N=20$ and $\lambda=1.7$ and the model is initialised from $\sim100$ points. Each point is used to produce  100 reads, with the peak of the probability distribution being identified as the predicted minimum for that setup.}
	\label{fig:potential_3_results}
\end{figure*}

Fig.~\ref{fig:potential_3_results} shows the distribution of results from the optimisation processes. It is clear that the NM and GD methods are very poorly equipped to deal with this potential. From Figs.~\ref{fig:potential_3_results}a and \ref{fig:potential_3_results}b  it appears that most of the results from these runs are no longer contained within the plotted grid, or rather they run away to the boundary. Thermal and quantum annealing both perform much better. Once again though, the quantum annealer performs the most consistently. 
%This consistency is shown again in Figure \ref{fig:initial_distances_potential_3}. 

\section{\label{Sec:conc}Conclusions}

Optimisation tasks are at the heart of many quantitative techniques across all sciences. Thus, various numerical optimisation algorithm have been proposed and implemented on classically-operating systems. 
As quantum computers, in particular quantum annealers, perform calculations in a fundamentally different way compared to classical computers, they are anticipated to have a qualitatively different and possibly transformative impact on the way optimisation tasks are realised.

In this work we have performed a detailed comparison between quantum annealing and three of the most popular classical optimisation algorithms, namely thermal annealing, Nelder-Mead and gradient descent. First we have shown that quantum annealing can quickly and reliably solve the 2-D Ising model, a highly complex latticised system. Indeed in many respects it displays a similar phase transition with respect to a change in overall scale of the potential, $\lambda$. For the quantum annealer this scaling plays a similar role to the temperature in a thermal anneal. 

Then, using the domain wall encoding, we were able to show that a quantum annealer can find the ground state of various continuous two-dimensional functions much more reliably than classical methods. That is it is far less dependent on the initial conditions of the optimisation algorithm and on the topographical profile of the function, as well as being much faster than classical algorithms. It is notable that scaling the function $\lambda U_i(\phi,\psi)$ up by increasing $\lambda$ improves the perfomance of the quantum annealer. This indicates that the quantum annealer samples the configuration space by tunnelling underneath potential barriers, for which a deeper well increases the tunnelling rate, rather than by sampling the configuration space by following the topographic structure of $U_i(\phi,\psi)$ in the lateral direction.

For many problems it can be very hard, even a posteriori, to know if a classical optimisation algorithm has settled into a local or a global minimum. Thus, finding the global minimum reliably is one of the biggest challenges for any optimisation task, which is often only heuristically addressed by rerunning the algorithms with different initial conditions multiple times, or by performing other partial scanning manoeuvres. However, the depth-awareness of a quantum algorithm seems to offer an entirely new approach to locating a global extrema, and may have implications for optimisation tasks in many areas.

\vspace{0.5cm}

\noindent {\bf \underline{Acknowledgements}:}  We would like to thank Nicholas Chancellor for helpful discussions. S.A. and M.S. are supported by the STFC under grant ST/P001246/1.

\FloatBarrier
\appendix

\section{Classical optimisation algorithms}
\label{appendix:classical_opt}

\subsection{Gradient Descent Method}
\label{appendix:gdm}

Simple gradient descent optimisation techniques can face difficulties converging to the correct solution. An example of this is when they are getting close to the bottom of the valley they are descending into. By following the steepest gradient, depending on the step size, it is possible to step entirely over the solution. This results in the algorithm oscillating over the correct answer but never finding it. 

Conjugate gradient descent aims to resolve this by accounting for previous gradients when calculating its next step \cite{Hestenes1952MethodsOC, 10.5555/865018}. Gradient descent methods, generally, follow the form 
\begin{equation}
x_{i+1} = x_i + \alpha d_i,
\end{equation}
where $x_i$ and $x_{i+1}$ is your current location and next location, respectively. The form also depends on $\alpha$, a step size, and $d_i$, a vector describing the direction in which you will move. For a standard gradient descent method, $d_i$ will typically be $-\nabla f(x_i)$, the negative gradient of the function you are finding the minimum of. 

In the conjugate gradient method, $d_i$ will be a combination of gradients. Here, it will depend on $-\nabla f(x_i)$ and $-\beta_i \nabla f(x_{i-1})$. The term $\beta_i$ is designed such that the new directional vector is conjugate to the previous. By accounting for the previous gradient when finding the next step, and using an adaptive step size, the likelihood of taking a step in the direction you have previously come from is reduced. This results in the optimisation performance increasing.

\subsection{Nelder-Mead Method}
\label{appendix:nmm}

Nelder-Mead (NM) is an optimisation technique that uses a geometric structure known as a simplex \cite{10.1093/comjnl/7.4.308}. The simplex, through the optimisation procedure, traverses the function space to find a minimum. For a $N$ dimensional feature space a simplex with ($N+1$) vertices can be initialised. The vertices $x_i$ will each have a corresponding value found from evaluating the function $f(x_i)$. At each iteration, these vertices are shifted according to a set of rules. The process repeats until either a maximum number of iterations is reached or the standard deviation of the points decreases under a threshold.

Specifically, we can describe the original NM optimisation method with three repeating steps. In step (i), $f(x)$ is evaluated for each simplex vertex. The vertices are then put in ascending order ($x_0$, $x_1$, ..., $x_{N+1}$) based on this evaluation. A centroid, $x_c$, is then calculated in step (ii). This is simply the mean of all the points evaluated in step (i), excluding $x_{N+1}$.  At step (iii) the algorithm will shift the simplex. We will consider a few important vertices of the simplex. Namely, the "best" point $x_0$, the second-worst point $x_N$, the worst point $x_{N+1}$ and the centroid. The updates are done using one of four rules:

\begin{itemize}
	\item Reflection: A new point, $x_R$ is found from $x_R=x_c + \alpha(x_c - x_{N+1})$, where $\alpha>0$. If $x_R$ is better than $x_N$, but worse than $x_0$, then $x_{N+1}$ (the worst point in the simplex) is replaced with $x_R$.
	\item Expansion: If $x_R$ was less than $x_0$ and therefore the new best point, another point $x_E =x_c + \gamma(x_R - x_c)$, where $\gamma>1$, can be found. Here, we aim to check if another, even better, point can be found. If $x_E$ is the new best point it can replace $x_{N+1}$, otherwise $x_{N+1}$ will be replaced by $x_R$.
	\item Contraction: This is the case where $x_R \geq x_N$. A new point $x_{cont}=x_c + \beta(x_{N+1} - x_c)$, where $0 < \beta \leq 0.5$, is found. If this new value $x_{cont}$ is better than the worst point it can replace $x_{N+1}$.
	\item Shrink: In the case the previous three checks fail, and $x_{cont}$ is worse than $x_{N+1}$, the shrink procedure can be performed. Here, all points are replaced, except $x_0$. A point $x_i$ is shifted such that $x_i=x_0 + \sigma(x_i - x_0)$, where $\sigma$ is typically 0.5.
\end{itemize}

\begin{figure}[tbh]
	\centering
	\begin{subfigure}{0.45\linewidth} \centering
		\includegraphics[width=\textwidth]{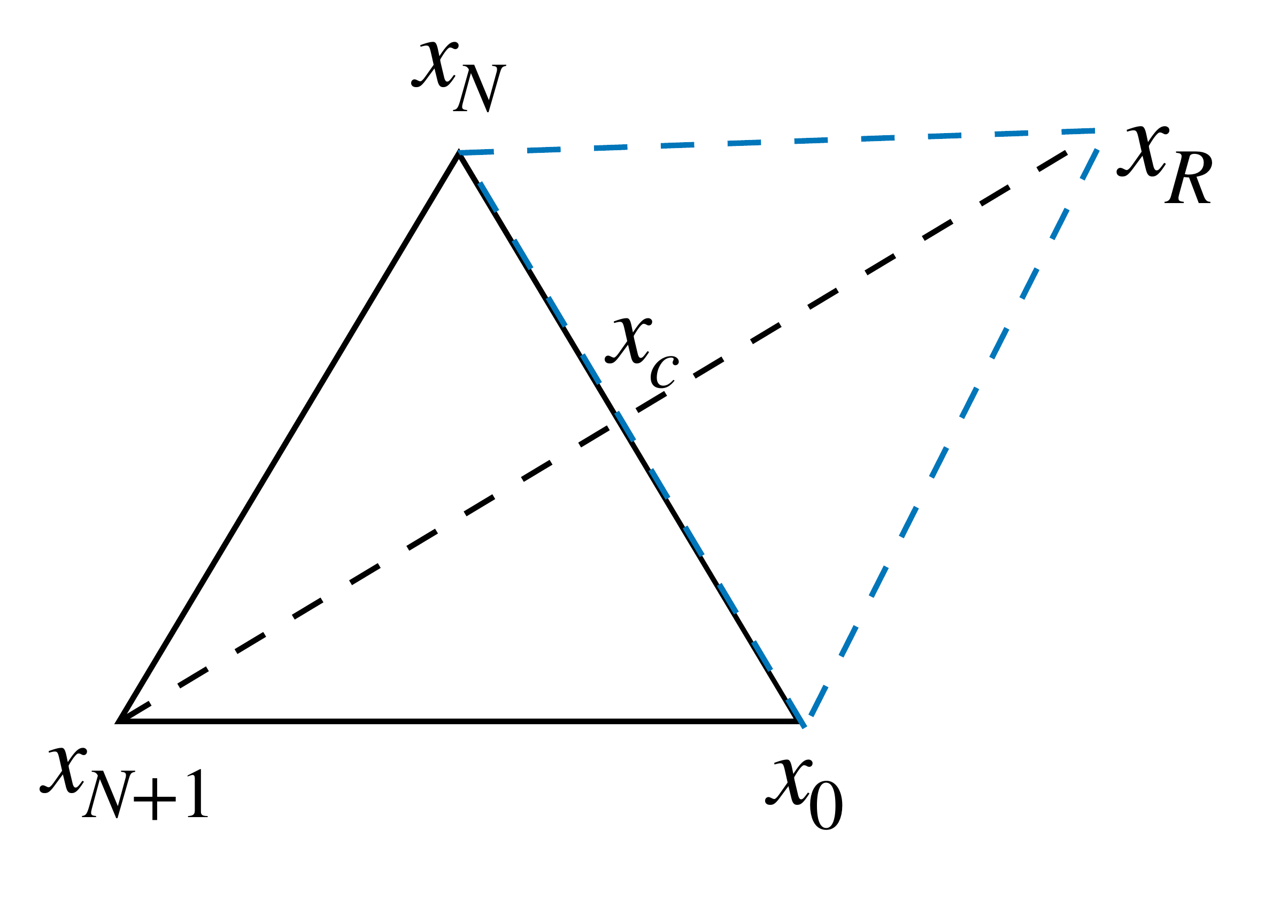}
		\caption{}
	\end{subfigure}
	\begin{subfigure}{0.45\linewidth} \centering
		\includegraphics[width=\textwidth]{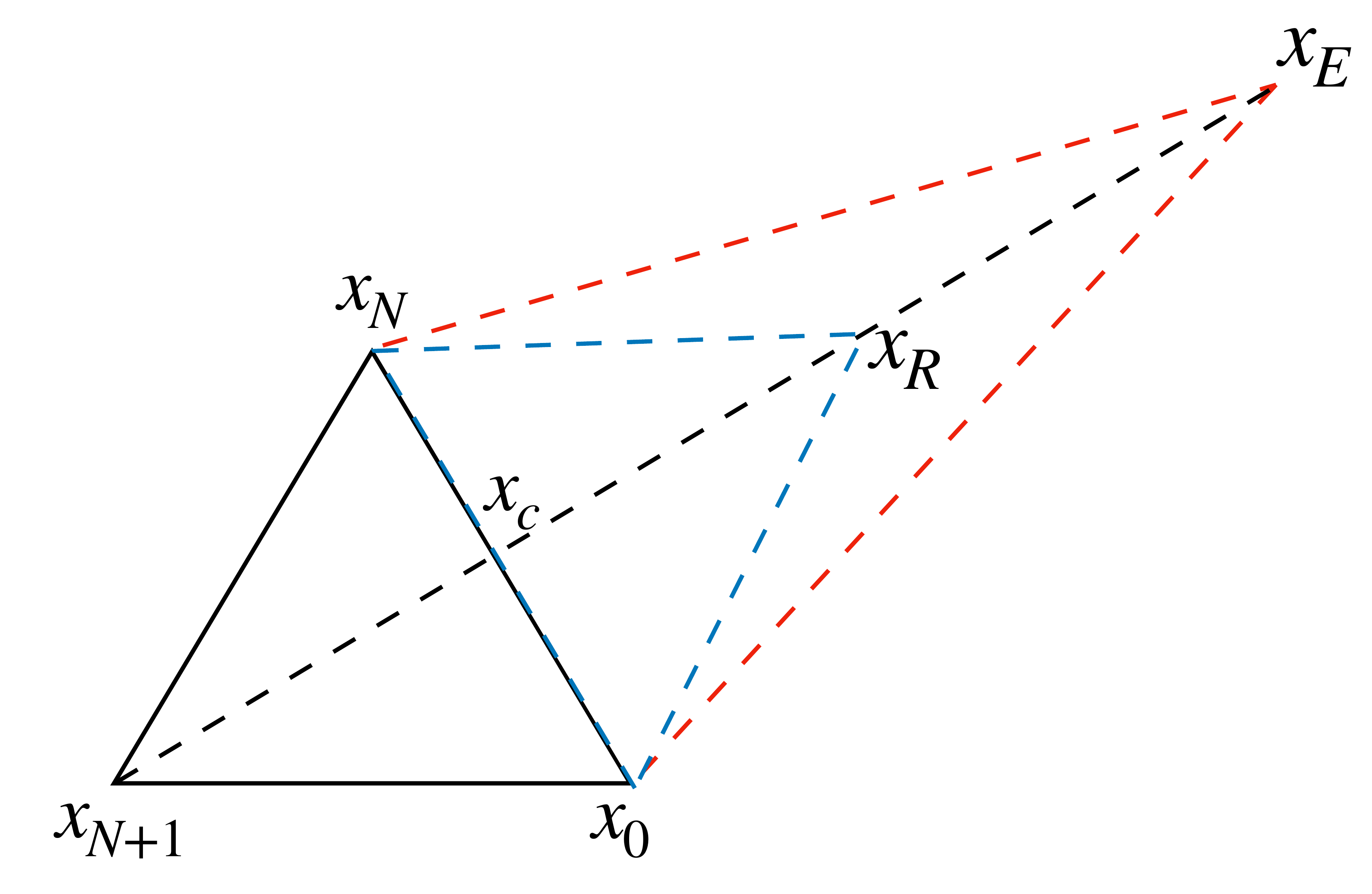}
		\caption{}
	\end{subfigure}
	\begin{subfigure}{0.45\linewidth} \centering
		\includegraphics[width=\textwidth]{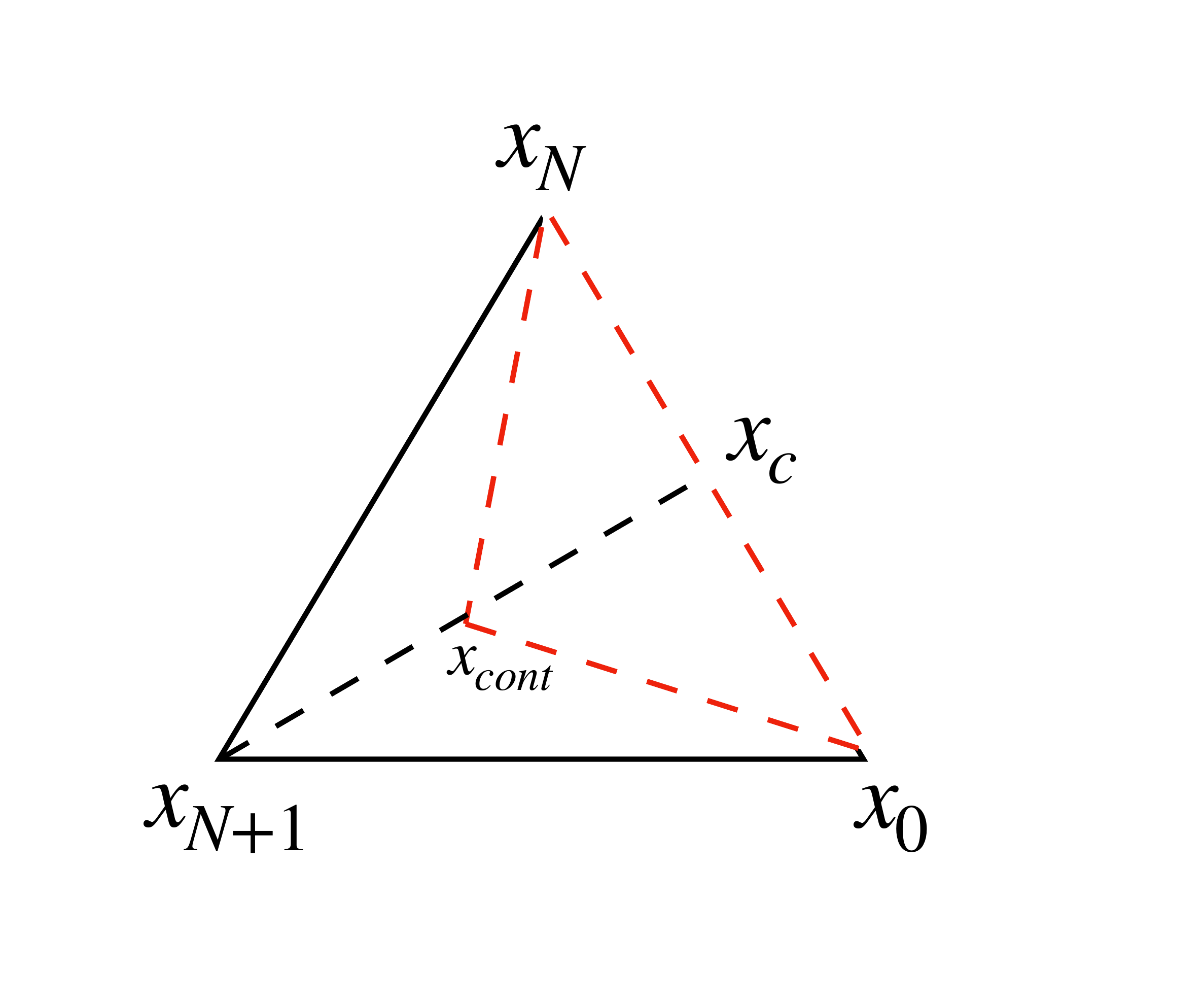}
		\caption{}
	\end{subfigure}
	\begin{subfigure}{0.45\linewidth} \centering
		\includegraphics[width=\textwidth]{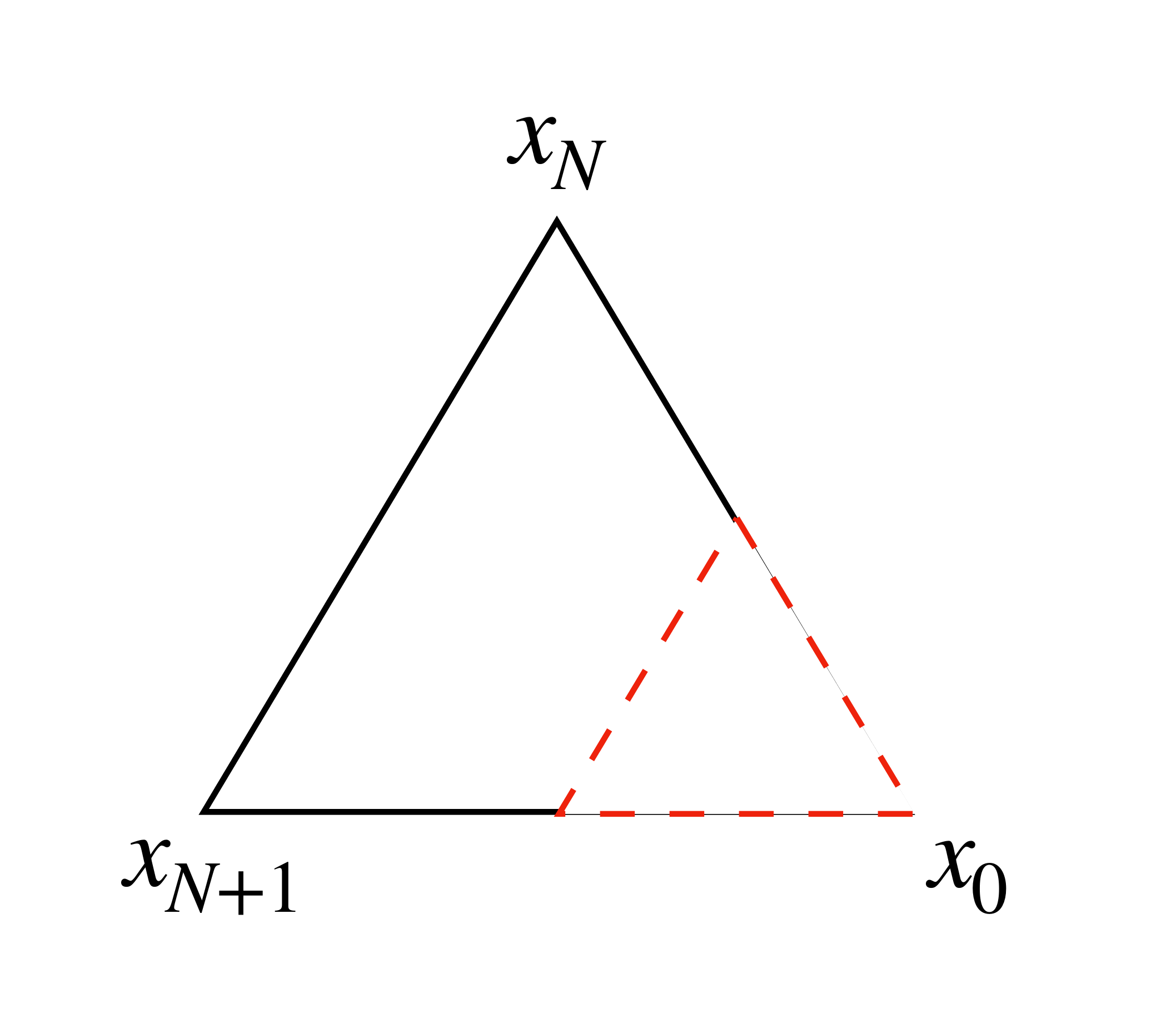}
		\caption{}
	\end{subfigure}
	\caption{The four variations of how the simplex vertices can be updated during a Nelder-Mead optimisation iteration: a) Reflection, b) Expansion, c) Contraction and d) Shrink.}
	\label{fig:NM}
\end{figure}

For a low-dimensional example, the simplex manipulations are shown in Figure \ref{fig:NM}. After the appropriate update is performed the iteration is complete and the process begins again. After step (i) the algorithm has the chance to be terminated. This depends on the standard deviation of the values found from evaluating the vertex values. If this is under a certain threshold, or a maximum number of iterations is reached, then the optimisation ends. The smallest value found from evaluating the vertices is returned as the minimum of the function. 

\subsection{Thermal annealing with Metropolis Algorithm}
\label{appendix:tam}

We can consider our potential as being a system that contains various states. These states, each representing different points in the potential, will have an associated energy. The minimum of the potential will be the state with the lowest energy.  Thermal annealing aims to optimise a system by finding this state \cite{Metropolis1953EquationOS}. At each iteration of the algorithm, the energy of the system can be measure and compared against the energy of a neighbouring state. Depending on the energy of the neighbour there is a probability it will be accepted. If accepted, this new configuration replaces the current state and the next iteration can begin. 

By beginning in a state $A$ an associated energy $E_A$ can be found. This state can be changed to create $B$, a potential new state with an energy $E_B$. If $E_B<E_A$ then the new state has lower energy and can be accepted as our new state. If $E_B>E_A$ then the candidate state is not energetically preferable. However, there is still a chance to accept it with a probability $e^{-(E_B-E_A)/T}$. This probability, and hence the likelihood of the algorithm accepting a new configuration, is guided by a temperature $T$. By being allowed to take steps away from a minimum reduces the likelihood of falling into local minima.

The choice of temperature is an important factor in thermal annealing. A higher temperature will result in more configurations being accepted. It is typically then to perform the annealing process with a schedule. By starting at a high temperature space can be quickly explored. After a select number of iterations, the temperature can be reduced until the algorithm settles into the global minimum.

\section{Performance comparison between optimisers with distinct initial conditions}
\label{appendix:starting}

\begin{figure*}[htb]
	\centering
	\begin{subfigure}{0.24\linewidth} \centering
		\includegraphics[width=\textwidth]{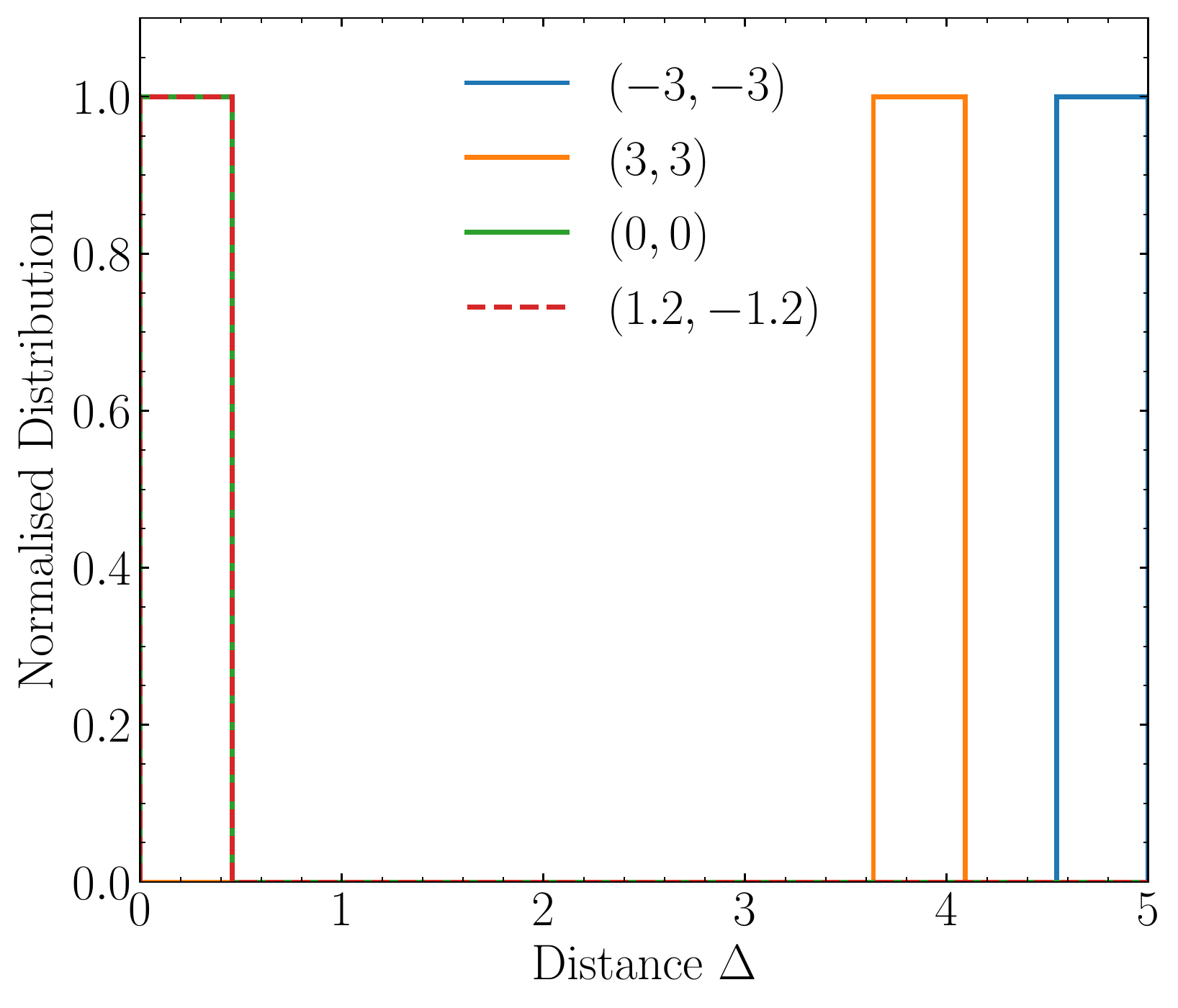}
		\caption{}
	\end{subfigure}
	\begin{subfigure}{0.24\linewidth} \centering
		\includegraphics[width=\textwidth]{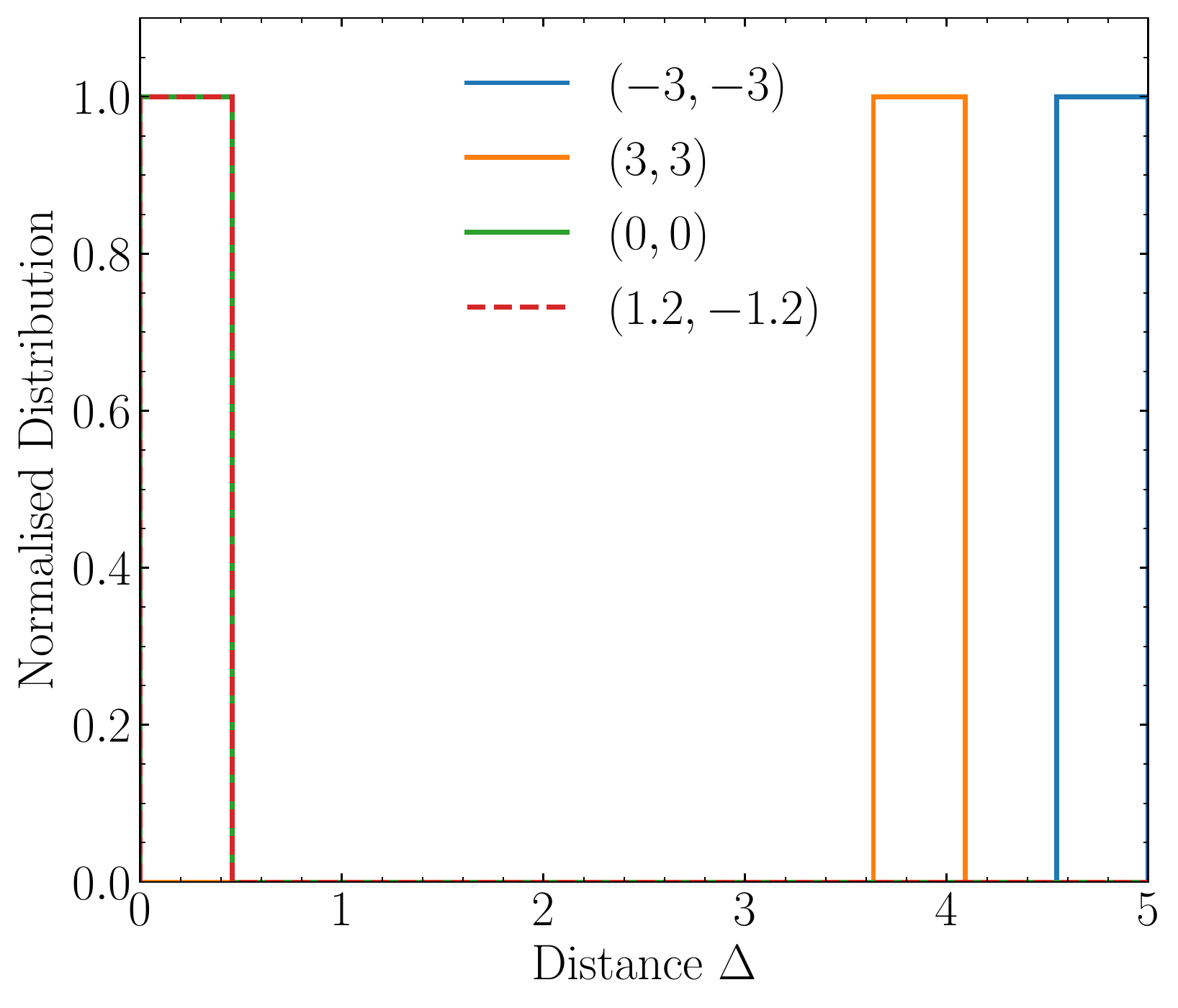}
		\caption{}
	\end{subfigure}
	\begin{subfigure}{0.24\linewidth} \centering
		\includegraphics[width=\textwidth]{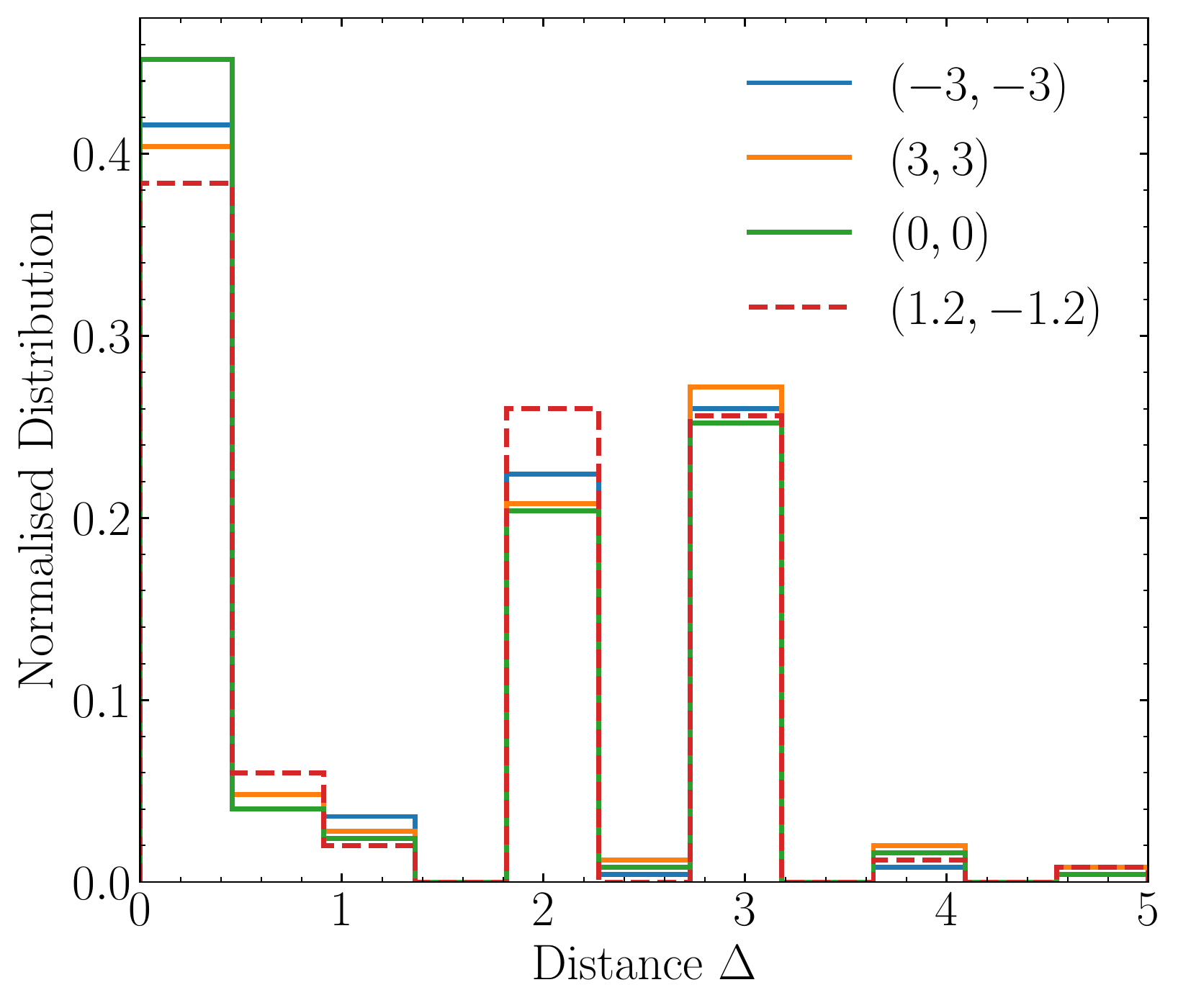}
		\caption{}
	\end{subfigure}
	\begin{subfigure}{0.24\linewidth} \centering
		\includegraphics[width=\textwidth]{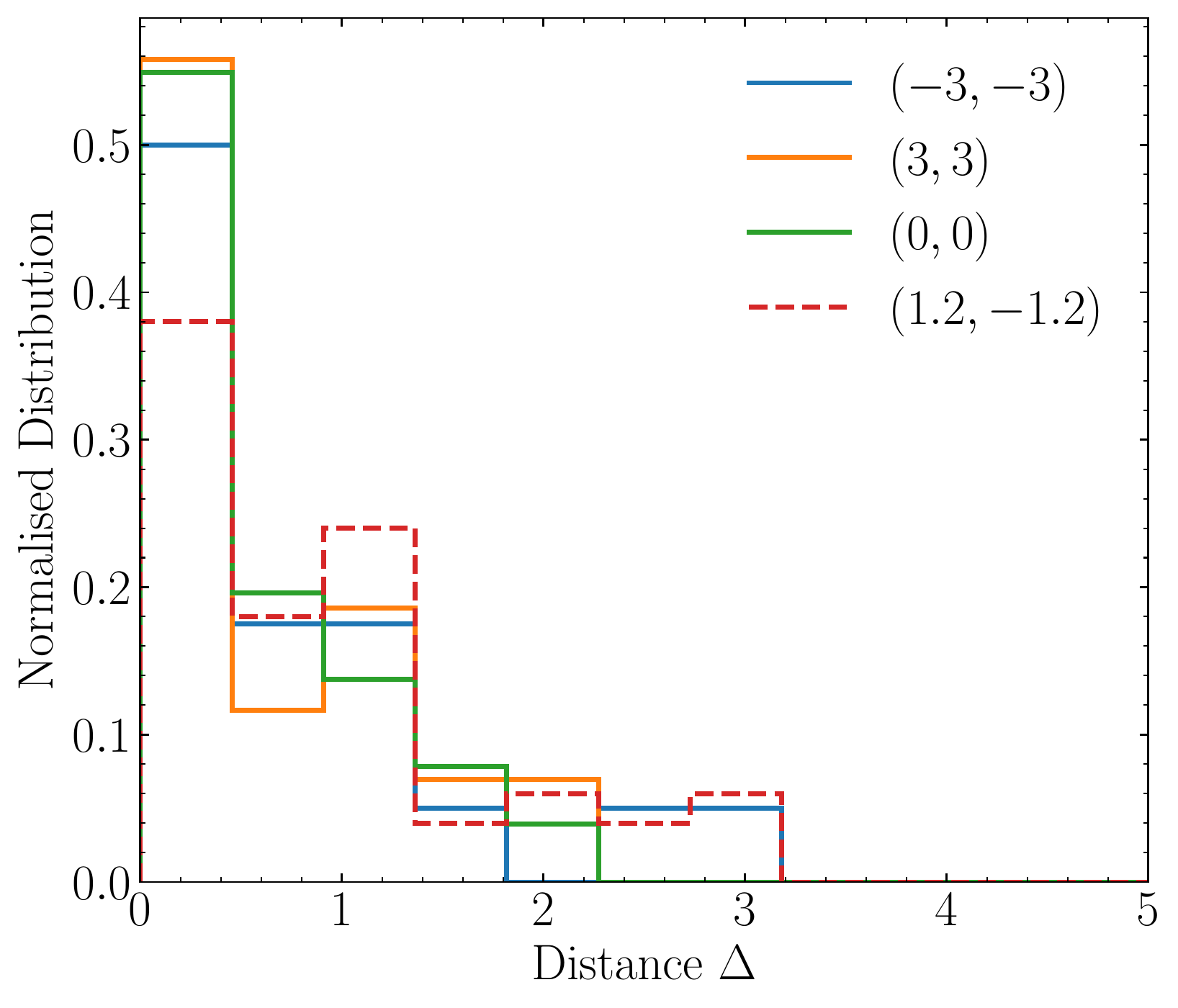}
		\caption{}
	\end{subfigure}
	\caption{A comparison of the distance a predicted point is away from the true minimum, for the four initial starting points in the legend,  using potential $U_1$. Four optimisation techniques are shown: Nelder-Mead (a), gradient descent (b), thermal annealing (c) and quantum annealing (d). Each initial condition was run 100 times.}
	\label{fig:initial_distances}
\end{figure*}

\begin{figure*}[htb]
	\centering
	\begin{subfigure}{0.24\linewidth} \centering
		\includegraphics[width=\textwidth]{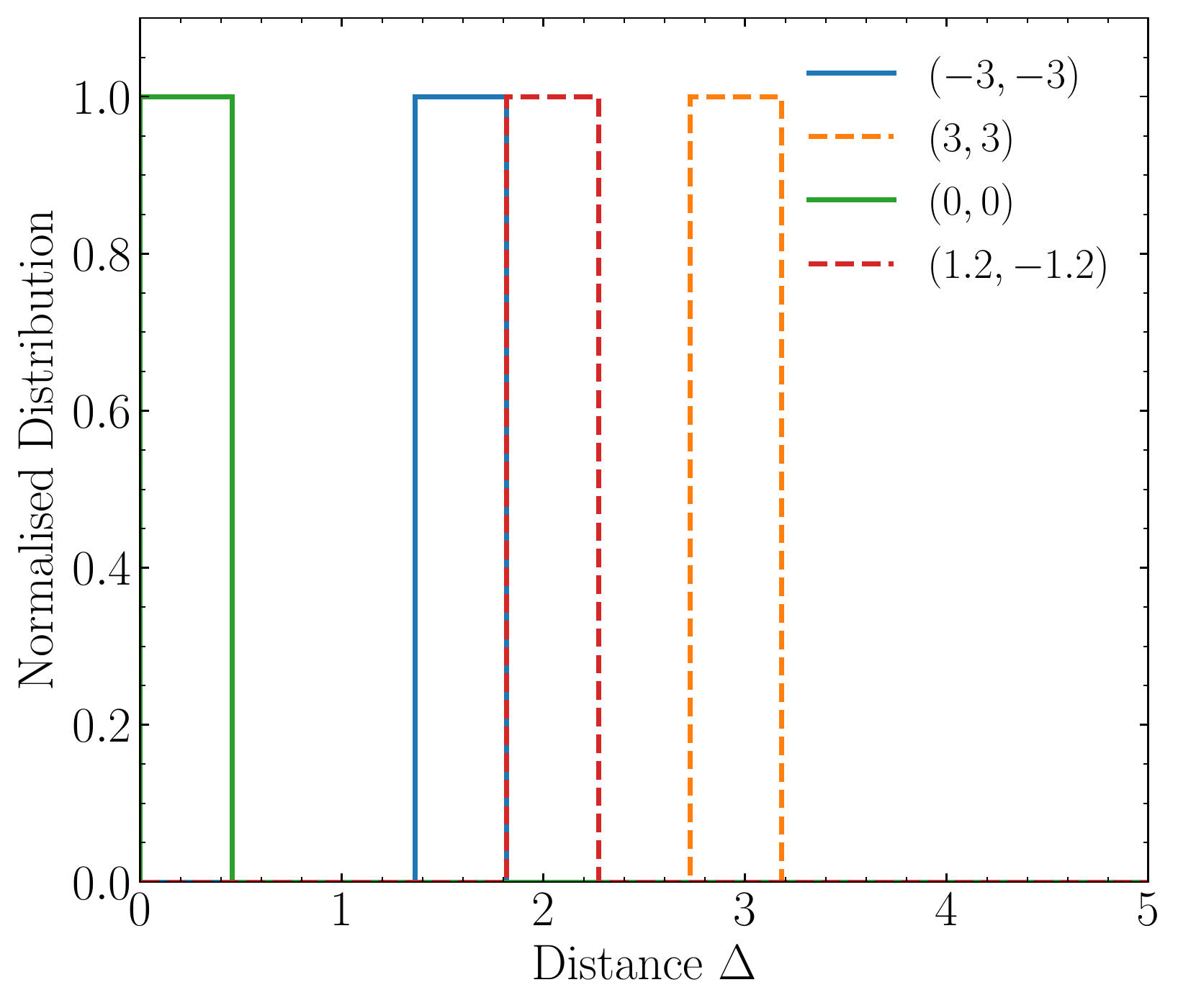}
		\caption{Nelder-Mead}
	\end{subfigure}
	\begin{subfigure}{0.24\linewidth} \centering
		\includegraphics[width=\textwidth]{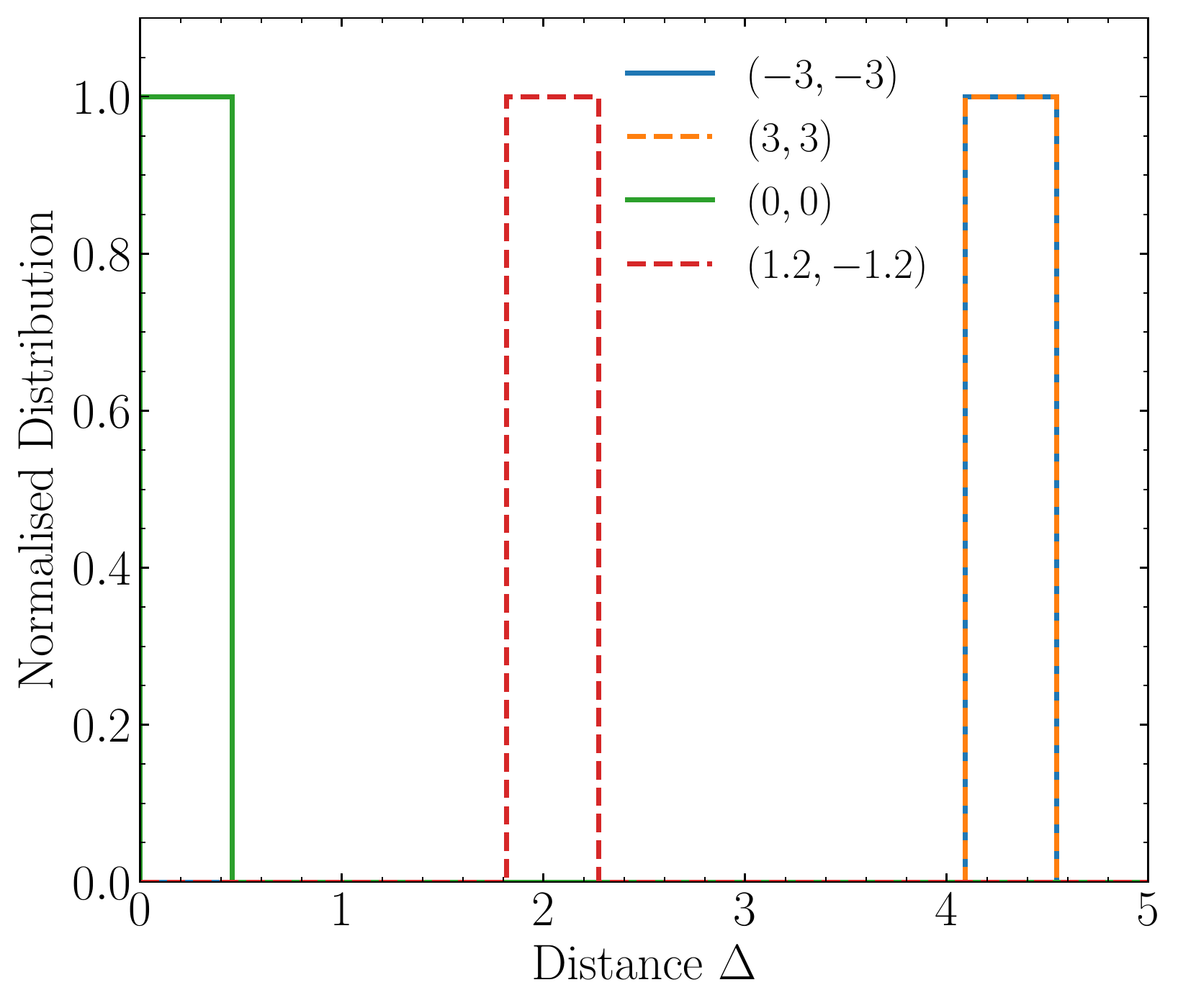}
		\caption{Gradient descent}
	\end{subfigure}
	\begin{subfigure}{0.24\linewidth} \centering
		\includegraphics[width=\textwidth]{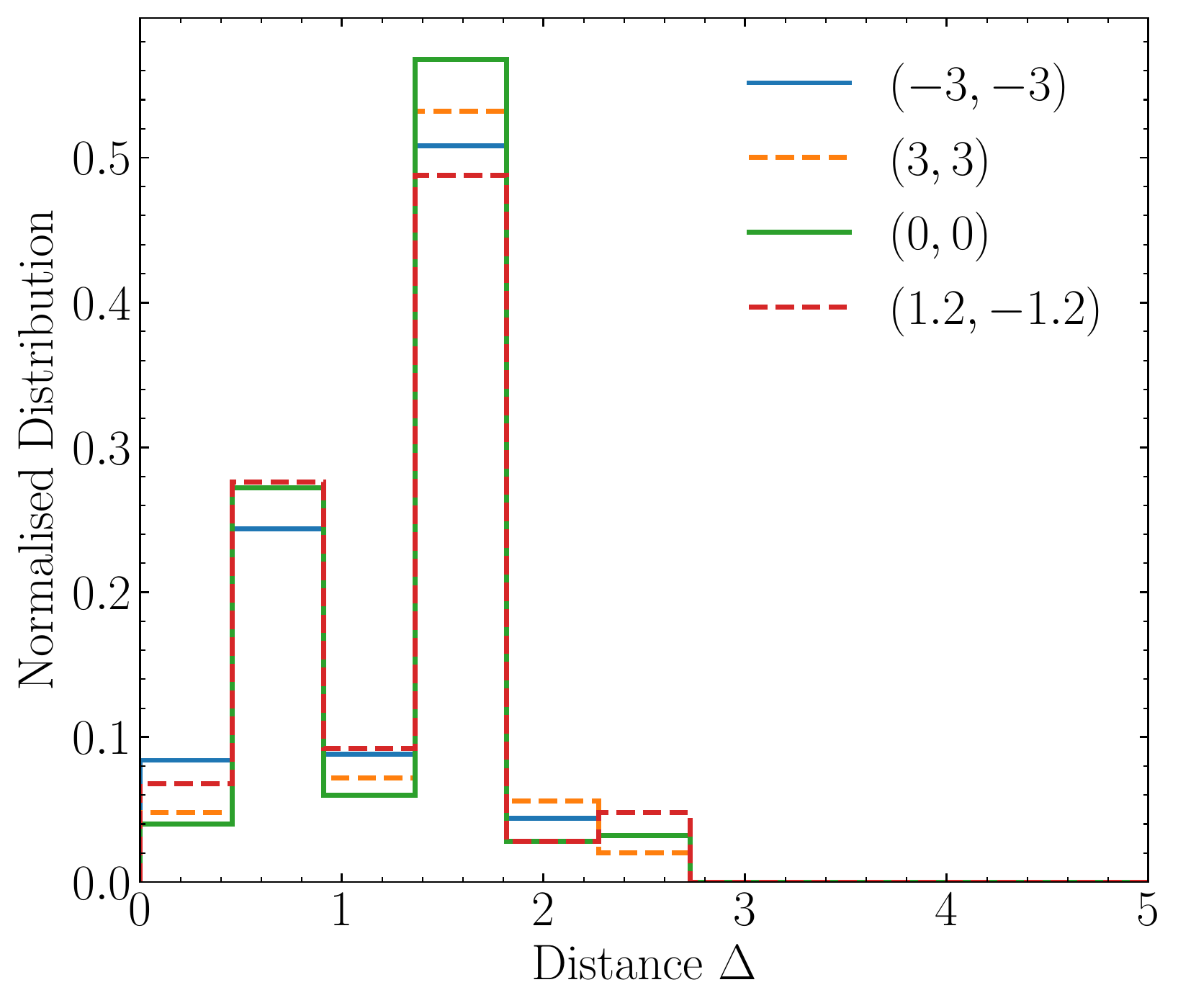}
		\caption{Thermal annealing}
	\end{subfigure}
	\begin{subfigure}{0.24\linewidth} \centering
		\includegraphics[width=\textwidth]{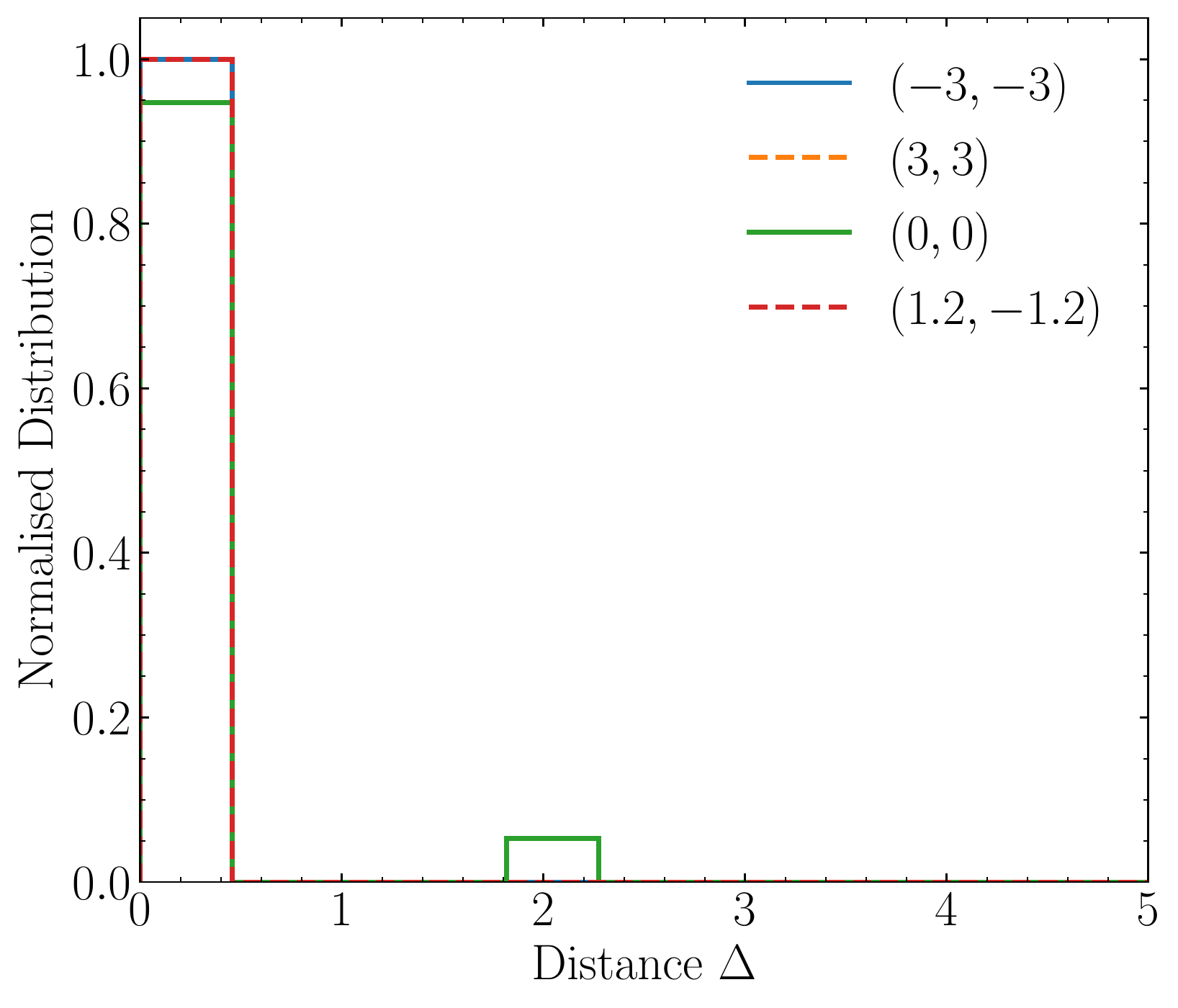}
		\caption{Quantum annealing}
	\end{subfigure}
	\caption{A comparison of the distance a predicted point is away from the true minimum. We choose to do this for four initial starting points using the multiwell potential (potential 2). Four optimisation techniques are shown: Nelder-Mead (a), gradient descent (b), thermal annealing (c) and quantum annealing (d). Each initial condition is ran 100 times.}
	\label{fig:initial_distances_tanh}
	%\label{fig:potential_2_hist}
\end{figure*}

\begin{figure*}[htb]
	\centering
	\begin{subfigure}{0.24\linewidth} 
		\includegraphics[width=\textwidth]{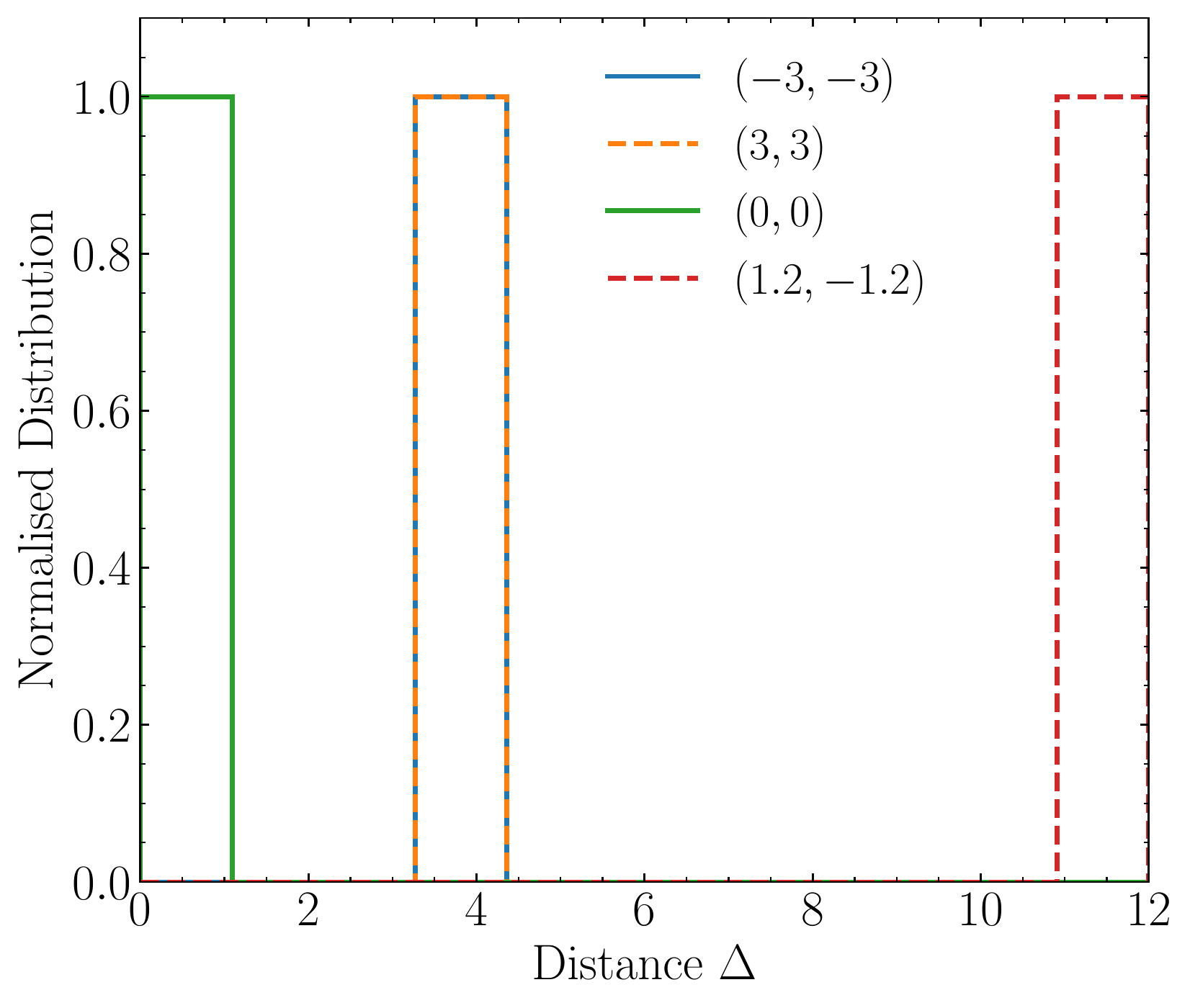}
		\caption{Nelder-Mead}
	\end{subfigure}
	\begin{subfigure}{0.24\linewidth} \centering
		\includegraphics[width=\textwidth]{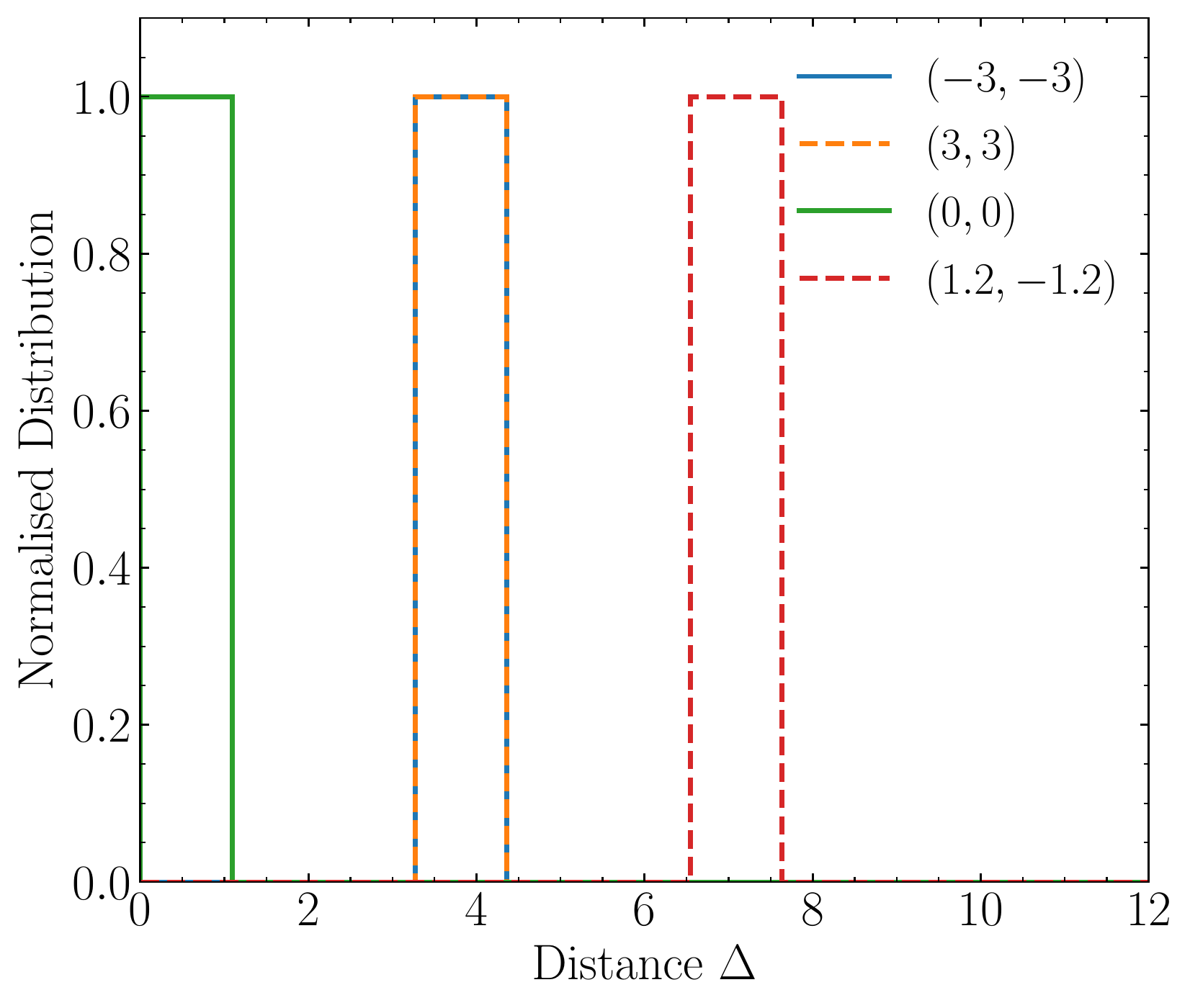}
		\caption{Gradient descent}
	\end{subfigure}
	\begin{subfigure}{0.24\linewidth} \centering
		\includegraphics[width=\textwidth]{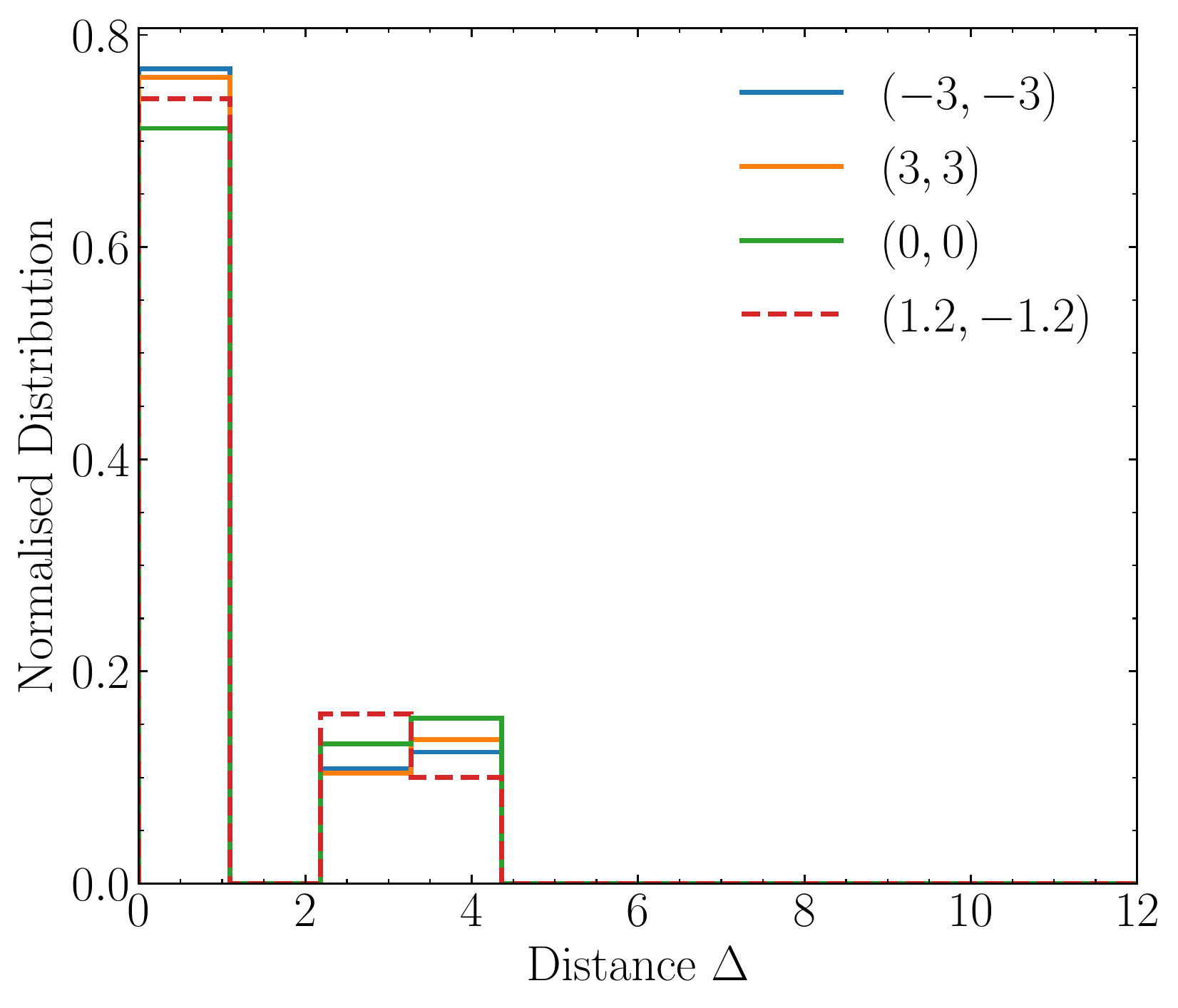}
		\caption{Thermal annealing}
	\end{subfigure}
	\begin{subfigure}{0.24\linewidth} \centering
		\includegraphics[width=\textwidth]{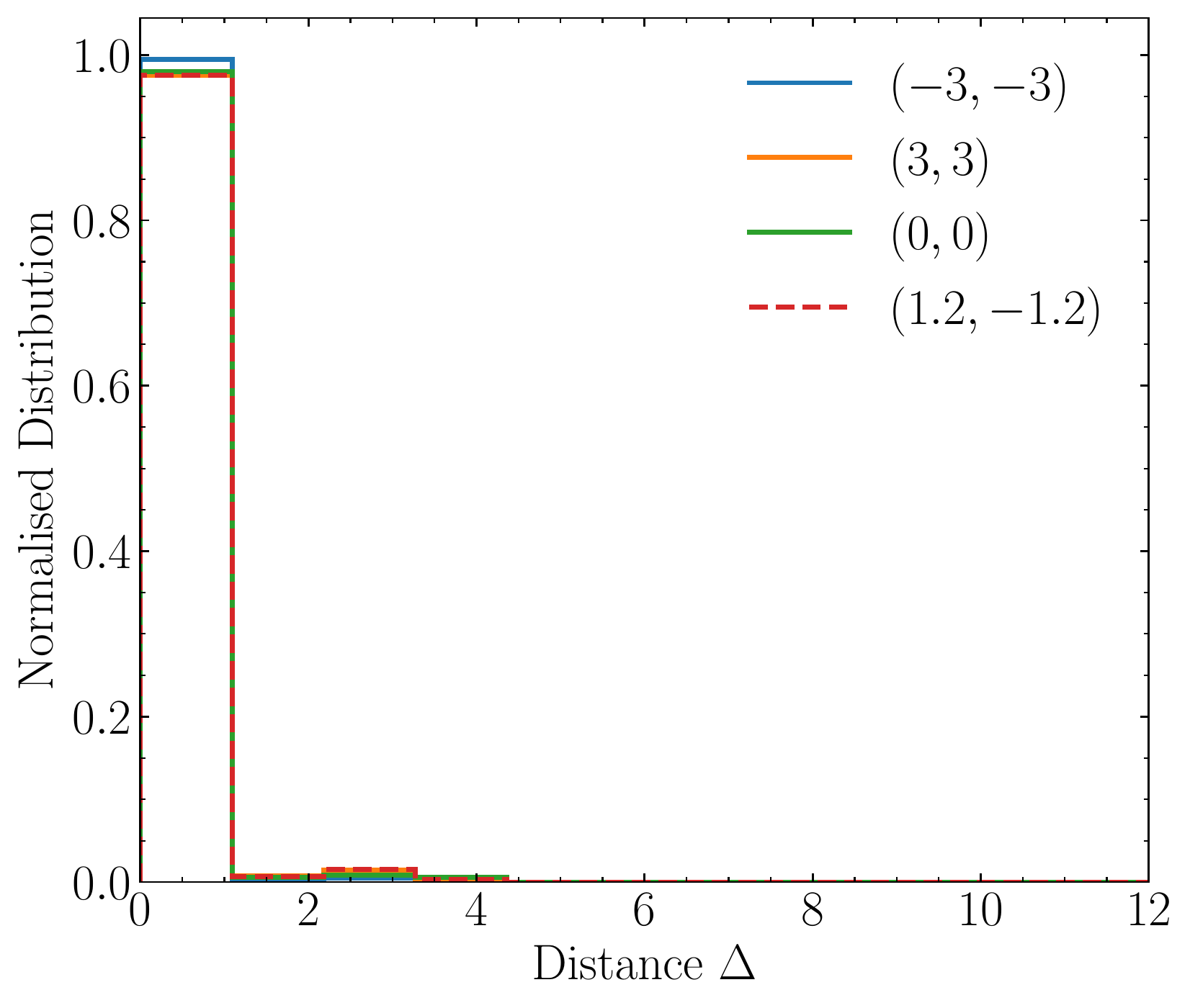}
		\caption{Quantum annealing}
	\end{subfigure}
	\caption{A comparison of the distance a predicted point is away from the true minimum. We choose to do this for four initial starting points using the volcano crater potential (potential 3). Four optimisation techniques are shown: Nelder-Mead (a), gradient descent (b), thermal annealing (c) and quantum annealing (d). Each initial condition is ran 100 times.}
	\label{fig:initial_distances_potential_3}
\end{figure*}

To visualise the performance differences of the four optimisers we choose four distinct initial conditions of $(\phi_{init}, \psi_{init})$, run each of them multiple times and plot the distribution of distances $\Delta$ in Figs.~\ref{fig:initial_distances}, \ref{fig:initial_distances_tanh} and \ref{fig:initial_distances_potential_3}. That is, similarly to what was done previously for each of the potentials in Figs.~\ref{fig:potential_1_results}, \ref{fig:potential_2_results} and \ref{fig:potential_3_results}. We again see that the classical methods are more dependent on their initial starting point, and the behaviour is in accord with what was concluded before: namely NA and GD {\it always} get stuck in the wrong minimum for some of the points, and only (but always) find the true minimum if they happen to start in its basin of attraction. On the other hand TA is reasonably good at jumping into the deepest well but gets stuck some of the time, while the QA almost always finds the deepest well but, at least for the limited $N$ system available to us, retains a small spread.  We would anticipate that larger $N$ would allow us to eliminate this final spread.

\bibliographystyle{inspire}
\bibliography{references,referencesSAMS}

\end{document}